\documentclass[a4paper,12pt]{report}  
\usepackage[latin1]{inputenc}   
\usepackage[brazil]{babel}      
\usepackage[dvips]{graphicx,color}
\usepackage{epsf}
\usepackage{color}
\usepackage{amsmath}
\usepackage{amssymb}
\usepackage{latexsym}
\newtheorem{postulado}{Postulado}

\newcommand{\be}{\begin{equation}}
\newcommand{\ee}{\end{equation}}
\newcommand{\beq}{\begin{eqnarray}}
\newcommand{\eeq}{\end{eqnarray}}
\newcommand{\beqa}{\begin{eqnarray*}}
\newcommand{\eeqa}{\end{eqnarray*}}
\newcommand{\bmt}{\begin{array}}
\newcommand{\emt}{\end{array}}

\begin{document}
\hyphenation{e-vi-den-te}

\thispagestyle{empty}
\centerline{\large{UNIVERSIDADE DE SÃO PAULO}}
\centerline{\large{INSTITUTO DE FÍSICA}}
\vskip 2.0cm
\centerline{\Large{\bf  Perturbação de Spin Zero no Espaço-Tempo  }}
\centerline{\Large{\bf de Kerr-Randall-Sundrum }} 
\vskip 1.0cm
\centerline{\large{}}
\vskip 0.3cm
\centerline{\textbf{ Jeferson de Oliveira\footnote{jeferson@fma.if.usp.br}}}
\vskip 1.0cm
\parbox{2.4in}{\mbox{}}

\vskip 0.05cm

\hfill \parbox{4in}{
Disserta\c c\~ao apresentada ao Instituto de Física da Uni\-versidade de São Paulo, 
para a obtenção do título de Mestre em Ciências.}
\vskip 1cm
\hfill {\bf Orientador:} {\textbf{Prof. Dr. Elcio Abdalla}}

\begin{flushleft}
{\bf{Comissão Examinadora:}}\\
Prof. Dr. Elcio Abdalla (IF-USP)\\
Prof. Dr. Paulo Teotônio Sobrinho (IF-USP)\\
Prof. Dr. Frank Michael Forger (IME-USP)
\end{flushleft}

\vskip 3.0cm
\centerline{São Paulo} 
\centerline{2006}

\newpage
\centerline{\large{RESUMO}}
\vskip 2.0cm
Esta dissertação visa realizar um estudo acerca dos modelos de mundo brana no contexto proposto por Randall e Sundrum. O trabalho focaliza as perturbações de spin-0 no espaço-tempo de Kerr tomado como um mundo brana $4-$dimensional. 

Para isso apresentamos os principais aspectos da Relatividade Geral de Einstein, bem como perturbações em métricas que descrevem buracos negros. Fizemos uma revisão dos modelos de Randall-Sundrum, suas motivações e tentativas de descrever buracos negros na brana. Por fim a perturbação escalar da corda negra em rotação (Kerr-Randall-Sundrum) e o fenômeno de super-radiação são analisados.

\newpage

\centerline{\large{ABSTRACT}}

\vskip 2.0cm
This dissertation aims at studying the braneworld models in the context proposed by Randall and Sundrum. The focus is on the spin-0 perturbations in the Kerr space-time as a $4-$dimensional braneworld.

The work deals the main aspects of Einstein General Relativity as well as perturbations of black holes metrics. We also review the Randall-Sundrum models and their motivations and attempts to describe braneworld black holes. In the end the Kerr-Randall-Sundrum black string scalar perturbation and superradiance are obtained.

\vskip 2.0cm

\newpage
\begin{flushright}
\end{flushright}  
\vspace{14cm}
\begin{flushright}
\begin{minipage}[t]{10cm}
{\it{Aos meus amores Sirlei, Ariana e João.}}
\end{minipage}  
\end{flushright}

\newpage

\centerline{\large{AGRADECIMENTOS}}
\vskip 2.0cm
Foram inúmeras as pessoas que contribuiram, seja de forma consciente ou não, para que eu pudesse terminar este trabalho. 

Ao meu orientador Prof. Elcio Abdalla, agradeço pelas conversas esclarecedoras e pela orientação ao longo desses dois últimos anos. 

À minha grande amiga Michele, pela ajuda inical quando em São Paulo cheguei. 

Ao {\it{povo}} do departamento de Física Matemática: Marcelo, Júlia, à minha inestimável amiga Arlene pelas nossas extasiantes conversas, Karlúcio, Carlos Molina e Roman Konoplya pelas boas discussões de trabalho. Agradeço em especial todos os que passaram pela sala $319$ nesses dois últimos anos, Flávio Henrique, Carlos Eduardo Pellicer, Rodrigo Dal Bosco Fontana e Alan Pavan, que sem a amizade e paciência nos dias de mau humor, que não foram poucos, o caminho para a conclusão deste trabalho teria sido muito mais difícil. Às meninas da secretaria do departamento: Amélia, Simone e Bety, que sem a ajuda, o trabalho na física matemática não seria tão bom. Aos Professores: Josif Frenkel, Raul Abramo e Coracci Pereira Malta, que contribuíram para minha formação durante minha passagem pelo Instituto de Física.

Agradecimentos especiais aos amigos do Crusp: Michela, João Basso, sobretudo aqueles que passaram pelo 210C: Admar Mendes e Leandro Ibiapina.    

Finalmente, dedico este trabalho àquelas pessoas que nunca desistiram de mim, nas situações mais críticas, onde parecia que nada parecia dar certo: aos meus pais João, Sirlei, minha irmã Ariana e meus tios Mathias e Cloé o meu sincero muito obrigado.

Agradeço à FAPESP pelo apoio financeiro.

\tableofcontents
\listoffigures
\listoftables
\chapter{Introdução}

A Teoria da Relatividade Geral de Einstein é a teoria física mais bem sucedida em explicar a estrutura do nosso Universo em larga escala. Em síntese, ela relaciona o conteúdo de energia dos campos com a configuração da curvatura do espaço-tempo onde são definidos esses campos. A equação que rege essa dinâmica é a equação de Einstein
\beq\label{0.1}
R_{\mu\nu}-\frac{1}{2}Rg_{\mu\nu}=\frac{8\pi G}{c^{4}}T_{\mu\nu},
\eeq 
sendo o conteúdo de energia representado pelo no lado direito da equação e o espaço-tempo correspondente no lado esquerdo. A descrição do Universo em larga escala é apenas uma das consequências dessas equações. Temos ainda, o desvio de raios de luz em regiões com campo gravitacional, a previsão da existência de ondas gravitacionais e de buracos negros. Neste texto, abordaremos as idéias fundamentais da Relatividade Geral e as soluções tipo buraco negro.

O tema central desta dissertação são os modelos de mundo brana propostos por Randall e Sundrum. Estes modelos são considerados como um teste para idéias fenomenológicas da teoria de supercordas, sobretudo a conjectura da teoria M, que considera as diversas teorias de cordas como diferentes limites de uma única teoria, batizada de teoria M. No trabalhos de Randall e Sundrum \cite{rs1} \cite{rs2}, o intuito foi de resolver o problema da hierarquia entre a escala eletrofraca e a gravitacional. Para isso, consideraram o nosso Universo como uma hipersuperfície $4-$dimensional, chamada de $3-$brana, mergulhada num espaço-tempo Anti-De Sitter $5-$dimensional ($AdS_{5}$), chamado de bulk. A métrica considerada é não fatorável e depende da coordenada da dimensão extra $y$ por uma função que varia rapidamente. Tal métrica é dada por
\beq\label{0.2}
ds^{2}=e^{\pm 2\kappa |y|}\eta_{\mu\nu}dx^{\mu}dx^{\nu}+dy^{2},
\eeq
sendo $\eta_{\mu\nu}$ é a métrica de Minkowski.

A escala $\kappa=1/l$, onde $l$ é a escala de curvatura do bulk, é determinada pela constante cosmológica $5-$dimensional e pela massa de Planck $5-$dimensional $M_{5}$,
\beq\label{0.3}
\kappa=\sqrt{-\frac{\Lambda_{5}}{6}},
\eeq
onde observamos que, necessariamente para este cenário, a constante cosmológica $5-$dimensional deve ser negativa.

Outro ponto desta dissertação, apresentado no último capítulo, é a análise das perturbações de campos não massivos de spin zero na métrica (\ref{0.2}), com $\eta_{\mu\nu}$ substituida pela métrica de Kerr. Este sistema descreve uma corda negra $5-$dimensional que intercepta a $3-$brana no espaço-tempo de Kerr $4-$dimensional. Mostraremos que estas perturbações tornam este cenário instável.


\chapter{Relatividade Geral e Buracos Negros}

Neste capítulo apresentaremos um resumo dos principais pontos da teoria física que melhor descreve a gravidade, que é a Relatividade Geral de Einstein. Pretendemos estudar as principais características dessa teoria, bem como as principais soluções tipo buraco negro previstas. 

\section{O papel da gravidade}

Até o presente momento, podemos descrever todos os sistemas físicos através de quatro interações fundamentais. São elas, em ordem de magnitude, a força nuclear forte, a força nuclear fraca, a força eletromagnética e a gravidade. A unificação da interação eletromagnética e da força nuclear fraca foi operada pela teoria de Salam-Weinberg. A unificação das interações restantes, num cenário coerente, talvez posssa ser realizada pela teoria de supercordas, num futuro não muito distante.
Quanto à magnitude das interações, a força gravitacional é de longe a mais fraca, tanto que a razão entre a interação gravitacional e a força elétrica entre dois elétrons é da ordem de $10^{-40}$. Apesar disso, tal interação é dominante sobre as outras no regime de larga escala espacial e é a força que afeta todas as partículas da mesma forma. Esta universalidade da interação gravitacional foi pela primeira vez estudada por Galileo, que descobriu que quaisquer dois corpos em queda livre têm a mesma velocidade. Esta descoberta levou Einstein a formular o princípio da equivalência, que é a base da Relatividade Geral. Existem duas formulações de tal princípio, a forma fraca apenas identifica a massa inercial com a massa gravitacional para partículas teste. Este princípio é verificado experimentalmente com grande precisão. A versão mais restritiva deste princípio, segundo  Weinberg \cite{weinberg}, diz que em qualquer ponto do espaço-tempo em um campo gravitacional arbitrário, é possível escolher localmente um sistema de coordenadas que pode ser chamado de inercial, tal que se considerarmos uma região suficientemente pequena em torno dos pontos em questão, as leis da natureza terão a mesma forma que as observadas em um sistema de coordenadas cartesiano em repouso na ausência de gravitação. 

Também é experimentalmente verificado que raios de luz são defletidos por campos gravitacionais, sendo que nenhum sinal pode viajar com velocidade maior que a velocidade de luz no vácuo, então a gravidade determina a estrutura causal do espaço-tempo. A interação gravitacional determina que eventos do espaço-tempo podem ser causalmente relacionados. Estas propriedades da gravidade criam situações de grande interesse, como quando houver uma grande quantidade de matéria numa pequena região. O primeiro estudo acerca de tal situação, no contexto da gravitação newtoniana, foi realizado por Laplace, cujo trabalho traduzido se encontra na referência \cite{hawking1}. Neste trabalho Laplace mostra que para um corpo com um raio 250 vezes maior que o do Sol com a mesma densidade, o campo gravitacional produzido seria tão forte que a velocidade de escape na superfície desse corpo seria maior que a velocidade da luz. Portanto esta ficaria aprisionada na superfície do corpo. Tal sistema foi batizado de buraco negro por John A. Wheleer na década de 1960, mas no contexto da Relatividade Geral. Discutiremos os buracos negros na próxima seção.

É importante, antes de prosseguirmos, definirmos alguns conceitos da geometria dife\-rencial, que serão de grande valia na formulação das quantidades físicas segundo a visão da Relatividade Geral, já que esta abdica do conceito de força gravitacional, substituíndo-a por uma configuração da curvatura do espaço-tempo. Seguiremos basicamente os textos clássicos sobre este tema \cite{hawking1} \cite{chandra} \cite{joshi}. 

 
\section{Variedade Espaço-Temporal e Equações de Eins\-tein}

Uma variedade $\mathcal{M}$ corresponde à idéia intuitiva que temos da continuidade do espaço-tempo. Esta continuidade é bem estabelecida, segundo experimentos de espalhamento de píons, até distâncias da ordem de $10^{-15}cm$. Apenas a noção de variedade não é suficiente para descrever o conjunto de todos os eventos que forma o espaço-tempo, ela será um dos ingredientes fundamentais do modelo. A outra quantidade fundamental é o objeto matemático que usamos para medir distâncias entre pontos de $\mathcal{M}$, o tensor de ordem 2 não degenerado simétrico chamado de tensor métrico ${\bf{g}}$. Então ao par $(\mathcal{M},{\bf{g}})$ chamamos de espaço-tempo.

Formalmente, uma variedade é um espaço que possui a propriedade de não distinguir sistemas de coordenadas, permitir a definição da diferenciação e localmente ser similar ao espaço euclidiano. Esta variedade deve ser do tipo Haussdorff, isto é, satisfazer o axioma da separação de Hausdorff: se $p, q$ são dois pontos distintos quaisquer em $\mathcal{M}$, então existem dois abertos disjuntos $\mathcal{U}$ e $\mathcal{V}$ em $\mathcal{M}$, tal que $p\in \mathcal{U}$, $q\in \mathcal{V}$. Além disso, $\mathcal{M}$ deve ser orientável, o que significa que existe apenas um parâmetro de tempo. Deve ser paracompacta, o que implica que a intersecção de dois abertos da variedade tenha tamanho finito.

Dadas estas características, podemos construir de maneira natural campos de funções, campos vetoriais e campos tensoriais. Uma função $f$ em $\mathcal{M}$ é definida como um mapa de $\mathcal{M}$ para a reta real $\mathbb{R}$. Campos tensoriais são equivalentes a um tensor definido em cada ponto da variedade. Uma forma de obter estas quantidades é partir do conceito de vetor em cada ponto de $\mathcal{M}$.

Dado o parâmetro $t$, podemos definir uma curva diferenciável $\lambda(t)$ como um mapa de um intervalo pertencente a reta real $\mathbb{R}$ em $\mathcal{M}$. Identificamos o mapa que leva uma função $f$ da curva $\lambda(t_{0})$ no número $\left(\partial f/ \partial t\right)_{\lambda}|_{t_{0}}$ ao objeto chamado vetor contravariante. De forma explicita,
\be\label{2.1}
\left(\frac{\partial f}{\partial t}\right)_{\lambda(t)}=\lim_{\epsilon\rightarrow 0} \frac{1}{s}\left[f(\lambda(t+s))-f(\lambda(t))\right]\quad,
\ee
é a derivada direcional. Esta expressão representa a derivada da função $f$ na direção da curva $\lambda(t)$ em relação ao parâmetro $t$. Podemos mostrar que o conjunto dos vetores contravariantes formam um espaço vetorial no ponto $p$ de $\mathcal{M}$ chamado de espaço tangente, denotado por $T_{p}$. Para realizar isto, consideremos um sistema de coordenadas locais no aberto que define a vizinhança do ponto $p$ de $\mathcal{M}$, dado pelo conjunto $\left(x^{1},\cdots, x^{n}\right)$, então
\be\label{2.2}\nonumber
\left(\frac{\partial f}{\partial t}\right)_{\lambda(t)}=\frac{d}{dt}x^{i}(\lambda(t))\arrowvert_{t_{0}}\frac{\partial f}{\partial x^{i}}\arrowvert_{\lambda(t_{0})}=\frac{dx^{i}}{dt}\frac{\partial f}{\partial x^{i}}\quad.
\ee
Podemos tomar como base para escrever qualquer vetor deste espaço tangente as derivadas em relação as coordenadas locais no ponto p. Então $T_{p}$ representa um espaço vetorial $n-$dimensional. Um vetor $V\in T_{p}$ pode ser tomado como uma flecha apontando na direção da curva $\lambda(t)$. 

Usando o espaço tangente $T_{p}$ podemos construir o espaço cotangente ou dual $T^{*}_{p}$ onde podemos definir os vetores covariantes. Tais vetores são funções lineares reais dos vetores de $T_{p}$, chamadas de 1-formas, denotadas por $\boldsymbol{w}$. Se $\mathbf{X}$ é um vetor em $p$, o número em que $\boldsymbol{w}$ mapeia $\mathbf{X}$ é escrito como $<\boldsymbol{w},\mathbf{X}>$, cuja linearidade de $\boldsymbol{w}$ implica em, sendo $\alpha$ e $\beta$ números reais,
\be\label{2.3}
<\boldsymbol{w},\alpha\mathbf{X}+\beta\mathbf{Y}>=\alpha<\boldsymbol{w},\mathbf{X}>+\beta<\boldsymbol{w},\mathbf{Y}>\quad .
\ee

Dada uma base ${e_{i}}$ em $p$, podemos definir um conjunto de $n$ 1-formas ${e^{i}}$, desde que $e^{i}$ mapeie qualquer vetor $\mathbf{X}$ no número $X^{i}$, que é a $i-$ésima componente do vetor $\mathbf{X}$ em relação a base ${e_{i}}$. Se escolhermos então
\be\label{2.4}
<e^{i},e_{j}>=\delta^{i}_{j}\quad,
\ee
e  se definirmos a combinação linear de 1-formas pela regra
\be\label{2.5}
<\alpha\boldsymbol{\omega}+\beta\boldsymbol{\eta},\mathbf{X}>=\alpha<\boldsymbol{\omega},\mathbf{X}>+\beta<\boldsymbol{\eta},\mathbf{X}>\quad,
\ee
teremos o conjunto ${e^{i}}$ como a base das 1-formas, isto é, qualquer 1-forma $\boldsymbol{\omega}$ em $p$ pode ser expressa como
\be\label{2.6}\nonumber
\boldsymbol{\omega}=<\boldsymbol{\omega}, e_{i}>e^{i}\quad .
\ee

Do conjunto de todas as 1-formas definidas no ponto $p$ de $\mathcal{M}$, formamos o espaço vetorial $n-$dimesional chamado de espaço cotangente $T^{*}_{p}$ ou também chamado de espaço dual a $T_{p}$. A base ${e_{a}}$ das 1-formas é a base dual a base ${e^{b}}$ dos vetores contravariantes. Usando estas construções podemos definir o diferencial $df$ de uma função $f$ como $1-$forma, que para qualquer vetor $\textbf{X}$, obedece à relação
\be\label{2.7}\nonumber
<df,\textbf{X}>=Xf\quad .
\ee
 Se usarmos um sistema de coordenadas locais $(x^{1},\cdots,x^{n})$ da mesma forma que o fizemos definindo derivada direcional, o conjunto de diferenciais $\left(dx^{1},\cdots.dx^{n}\right)$ no ponto $p$ forma uma base de $1-$formas, dual a base $(\partial/\partial x^{1},\cdots, \partial/\partial x^{n})$ dos vetores em $p$, sendo
\be\label{2.8}\nonumber
<dx^{i},\partial/\partial x^{j}>=\frac{\partial x^{i}}{\partial x^{j}}=\delta^{i}_{j}\quad,
\ee
como relação direta, obtemos a diferencial $df$, em termos da base, como
\be\label{2.9}\nonumber 
df=\frac{\partial f}{\partial x^{i}}dx^{i}\quad.
\ee

Partindo do espaço vetorial $T_{p}$ dos vetores contravariantes em $p$ e do seu espaço dual definido pelo espaço $T^{*}_{p}$ das $1-$formas ou vetores covariantes também em $p$, podemos formar o produto cartesiano $\Pi^{s}_{r}$ de $r$ espaços $T^{*}_{p}$ com $s$ espaços tangentes $T_{p}$ como
\be\label{3}
\Pi^{s}_{r}=\underbrace{T^{*}_{p}\times T^{*}_{p}\times\cdots\times T^{*}_{p}}_{r\hspace{0.1cm} termos}\times\underbrace{T_{p}\times T_{p}\times\cdots\times T_{p}}_{s\hspace{0.1cm} termos}\quad.
\ee
O produto cartesiano pode ser colocado como um conjunto ordenado de vetores contravariantes e covariantes $\left(\boldsymbol{\eta^{1}},\boldsymbol{\eta^{2}},\cdots,\boldsymbol{\eta^{r}},\mathbf{Y_{1}},\cdots,\mathbf{Y_{s}}\right)$.

Definimos como tensor do tipo $(r,s)$ em um ponto $p$ de $\mathcal{M}$,  a função de $\Pi^{r}_{s}$ que é linear em cada um dos seus argumentos. O número que o tensor $\mathbf{T}$ do tipo $(r,s)$ mapeia o elemento $\left(\boldsymbol{\eta^{1}},\boldsymbol{\eta^{2}},\cdots,\boldsymbol{\eta^{r}},\mathbf{Y_{1}},\cdots,\mathbf{Y_{s}}\right)$ de $\Pi^{r}_{s}$ é $T\left(\boldsymbol{\eta^{1}},\boldsymbol{\eta^{2}},\cdots,\boldsymbol{\eta^{r}},\mathbf{Y_{1}},\cdots,\mathbf{Y_{s}}\right)$\quad. 

Da lienaridade de $\mathbf{T}$, para todo $\alpha$, $\beta\in R^{1}$ e todo $\mathbf{X}$, $\mathbf{Y}\in T_{p}$, temos
\beq\label{2.10}\nonumber
T\left(\boldsymbol{\eta^{1}},\cdots,\boldsymbol{\eta^{r}},\alpha\mathbf{X}+\beta\mathbf{Y},\mathbf{Y_{2}},\cdots,\mathbf{Y_{s}}\right)&=&\alpha . T\left(\boldsymbol{\eta^{1}},\cdots,\boldsymbol{\eta^{r}},\mathbf{X},\mathbf{Y_{2}}\cdots,\mathbf{Y_{s}}\right)\\\nonumber
&+&\beta . T\left(\boldsymbol{\eta^{1}},\cdots,\boldsymbol{\eta^{r}},\mathbf{Y}, \mathbf{Y_{2}},\cdots,\mathbf{Y_{s}}\right).
\eeq

A operação chamada de produto tensorial $\otimes$  definida como
\be\label{2.11}\nonumber
T^{r}_{s}(p)=\underbrace{T_{p}\otimes\cdots\otimes T_{p}}_{r\hspace{0.1cm}termos}\otimes\underbrace{ T^{*}_{p}\otimes\cdots\otimes T^{*}_{p}}_{s\hspace{0.1cm}termos}\quad,
\ee
forma uma álgebra cujos elementos são todos espaços tensoriais definidos em um ponto $p\in\mathbf{M}$.

Outras operações algébricas podem ser construídas de forma natural, tal como a adição de tensores $(\mathbf{T}+\mathbf{T'})$
\beq\label{2.12}\nonumber
 (\mathbf{T}+\mathbf{T'})\left(\boldsymbol{\eta^{1}},\cdots,\boldsymbol{\eta^{r}},\mathbf{Y_{1}},\cdots,\mathbf{Y_{s}}\right)&=&T\left(\boldsymbol{\eta^{1}},\cdots,\boldsymbol{\eta^{r}},\mathbf{Y_{1}},\cdots,\mathbf{Y_{s}}\right)\\
&+& T^{'}\left(\boldsymbol{\eta^{1}},\cdots,\boldsymbol{\eta^{r}},\mathbf{Y_{1}},\cdots,\mathbf{Y_{s}}\right)\quad,
\eeq
a multiplicação de tensores por um escalar $\alpha\in R^{1}$
\be\label{2.13}
(\alpha T)\left(\boldsymbol{\eta^{1}},\cdots,\boldsymbol{\eta^{r}},\mathbf{Y_{1}},\cdots,\mathbf{Y_{s}}\right)\quad.
\ee
Com as operações (\ref{2.12}) e (\ref{2.13}), o produto tensorial $T^{r}_{s}(p)$ forma um espaço vetorial de dimensão $n^{r+s}$ sobre a reta real $R^{1}$. 

Se $\{e_{a}\}$, $\{e^{a}\}$ são bases de $T_{p}$ e  $T^{*}_{p}$, respectivamente, então os produtos
\be\label{2.14}\nonumber
\left\{e_{a_{1}}\otimes\cdots\otimes e_{a_{r}}\otimes e^{b_{1}}\otimes\cdots\otimes e^{b_{s}}\right\}\quad,
\ee
formam uma base para o espaço do produto tensorial $T^{r}_{s}(p)$. Então qualquer tensor $\mathbf{T}\in T^{r}_{s}(p)$ pode ser expresso em termos dessa base como
\be\label{2.15}\nonumber
\mathbf{T}=T^{a_{1}\cdots a_{r}}_{b_{1}\cdots b_{s}}e_{a_{1}}\otimes\cdots\otimes e_{a_{r}}\otimes e^{b_{1}}\otimes\cdots\otimes e^{b_{s}}\quad,
\ee
onde $\{T^{a_{1}\cdots a_{r}}_{b_{1}\cdots b_{s}}\}$ são as componentes de $\mathbf{T}$ com respeito as bases $\{e_{a}\}$, $\{e^{a}\}$. 

Outra importante operação algébrica usando tensores, é a contração de índices. Esta operação é independente da base usada. Dado um tensor $\mathbf{T}$ do tipo $(r,s)$ com componentes $\{T^{ab\cdots d}_{ef \cdots g}\}$, a contação no primeiro índice contravariante e no primeiro índice covariante é definida como o tensor $B^{1}_{1}(\mathbf{T})$ do tipo $(r-1,s-1)$, cujas componentes com relação as bases $\{e_{a}\}$, $\{e^{a}\}$, são $T^{ab\cdots d}_{af\cdots g}$, então
\be\label{2.16}\nonumber
B^{1}_{1}(\mathbf{T})=T^{ab\cdots d}_{af\cdots g}e_{b}\otimes\cdots\otimes e_{d}\otimes e^{f}\otimes\cdots\otimes e^{g}\quad.
\ee

Denotamos a parte simétrica das componentes do tensor $\mathbf{T}$ por
\be\label{2.17}\nonumber
T^{(ab)}=\frac{1}{2!}\left\{ T^{ab}+T^{ba}\right\}\quad,
\ee
e a parte anti-simétrica
\be\label{2.17}
T^{[ab]}=\frac{1}{2!}\left\{ T^{ab}-T^{ba}\right\}\quad.
\ee

Um subconjuntto do espaço tensorial, é aquele formado pelos tensores do tipo $(0,q)$, totalmente antisimétricos nas posições do índice $q$ (com $q\leq n$). Estes tensores são chamados de $q-$formas. Se $\mathbf{A}$ e $\mathbf{B}$ são $p-$ e $q-$formas respectivamente, podemos construir, a partir destes, a $(p+q)-$forma $\mathbf{A}\wedge\mathbf{B}$, onde $\wedge$ representa a contraparte anti-simétrica do produto tensorial $\otimes$. O tensor $\mathbf{A}\wedge\mathbf{B}$ é do tipo $(0,p+q)$ com componentes dadas por
\be\label{2.18}\nonumber
\left(A\wedge B\right)_{a\cdots bc\cdots f}=A_{[a\cdots b}B_{c\cdots f]}\quad.
\ee
Esta expressão implica que
\be\label{2.19}
\left(\mathbf{A}\wedge \mathbf{B}\right)=(-1)^{pq}A_{[a\cdots b}B_{c\cdots f]}\quad.
\ee
O conjunto de todas $p-$ e $q-$formas munidas da operação (\ref{2.19}) constituem uma álgebra de Grassmann para as formas.

Até aqui definimos as mais importantes operações algébricas dos espaços tensoriais. Para que possamos usar tais objetos para construir quantidades dinâmicas precisamos estudar operadores diferenciais na variedade. Neste trabalho não pretendemos esgotar este assunto, para mais detalhes ver \cite{lovelock}. Estudaremos, três operadores de particular interesse, dois deles, a derivada exterior e a derivada de Lie, só necessitam para serem definidos da própria variedade $\mathcal{M}$. O terceiro operador, a derivada covariante, precisa de uma estrutura adicional que é a conexão $\nabla$ de $\mathcal{M}$. 

O operador $d$ que determina a operação chamada de derivada exterior, mapeia as $r-$formas de maneira linear em $(r+1)-$ formas. Se operarmos com $d$ na $r-$forma 
\be\label{2.20}\nonumber
\mathbf{A}=A_{ab\cdots d}dx^{a}\wedge dx^{b}\wedge\cdots\wedge dx^{d},
\ee
obtemos
\be\label{2.21}\nonumber
d\mathbf{A}=dA_{ab\cdots d}\wedge dx^{a}\wedge dx^{b}\wedge\cdots\wedge dx^{d},
\ee
que é uma $(r+1)-$forma independente do sistema de coordenadas $\{x^{a}\}$ usado. Da definição do produto $\wedge$, temos que
\be\label{2.22}
d\left(d\mathbf{A}\right)=0,
\ee
para toda $r-$forma $\mathbf{A}$. 

O segundo tipo de diferenciação que é definido naturalmente da estrutura da variedade é a derivada de Lie. Antes de passar a esta operação, devemos definir o parênteses de Lie.

Dados dois vetores, $\mathbf{X}$ e $\mathbf{Y}$, seu parênteses de Lie, $[\mathbf{X},\mathbf{Y}]$, é definido pela ação deste numa $0-$forma  da seguinte maneira
\be\label{2.24}\nonumber
[\mathbf{X},\mathbf{Y}]f=\left(\mathbf{X}\mathbf{Y}-\mathbf{Y}\mathbf{X}\right)f=\mathbf{X}\left(\mathbf{Y}f\right)-\mathbf{Y}\left(\mathbf{X}f\right).
\ee
Sendo $f$ e $g$ $0-$formas, $\alpha$ e $\beta$ números reais, o parênteses de Lie de dois vetores do espaço tangente aplicado numa combinação das $0-$formas, tem como resultado também vetores deste espaço, isto é,
\beq\label{2.25}
[\mathbf{X},\mathbf{Y}](\alpha f+ \beta g)&=&\alpha[\mathbf{X},\mathbf{Y}]f + \beta[\mathbf{X},\mathbf{Y}]g,\\\label{2.26}
[\mathbf{X},\mathbf{Y}](fg)&=&g[\mathbf{X},\mathbf{Y}]f+f[\mathbf{X},\mathbf{Y}]g
\eeq
A relaçao (\ref{2.25}) estabelece o parênteses de Lie como um operador linear, e a relação (\ref{2.26}) como uma operação de diferenciação. 
Podemos expressar esta operação em forma de componentes, já que o parênteses de Lie de vetores no espaço tangente também é um vetor no espaço tangente. Aplicando no sistema de coordenadas local $\{x^{a}\}$, obtemos
\be\label{2.27}\nonumber
[\mathbf{X},\mathbf{Y}]x^{a}=X^{b}Y^{a}{}_{,b}-Y^{b}X^{a}{}_{,b}.
\ee
Consideremos o parênteses de Lie $[\mathbf{X},\mathbf{Y}]$ como uma operação de diferenciação, chamada de derivada de Lie do vetor $\mathbf{Y}$ na direção de $\mathbf{X}$, denotada por
\be\label{2.28}
\mathfrak{L}_{\mathbf{X}}\mathbf{Y}=[\mathbf{X},\mathbf{Y}]=-[\mathbf{Y},\mathbf{X}]=-\mathfrak{L}_{\mathbf{Y}}\mathbf{X}.
\ee

A derivada de Lie $\mathfrak{L}_{\mathbf{X}}\mathbf{T}$ de um tensor $\mathbf{T}$ do tipo $(r,s)$ em relação a $\mathbf{X}$ é um tensor também do tipo $(r,s)$. Esta diferenciação, além de preservar o tipo dos tensores, também preserva a contração e obedece à regra de Leibniz do cálculo 
\be\label{2.30}
\mathfrak{L}_{\mathbf{X}}\left(\mathbf{S}\otimes\mathbf{T}\right)=\mathfrak{L}_{\mathbf{X}}(\mathbf{S})\otimes\mathbf{T}+ \mathbf{S}\otimes\mathfrak{L}_{\mathbf{X}}\mathbf{T}
\ee
As componentes da derivada de Lie das $1-$formas $\omega$ pode ser obtida contraindo a relação (\ref{2.30}), obtendo
\be\label{2.31}
\mathfrak{L}_{\mathbf{X}}<\boldsymbol{\omega},\mathbf{Y}>=<\mathfrak{L}_{\mathbf{X}}\boldsymbol{\omega},\mathbf{Y}>+<\boldsymbol{\omega},\mathfrak{L}_{\mathbf{X}}\mathbf{Y}>.
\ee
Escolhendo $\mathbf{Y}$ com a base vetorial $e_{a}=\partial/\partial x^{a}$, temos em forma de componentes,
\be\label{2.32}\nonumber
\left(\mathfrak{L}_{\mathbf{X}}\boldsymbol{\omega}\right)_{a}=\left(\frac{\partial \omega_{a}}{\partial x^{b}}X^{b}\right)+\omega_{b}\left(\frac{\partial X^{b}}{\partial x^{j}}\right),
\ee
já que
\be\label{2.33}\nonumber
\left(\mathfrak{L}_{\mathbf{X}}\partial/\partial x^{a}\right)^{b}=-\frac{\partial X^{b}}{\partial x^{a}}.
\ee
Seguindo a mesma receita, podemos obter as componentes da derivada de Lie de qualquer tensor $\mathbf{T}$ do tipo $(r,s)$ usando a regra de Leibniz em
\be\label{2.34}\nonumber
\mathfrak{L}_{\mathbf{X}}\left(\mathbf{T}\otimes e^{a}\otimes\cdots\otimes e^{d}\otimes e_{e}\otimes\cdots\otimes e_{g}\right),
\ee
e então contraindo em todas posições, encontramos
\beq\label{2.35}\nonumber
\left(\mathfrak{L}_{\mathbf{X}}\mathbf{T}\right)^{ab\cdots d}{}_{ef\cdots g}&=&\left(\frac{\partial }{\partial x^{i}}T^{ab\cdots d}{}_{ef\cdots g}\right)X^{i}-T^{ib\cdots d}{}_{ef\cdots g}\frac{\partial X^{a}}{\partial x^{i}}-\cdots\\
&+&T^{ab\cdots d}{}_{if\cdots g}\frac{\partial X^{i}}{\partial x^{e}}+\cdots.
\eeq
A derivada de Lie comuta com o operador $d$ da derivada exterior, isto é
\be\label{2.36}\nonumber
d\left(\mathfrak{L}_{\mathbf{X}}\boldsymbol{\omega}\right)=\mathfrak{L}_{\mathbf{X}}\left(d\boldsymbol{\omega}\right)\quad .
\ee

Podemos interpretar a derivada de Lie $\mathfrak{L}_{\mathbf{X}}\mathbf{T}|_{p}$ de um tensor $\mathbf{T}$ do tipo $(r,s)$ no ponto $p$ de $\mathcal{M}$ não apenas como a derivada de $\mathbf{T}$ na direção do campo vetorial $\mathbf{X}$ no ponto $p$, mas também como a derivada direcional ao longo da {\it{vizinhança}} do ponto $p$.

Os dois operadores diferenciais que definimos pela estrutura da variedade, a derivada exterior e a derivada de Lie, são limitados para servirem de generalização para o conceito de derivada parcial. O operador $d$ age apenas sobre formas, enquanto que derivadas parciais ordinárias são derivadas direcionais que dependem apenas da direção no ponto em questão, diferentemente da derivada de Lie , que também depende da vizinhança do ponto considerado. 

A generalização da derivada parcial ordinária é obtida pela adição de uma nova estrutura em $\mathcal{M}$, chamada de conexão e denotada por $\nabla$. Esta estrutura nos permite definir a operação de derivação covariante.

A conexão $\nabla$ no ponto $p$ de $\mathcal{M}$ é uma receita que associa cada vetor $\mathbf{X}$ em $p$ a um operador diferencial $\nabla_{\mathbf{X}}$. Este operador mapeia todo vetor $\mathbf{Y}$ no tensor $\nabla_{\mathbf{X}}\mathbf{Y}$, isto é, para quaisquer  funções $f$, $g$ e vetores $\mathbf{X}$, $\mathbf{Y}$, $\mathbf{Z}$ temos
\be\label{2.37}
\nabla_{f\mathbf{X}+g\mathbf{Y}}\mathbf{Z}=f\nabla_{\mathbf{X}}\mathbf{Z}+g\nabla_{\mathbf{Y}}\mathbf{Z}\quad .
\ee
apresenta a propriedade de linearidade
\be\label{2.38}\nonumber
\nabla_{\mathbf{X}}\left(\alpha\mathbf{Y}+\beta\mathbf{Z}\right)=\alpha\nabla_{\mathbf{X}}\mathbf{Y}+\beta\nabla_{\mathbf{X}}\mathbf{Z}\quad ,
\ee
e
\be\label{2.39}\nonumber
\nabla_{\mathbf{X}}\left(f\mathbf{Y}\right)=X(f)\mathbf{Y}+f\nabla_{\mathbf{X}}\mathbf{Y}\quad .
\ee
Destas propriedades definimos $\nabla\mathbf{Y}$ como a derivada covariante de $\mathbf{Y}$, como um tensor do tipo $(1,1)$, que quando contraído com $\mathbf{X}$ produz o vetor $\nabla_{\mathbf{X}}\mathbf{Y}$. De (\ref{2.37}) temos
\be\label{2.40}
\nabla\left(f\mathbf{Y}\right)=df\otimes\mathbf{Y}+f\nabla\mathbf{Y}.
\ee
As componentes de $\nabla_{\mathbf{X}}\mathbf{Y}$ são obtidas se escolhemos como base as bases duais $e_{a}$ e $e^{a}$, logo
\be\label{2.41}
\nabla_{\mathbf{X}}\mathbf{Y}=\nabla_{\mathbf{X}}\left(Y^{a}e_{a}\right)=\left(\mathbf{X}Y^{a}\right)e_{a}+Y^{a}\nabla_{\mathbf{X}}e_{a}\quad,
\ee

Se escolhermos a base $e_{a}$ como um tensor do tipo $(1,0)$, a expressão acima pode ser escrita na forma
\be\label{2.42}
\nabla_{\mathbf{X}}e_{a}=\omega^{b}_{a}\left(\mathbf{X}\right)e_{b}\quad ,
\ee
sendo $\omega^{b}_{a}$ $1-$formas. Com isso (\ref{2.41}) fica como
\be\label{2.43}
\nabla_{\mathbf{X}}\mathbf{Y}=\left(\mathbf{X}Y^{a}\right)e_{a}+Y^{a}\omega^{b}_{a}(\mathbf{X})e_{b}\quad ,
\ee
ou
\beq\label{2.44}\nonumber
\nabla_{\mathbf{X}}\mathbf{Y}&=&\left(\mathbf{X}Y^{a}\right)e_{a}+Y^{a}X^{c}\nabla_{e_{k}}e_{a}\\
&=&\left(\mathbf{X}Y^{a}\right)e_{a}+Y^{a}X^{c}\omega^{b}_{ac}e_{b}\quad ,
\eeq
sendo $\omega^{b}_{a}e_{c}=\omega^{b}_{ac}$ os coeficientes de $e^{c}$ quando expandimos a $1-$forma $\omega^{c}_{b}$ na base $\{e^{c}\}$. Então temos que a conexão $\nabla$ no ponto $p$ de $\mathcal{M}$ é determinada por estes coeficientes $\omega^{c}_{ab}$.

Reescrevendo (\ref{2.43}) como
\beq\label{2.45}\nonumber
\nabla_{X}\mathbf{Y}=\left[\mathbf{X}Y^{a}+\omega^{a}_{c}(\mathbf{X})Y^{c}\right]e_{a}\quad ,
\eeq
concluimos que
\beq\label{2.46}\nonumber
\left(\nabla_{\mathbf{X}}\mathbf{Y}\right)^{a}=\mathbf{X}Y^{a}+\omega^{a}_{c}(\mathbf{X})Y^{c}\quad .
\eeq
Tomando novamente a  base local $(\partial/\partial x^{c})$, denotando $\omega^{c}_{ab}$ por $\Gamma^{c}_{ab}$ e o sinal de ponto-e-vírgula como a operação de derivada covariante, temos a forma padrão desta operação
\be\label{2.47}
Y^{a}{}_{;c}=Y^{a}{}_{,c}+Y^{b}\Gamma^{a}{}_{bc}\quad .
\ee
Generalizando para qualquer tensor $\mathbf{X}$ do tipo $(r,s)$, temos
\beq\label{2.48}\nonumber
X^{ab\cdots d}{}_{ef\cdots g;h}&=&X^{ab\cdots d}{}_{ef\cdots g,h}+\Gamma^{a}{}_{hj}X^{jb\cdots d}{}_{ef\cdots g}+\cdots\\
&-&\Gamma^{j}{}_{he}X^{ab\cdots}{}_{jf\cdots g}-\cdots\quad .
\eeq

Dada a conexão $\nabla$, para quaisquer vetores $\mathbf{X}$, $\mathbf{Y}$, podemos definir um tensor $\mathbf{T}$ do tipo $(1,2)$
\beq\label{2.55}\nonumber
\mathbf{T}(\mathbf{X},\mathbf{Y})=\nabla_{\mathbf{X}}\mathbf{Y}-\nabla_{\mathbf{Y}}\mathbf{X}-\left[\mathbf{X},\mathbf{Y}\right]\quad ,
\eeq
chamado de {\it{torção}}, cuja expressão na base coordenada é 
\beq\label{2.56}\nonumber
T^{i}{}_{jk}=\Gamma^{i}{}_{jk}-\Gamma^{i}{}_{kj}\quad .
\eeq

Em nosso estudo apenas usaremos conexões livres de torção, isto é, $\mathbf{T}=0$, o que implica na simetria das componentes da conexão em relação aos índices covariantes, $\Gamma^{i}{}_{jk}=\Gamma^{i}{}_{kj}$. Uma conexão é livre de torção se e apenas se $f_{;ij}=f_{;ji}$ para todas funções $f$.

Neste cenário de torção zero, podemos relacionar a derivada covariante com a derivadada de Lie e como a derivada exterior de um tensor $\mathbf{T}$ do tipo $(r,s)$. Estas relações, com respeito a base coordenada são, respectivamente
\beq\label{2.57}\nonumber
\left(\mathfrak{L}_{\mathbf{X}}\mathbf{T}\right)^{ab\cdots d}{}_{ef\cdots g}&=&T^{ab\cdots d}{}_{ef\cdots g;h}X^{h}-T^{jb\cdots d}{}_{ef\cdots g}X^{a}{}_{;j}\\
&-&\cdots + T^{AB\cdots d}{}_{if\cdots g}X^{j}{}_{;e}+\cdots \quad ,\\
(dA)_{a\cdots cd}&=&(-1)^{p}A_{[a\cdots c;d]}\quad .
\eeq
Uma aplicação da derivada covariante é a obtenção das expressões para o {\it{transporte paralelo}} de um tensor $\mathbf{X}$ e da equação para as {\it{curvas geodésicas}}, que são as equações de movimento de uma partícula em queda livre no espaço-tempo determinado pela variedade $\mathcal{M}$.

Seja o tensor $\mathbf{X}$ definido ao longo da curva $\lambda(t)$ de $\mathcal{M}$, então podemos definir a derivada covariante de $\mathbf{X}$ ao longo desta curva $\lambda(t)$, como $\nabla_{\partial/\partial t} \bar{X}$, sendo $\bar{X}$ a extensão de $\mathbf{X}$ nos abertos dos pontos que formam a curva $\lambda(t)$, isto é na vizinhança de $\lambda(t)$. Um caso particular ocorre quando consideramos tal derivada covariante independente desta extensão $\bar{X}$, denotamos esta operação por $DX^{a}/\partial t$, que é um tensor definido em toda curva $\lambda(t)$. Em termos das componentes, se $\mathbf{Y}$ é um vetor definido em $T_{p}$ ao longo de $\lambda(t)$, temos
\beq\label{2.49}\nonumber
DX^{a}/\partial t= \frac{\partial X^{a}}{\partial t}+\Gamma^{a}_{bc}X^{c}\frac{dx^{b}}{dt}\quad .
\eeq

Escolhendo como coordenadas para $\lambda(t)$ o conjunto $\{x^{a}(t)\}$ e $\mathbf{X}=dx^{a}/dt$ temos
\beq\label{2.50}
DX^{a}/\partial t=\frac{\partial X^{a}}{\partial t}+\Gamma^{a}_{bc}\frac{dx^{b}}{dt}\frac{dx^{c}}{dt}\quad .
\eeq

Diz-se que um tensor $\mathbf{X}$ foi transportado paralelamente ao longo de $\lambda(t)$ se
\beq\label{2.51}\nonumber
D\mathbf{X}/\partial t=0\quad .
\eeq
Se a curva $\lambda(t)$ começa no ponto $p$ e termina no ponto $q$, e se a conexão $\nabla$ é pelo menos diferenciável uma vez, então pela teoria das equações diferenciais deve-se obter um único tensor no ponto $q$ através do transporte paralelo de qualquer dado tensor em $p$ ao longo da curva $\lambda(t)$. O transporte paralelo é um mapa linear que leva os elementos de $T^{r}_{s}(p)$ em $T^{r}_{s}(q)$ preservando o produto tensorial e a operação de contração.

Uma curva $\lambda$ de $\mathcal{M}$ é uma curva geodésica, se o resultado do transporte paralelo de um vetor tangente a ela for um múltiplo deste mesmo vetor. Sendo $dx^{j}(\lambda(t))/dt$ um vetor tangente a $\lambda(t)$, então seu transporte paralelo, segundo a condição geodésica, fica
\beq\label{2.52}
\frac{d^{2}x^{j}}{dt^{2}}+\Gamma^{j}_{lk}\frac{dx^{l}}{dt}\frac{dx^{k}}{dt}=\phi(t)\frac{dx^{j}}{dt}\quad ,
\eeq
sendo $\phi(t)$ uma função bem definida de $t$. Em vez de $t$, se usarmos o parâmetro
\beq\label{2.53}\nonumber
s=\int^{t}dt''\exp\left[\int^{t''}dt'\phi(t')\right]\quad ,
\eeq
a equação (\ref{2.52}) fica 
\beq\label{2.54}
\frac{d^{2}x^{j}}{ds^{2}}+\Gamma^{j}{}_{lk}\frac{dx^{l}}{ds}\frac{dx^{k}}{ds}=0\quad ,
\eeq
sendo $s$ chamado de parâmetro afim. A única liberdade na escolha de $s$ é sua origem e sua escala.

Ao tensor que mede a não comutatividade dos operadores da segunda derivada covariante (aplicação da derivada covariante duas vezes), chamamos de {\it{tensor de Riemann}} ou {\it{tensor de curvatura}}. Se consideramos um dado ponto $p$ da curva fechada $\gamma$ como o ponto inicial do transporte paralelo do vetor $\mathbf{X}_{p}$, então o resultado disto será o vetor $\mathbf{X'}_{p}$, que em geral é diferente de $\mathbf{X}_{p}$, além disso, se considerarmos outra curva fechada $\gamma'$ passando por $p$ e realizarmos a mesma operação, o vetor obtido $\mathbf{X''}_{p}$ será diferente de $\mathbf{X}_{p}$ e $\mathbf{X'}_{p}$. Esta não integrabilidade do transporte paralelo é a manifestação da não comutatividade da segunda derivada covariante. Dada um a conexão $\nabla$, os vetores $\mathbf{X}$, $\mathbf{Y}$,$\mathbf{Z}$, definimos o tensor de Riemann como 
\beq\label{2.58}
\mathbf{R}(\mathbf{X},\mathbf{Y})\mathbf{Z}=\nabla_{\mathbf{X}}\left(\nabla_{\mathbf{Y}}\mathbf{Z}\right)-\nabla_{\mathbf{Y}}\left(\nabla_{\mathbf{X}}\mathbf{Z}\right)-\nabla_{[\mathbf{X},\mathbf{Y}]}\mathbf{Z}\quad .
\eeq

O tensor $\mathbf{R}(\mathbf{X},\mathbf{Y})\mathbf{Z}$ é linear em relação a $\mathbf{X}$, $\mathbf{Y}$,$\mathbf{Z}$, ou seja, depende apenas do valor destes vetores no ponto $p$ formando um tensor do tipo $(3,1)$. Reescrevendo (\ref{2.58}) em componentes, obtemos
\beq\label{2.59}\nonumber
R^{a}{}_{bcd}X^{c}Y^{d}Z^{b}&=&\left(Z^{a}{}_{;d}Y^{d}\right)_{;c}X^{c}-\left(Z^{a}{}_{;d}X^{d}\right)_{;c}Y^{c}\\\nonumber
&-&Z^{a}{}_{;d}\left(Y^{a}{}_{;d}X^{c}-X^{d}{}_{;c}Y^{c}\right)\\
&=&\left(Z^{a}{}_{;dc}-Z^{a}{}_{;cd}\right)X^{c}Y^{d}\quad .
\eeq
Já que $\mathbf{X}$ e $\mathbf{Y}$ são vetores arbitrários, podemos expressar a não comutatividade da segunda derivada covariante como
\beq\label{2.60}\nonumber
R^{a}{}_{bcd}Z^{b}=Z^{a}{}_{;dc}-Z^{a}{}_{;cd}\quad .
\eeq

Com relação às bases duais $\{e_{a}\}$, $\{e^{a}\}$, as componentes do tensor de Riemann são
\beq\label{2.61}\nonumber
R^{a}{}_{bcd}= <e^{a},\mathbf{R}(e_{c},e_{d})e_{b}>\quad .
\eeq
Escolhendo as bases como bases coordenadas, obtemos a expressão para o tensor de Riemann em termos das componentes coordenadas da conexão $\nabla$
\beq\label{2.62}
R^{a}{}_{bcd}=\frac{\partial\Gamma^{a}{}_{db}}{\partial x^{c}}-\frac{\partial\Gamma^{a}{}_{cb}}{\partial x^{d}}+\Gamma^{a}{}_{cf}\Gamma^{f}{}_{db}-\Gamma^{a}{}_{df}\Gamma^{f}{}_{cb}\quad .
\eeq

Destas expressões seguem as propriedades de simetria do tensor de Riemann
\beq\label{2.63}\nonumber
R^{a}{}_{b(cd)}=0,\\\nonumber
R^{a}{}_{[bcd]}=0\quad .
\eeq
Além disto, a derivada covariante primeira satisfaz as chamadas {\it{identidades de Bianchi}},
\beq\label{2.64}\nonumber
R^{a}{}_{b[cd;e]}=0\quad .
\eeq

O tensor de Ricci e o escalar de curvatura são obtidos por contrações do tensor de Riemann. Esses são dados, respectivamente, por
\beq\label{2.73}\nonumber
R_{ab}&=&g^{cd}R_{acbd}\quad ,\\\nonumber
R&=&g^{ab}R_{ab}\quad .
\eeq

O transporte paralelo de um vetor em uma curva fechada arbitrária é localmente integrável apenas se $R^{a}{}_{bcd}=0$ para todos os pontos da variedade $\mathcal{M}$. Neste caso, dizemos que a conexão é plana.  

Para construírmos consistentemente o espaço-tempo lorentziano, que é o objeto de estudo da Relatividade Geral, precisamos de mais um objeto geométrico, o tensor simétrico não singular do tipo $(0,2)$ chamado de {\it{tensor métrico}} ou simplesmente {\it{métrica}} e denotado por $\mathbf{g}$.
As componentes $g_{ab}$ da métrica $\mathbf{g}$ são dados simplesmente pelo produto escalar dos elementos da base vetorial $\{e_{a}\}$. Se usarmos a base coordenada $\{\partial/\partial x^{a}\}$ teremos
\beq\label{2.65}
\mathbf{g}=g_{ab}dx^{a}\otimes dx^{b}\quad .
\eeq

Os comprimentos definidos pela métrica estão em $T_{p}$ e estão relacionados com os comprimentos obtidos pela estrutura de $\mathcal{M}$. Considerando dois pontos $p$ e $q$ da curva $\gamma(t)$ na qual o vetor tangente $\partial/\partial t$ é tal que $g(\partial/\partial t,\partial/\partial t)$ tem o mesmo sinal em todos os pontos de $\gamma(t)$, o comprimento do caminho entre esses dois pontos é dado pela quantidade
\beq\label{2.66}
L=\int^{p}_{q}\sqrt{\left(|g(\partial/\partial t,\partial/\partial t)|\right)}dt\quad .
\eeq
Como é usual, podemos expressar (\ref{2.65}) e (\ref{2.66}) como um deslocamento infinitesimal
\beq\label{2.67}\nonumber
ds^{2}=g_{ij}dx^{i}dx^{j}\quad ,
\eeq
obtido pela mudança infinitesimal $x^{i}\rightarrow x^{i}+dx^{i}$.
Se para todo vetor $\mathbf{X},\mathbf{Y}\in T_{p}$ tivermos $g(\mathbf{X},\mathbf{Y})=0$ em $p$, então a métrica é dita não degenerada em $p$. Em termos das componentes de $\mathbf{g}$, isto quer dizer que a matriz $(g_{ab})$ é não singular. Trataremos sempre de métricas que não são degeneradas, isto é, definiremos de maneira única o tensor simétrico do tipo $(2,0)$ cujas componentes $g^{ab}$ em relação as bases duais $\{e_{a}\}$ e $\{e^{a}\}$ são dadas pela relação
\beq\label{2.68}
g^{ab}g_{bc}=\delta^{a}_{c}\quad ,
\eeq
ou seja, a matriz $(g^{ab})$ é a matriz inversa de $(g_{ab})$. Se adimitirmos sempre isso, $g^{ab}$ e $g_{ab}$
definem um isomorfismo entre qualquer tensor covariante e qualquer tensor contrava\-riante, de maneira mais simples, $g^{ab}$ e $g_{ab}$ são usadas para levantar e abaixar índices das quantidades geométricas definidas na mesma base.  Se $T_{ab}$ são as componentes do tensor $\mathbf{T}$ do tipo $(0,2)$, então podemos associar a elas unicamente as componentes tensorias $T^{a}{}_{b}=g^{ac}T_{cb}$, $T^{b}{}_{a}=g^{bc}T_{ac}$, $T^{ab}=g^{ac}g^{bd}T_{cd}$\quad .

O número de autovalores positivos da matriz $(g_{ab})$ é chamado de {\it{assinatura}} da métrica. Apenas se a métrica $\mathbf{g}$ for não degenerada e contínua, o que tomaremos como verdade, a assinatura será constante em toda variedade $\mathcal{M}$. Por uma certa escolha da base vetorial $\{e_{a}\}$, as componentes da métrica em qualquer ponto $p$ podem ser colocadas na forma 
\beq\label{2.69}\nonumber
g_{ab}=diag(\underbrace{+1, +1,\cdots, +1 }_{\frac{1}{2}(n+s)termos},\underbrace{-1, -1,\cdots, -1}_{\frac{1}{2}(n-s)termos})\quad ,
\eeq
na qual a assinatura de $\mathbf{g}$ e o número de dimensões de $\mathcal{M}$ são dados por $s$ e $n$ respectivamente. 

Uma métrica é chamada de {\it{positiva definida}} se o número de dimensões da variedade for igual à assinatura de $\mathbf{g}$, ou seja,
\beq\label{2.70}\nonumber
g_{ab}=diag(\underbrace{+1,\cdots, +1 }_{n\hspace{0.1cm}termos})\quad .
\eeq
A métrica positiva definida é propriamente o objeto que dá a {\it{disposição}} dos eventos no espaço, no sentido topológico da expressão.

A métrica cuja assinatura é $(n-2)$ é chamada de {\it{métrica lorentziana}}, cuja forma, por uma certa escolha da base vetorial, é
\beq\label{2.71}\nonumber
g_{ab}=diag(\underbrace{+1,\cdots, +1}_{(n-1)termos},-1)\quad .
\eeq
Se tomarmos um espaço-tempo dado por $(\mathcal{M},\mathbf{g})$, sendo $\mathbf{g}$ é uma métrica lorentziana, então vetores  em $p$ podem ser classificados como tipo tempo, tipo luz, ou do tipo espaço, se $g(\mathbf{X},\mathbf{X})$ é negativo, nulo ou positivo, respectivamente. Os vetores tipo luz formam um cone duplo em $T_{p}$ que separa os vetores do tipo espaço dos do tipo tempo.

Introduzimos separadamente em $\mathcal{M}$ o tensor métrico e a conexão. Apesar disso, é possível obtermos a conexão a partir de uma dada métrica, já que existe uma única conexão livre de torção em $\mathcal{M}$ dada pela condição de que a derivada covariante da métrica seja zero. Aplicando isto, obtemos as relações de Christtofel, que associam as componentes coordenadas de $\mathbf{\nabla}$ com as de $\mathbf{g}$, ou seja,
\beq\label{2.72}\nonumber
\Gamma^{a}{}_{bc}=\frac{1}{2}g^{da}\left[g_{cd,b}+g_{bd,c}-g_{cb,d}\right]\quad .
\eeq

Usando as propriedades de simetria do tensor de Riemann, e se o número de dimensões for $n=3$, o tensor de Ricci determina completamente a curvatura de $\mathcal{M}$, mas se $n>3$ aparecem outros termos de curvatura, aos quais o tensor de Ricci não descreve. Usamos o tensor de Weyl $C_{abcd}$ para descrever estes termos adicionais. As componentes do tensor de Weyl, são relacionadas com o tensor de Riemann, o tensor de Ricci e o escalar de curvatura  por
\beq\label{2.74}\nonumber
C_{abcd}=R_{abcd}+\frac{2}{n-2}\left[g_{a[d}R_{c]b}+g_{b[c}R_{d]a}\right]+\frac{2}{(n-1)(n-2)}g_{a[c}g_{d]b}R\quad ,
\eeq
e como consequência das simetrias de $R_{abcd}$,
\beq\label{2.75}\nonumber
C^{a}{}_{bad}=0\quad .
\eeq

Até aqui construímos o espaço-tempo $(\mathcal{M},\mathbf{g})$ e os objetos geométricos de interesse, tal como as componentes da curvatura representadas pelo tensor de Riemann, e a condição para o espaço-tempo ser plano, que descreve a relatividade restrita.

A resposta dada pela geometria do espaço-tempo à presença de um campo externo é modificar sua curvatura de acordo com a densidade de energia armazenada no campo. Tal situação dinâmica é descrita pelas equações de Einstein, que é um dos postulados fundamentais da Relatividade Geral. Podemos obter estas equações, supondo uma ação física que seja uma função escalar das componentes da métrica e suas derivadas. Neste caso, $g_{ab}$ são as únicas variáveis dinâmicas do sistema. Outra exigência para esta ação é que no limite não-relativístico, a teoria produzida recaia na teoria de Newton da gravitação. Para isso acontecer, na ação devem aparecer apenas derivadas até segunda ordem no tensor métrico de forma linear. 

A ação mais simples que obedece a estas exigências é a ação de Einstein-Hilbert, dada em unidades relativísticas ($c=G=\hbar=1$) por
\beq\label{2.76}
S_{g}=-\frac{1}{16\pi}\int R\sqrt{-g}d^{4}x\quad ,
\eeq
sendo $R$ o escalar de curvatura e $g$ o determinante da métrica. Aplicando o princípio de mínima ação $\delta S_{g}=0$, sendo a variação em relação as quantidades dinâmicas consideradas, que neste caso é o tensor métrico. Então a variação de (\ref{2.76}) tem como resultado
\beq\label{2.77}\nonumber
\delta S_{g}&=&-\frac{1}{16\pi}\int\delta\left(g^{ab}R_{ab}\sqrt{-g}\right)d^{4}x\\\nonumber
&=&-\frac{1}{16\pi}\int d^{4}x\bigg(R_{ab}\sqrt{-g}\delta g^{ab}+g^{ab}\delta R_{ab}\sqrt{-g}\\\nonumber&-&\frac{1}{2}g^{ab}R_{ab}g_{cd}\delta g^{cd}\sqrt{-g}
\bigg)\quad .
\eeq
A variação $\delta R_{ab}$ será feita na relação das componentes do tensor de Ricci com as da conexão $\nabla$ na base coordenada, que é obtida através da contação de (\ref{2.62}). Tal variação  fica
\beq\label{2.78}
\delta R_{ab}=\left(\delta\Gamma^{c}{}_{ab}\right)_{,c}-\left(\delta\Gamma^{c}{}_{ac}\right)_{,b}+\delta\Gamma^{d}{}_{ab}\Gamma^{c}{}_{dc}-\Gamma^{d}{}_{ab}\delta\Gamma^{c}{}_{dc}-\delta\Gamma^{d}{}_{ac}\Gamma^{c}{}_{db}-\Gamma^{d}{}_{ac}\delta\Gamma^{c}{}_{db}\quad .
\eeq
 A variação das componentes da conexão dá como resultado
\beq\label{2.79}\nonumber
\delta\Gamma^{a}{}_{bc}=\frac{1}{2}g^{ad}\left(\left(\delta g_{db}\right)_{;c} + \left(\delta g_{dc}\right)_{;b} - \left(\delta g_{bc}\right)_{;d}\right)\quad .
\eeq
Substituindo este resultado em (\ref{2.78}) chgamos à identidade de Palatini
\beq\label{2.80}
\delta R_{ab}=\left(\delta\Gamma^{c}{}_{ab}\right)_{;c}-\left(\delta\Gamma^{c}{}_{ac}\right)_{;b}\quad .
\eeq
A identidade de Palatini após a integração não dá nenuma contribuição a $\delta S_{g}$, enta~a variação da ação assume a forma
\beq\label{2.81}
\delta S_{g}=\frac{1}{16\pi}\int d^{4}x\sqrt{-g}\left[R_{ab}-\frac{1}{2}g_{ab}R\right]\delta g^{ab}\quad .
\eeq
Definindo $G_{ab}\equiv R_{ab}-\frac{1}{2}g_{ab}R$ como tensor de Einstein, e aplicando $\delta S_{g}=0$ para qualquer $\delta g^{ab}$, obtemos as equações de Eintein para o espaço-tempo no vácuo
\beq\label{2.82}\nonumber
G_{ab}=0\quad .
\eeq
Esta equação não só tem como solução a geometria para o espaço-tempo exterior à distribuições de energia e matéria, mas também nos informa que a geometria no caso de ausência quaisquer distribuições de matéria não é apenas a solução de Minkowski. Isto é a manifestação da não-linearidade da interação gravitacional. O próprio campo gravitacional, pelo seu conteúdo de energia, produz mais curvatura, o que se convêm chamar de retroação. Estas soluções descrevem campos gravitacionais auto-sustentáveis. 

Agora, se ao invés de termos apenas $S_{g}$, adicionarmos a esta, a ação $S_{m}$, que representa outros campos (campo escalar, eletromagnético, femiônico, etc.) presentes na geometria, teremos 
\beq\label{2.83}\nonumber
S=S_{g}+S_{m}\quad .
\eeq
Aplicando o princípio de mínima ação $\delta S=0$, obtemos as equações de Einstein  para qualquer espaço-tempo, que tomaremos como o postulado 1 da teoria.
\begin{postulado}{: Equações de Einstein}
\beq\label{2.84}
G_{ab}=8\pi T_{ab},
\eeq 
\end{postulado}
sendo $T_{ab}$ o tensor energia-momento. Neste objeto está representado o conteúdo de energia de uma certa região do espaço-tempo. Dado isto, em princípio resolvendo (\ref{2.84}), teremos as componentes do tensor de Einstein, ou seja, teremos o perfil da curvatura espaço-temporal gerada por $T_{ab}$.

As equações de Einstein são dez equações diferenciais parciais não-lineares de segunda ordem acopladas das componentes do tensor métrico. Pelas identidades de Bianchi $G_{ab;b}=0$ e tendo a divergência $T_{ab;b}$ nula, o número de equações se torna seis.

Uma implicação  das equações de Einstein é o aparecimento de contribuições não-locais de distribuições de energia na definição local de curvatura. Pela equação (\ref{2.74}), temos que o tensor de Riemann é dado localmente pela composição do tensor de Weyl mais o tensor de Ricci e o escalar de curvatura. Se considerarmos uma situação de vácuo, por (\ref{2.82}) $R_{ab}=0$, então a curvatura terá apenas a contribução do tensor de Weyl. Esta contribuição será não nula desde que haja uma distibuição não homogênea de energia no espaço-tempo. Tomemos o divergente de $C^{ab}{}_{cd}$
\beq\label{2.85}\nonumber
C^{ab}{}_{cd;b}&=&\frac{1}{2}\left[\left(R^{a}{}_{c}-\frac{1}{6}R\delta^{a}_{c}\right)_{;d}-\left(R^{a}{}_{d}-\frac{1}{6}R\delta^{a}_{d}\right)_{;c}\right]\\
&=&8\pi\left[T^{a}{}_{c;d}-T^{a}{}_{d;c}+\frac{1}{3}\left(\delta^{a}_{d}T_{,c}-\delta^{a}_{c}T_{,d}\right)\right]\quad ,
\eeq  
sendo $T$ o traço do tensor energia-momento. Desta equação, observamos que as fontes do tensor de Weyl são os gradientes do tensor energia-momento. Tal equação pode ser interpretada como análoga as equações de Maxwell, tomado $T$ como fonte.

Como foi dito as equações de Einstein são apenas um dos postulados fundamentais da Relatividade Geral. Existem outros dois, cujos expedientes são sobre a causalidade local do espaço-tempo e a conservação local do tensor energia-momento.
\begin{postulado}{Causalidade local}
\end{postulado}
As equações que governam a dinâmica dos campos de matéria no espaço-tempo, devem ser tais que, se $\mathcal{U}$ for a vizinhança que contém os pontos $p$ e $q$, então um sinal pode ser enviado em $\mathcal{U}$ entre os pontos $p$ e $q$ se e apenas se for possível ligar estes pontos por uma curva totalmente em $\mathcal{U}$, cujo vetor tangente é sempre diferente de zero, do tipo-tempo ou do tipo luz, isto é, devemos ter uma curva ligando estes pontos que não seja do tipo-espaço. Se $\{x^{a}\}$ forma um conjunto de coordenadas em $\mathcal{U}$ sobre o ponto p, então pontos que podem ser ligados a $p$ através de curvas não tipo-espaço, são aqueles cujas coordenadas satisfazem
\beq\label{2.86}
(x^{1})^{2}+(x^{2})^{2}+(x^{3})^{2}-(x^{4})^{2}\leq 0\quad , 
\eeq 
isto é, que estejam dentro do cone duplo em $T_{p}$ formado pelos vetores tipo-luz no ponto $p$. 
Empiricamente, a causalidade local é observada pelo fato de que nenhum sinal se deslocando mais rápido que 
a luz, que viaja em geodésicas tipo-luz, ter sido observado. Estas relações de causalidade podem ser usadas para determinar a estrutura topológica de $\mathcal{M}$. 

\begin{postulado}{Conservação local da energia e momento}
\end{postulado}

As equações da dinâmica dos campos de matéria são tais que existe um tensor simétrico chamado tensor energia-momento, que depende dos campos, de suas derivadas covariantes e métrica, e que obedeça as propriedades:

$(i)$ $T_{ab}$ se anula num aberto $\mathcal{U}$ se e apenas se os campos de matéria também se anu\-larem. Isto expressa o fato de que todos campos têm energia e portanto são afetados gravitacionalmente

$(ii)$ $T_{ab}$ obedece à equação
\beq\label{2.87}\nonumber
T^{ab}{;b}=0\quad .
\eeq

Estes tês postulados, em síntese, formam a base da Teoria da Relatividade Geral.
\section{Buracos Negros}

Buracos negros são uma das mais interessantes previsões da Relatividade Geral. Estes objetos produzem no espaço-tempo uma curvatura extrema, de tal modo que nem sinais luminosos, uma vez dentro  do horizonte de eventos do buraco negro, podem escapar. A prova da existência de buracos negros e a análise de suas propriedades, são de grande interesse não só para a  astrofísica, que especula que grandes emissões de raios-X nos centros galáticos sejam devido a presença de buracos negros supermassivos, mas também para testar nossa noção de espaço-tempo em regiões cujo campo gravitacional é tão intenso. Um buraco negro pode ser formado quando um corpo de massa $M$ contrai seu tamanho num raio menor que o {\it{raio gravitacional}} $2GM/c^{2}$. A velocidade de escape nesta situação é igual a velocidade da luz, isto é, nenhum campo ou partícula pode escapar do buraco negro. Esta conclusão vem de maneira natural da relatividade geral, pois a interação gravitacional tem a característica de ser universal, já que a carga gravitacional, ou massa, tem valor proporcional a energia total do sistema. Apesar das equações de Einstein, que descrevem os buracos negros, serem não lineares, buracos negros logo após sua formação se tornam estacionários e os únicos parâmetros que os  descrevem são sua carga, massa e momento angular, que são seus únicos {\it{cabelos}}, segundo o {\it{teorema no hair}}. A superfície do buraco negro é chamada de horizonte de eventos. Esta é uma superfície tipo-luz, que determina, juntamente com a singularidade espaço-temporal além dessa superfície, uma estrutura causal não trivial no epaço-tempo.

Possíveis aspectos observacionais de buracos negros podem ser obtidos pelo estudo do movimento de partículas e propagação de campos em espaços-tempos de buracos negros estacionários. Este problema é predominatemente matemático e envolve a integração das equações da geodésica, e a obtenção da solução da equação de onda nesses espaços. 

A constatação teórica de que buracos negros irradiam, contrariando a definição clássica dada pela relatividade geral desses objetos, foi feita por Hawking \cite{hawking2}. Este resultado é devido à instabilidade do vácuo gerada pela curvatura do buraco negro, que se torna fonte de radiação quântica. A propriedade mais interessante dessa radiação é que o seu espectro é precisamente térmico.

A colisão de um buraco negro com uma estrela de nêutrons ou a coalescência de um par de buracos negros  num sistema binário, forma uma fonte de radiação gravitacional de grande intensidade, que poderiam ser medidos por experimentos de interferometria de ondas gravitacionais. Para que se efetue tal medição, deve-se saber com detalhes o campo gravitacional do buraco negro durante a colisão. Tal solução é possível através de simulações numéricas dessas colisões.

Buracos negros funcionam como laboratórios teóricos para testar idéias sobre como seria a  teoria unificada das quatro interações fundamentais. A teoria de supercordas, que parece ser a mais promissora teoria para a unificação, explica a gravidade como uma coleção de estados da excitação de uma corda fundamental fechada. Esta teoria recupera a Relatividade Geral no limite de campo fraco. 

As  equações de Einstein $4-$dimensionais sem constante cosmológica, possuem quatro famílias de soluções exatas que descrevem buracos negros. A solução de Schwarzschild descreve um buraco negro esfericamente simétrico de massa $M$. A solução de Reissner-Nordström, generaliza a solução de Schwarschild para o caso em que além da massa, o buraco negro contém carga elétrica $Q$. O buraco negro de Kerr é a solução que descreve um buraco negro de massa $M$ e momento angular total $J$. A solução tipo buraco negro mais geral é a solução de Kerr-Newman, que depende dos três parâmetros do teorema no hair, massa $M$, carga elétrica $Q$ e momento angular $J$.

\section{Propriedades gerais dos buracos negros}

O nosso objetivo nesta seção é apresentar de forma geral propiedades inerentes a todas soluções das equações de Einstein que descrevem buracos negros sem constante cosmológica. Para que possamos definir um buraco negro em termos do par $(\mathcal{M},\mathbf{g})$, prescisamos de conceitos acerca da estrutura causal de $(\mathcal{M},\mathbf{g})$, tais como orientabilidade no tempo, curvas causais, condições de causalidade e o diagrama de Penrose-Carter.

Um espaço-tempo é tido como {\it{temporalmente orientado}} se em todos seus pontos pu\-dermos separar continuamente os vetores não tipo-espaço em duas classes distintas:os direcionados ao futuro (vetor futuro) e os direcionados ao passado (vetor passsado), do contrário teremos espaços-tempo não temporalmente orientados. Estas situações não permitem o aparecimento de singularidades, seja em buracos negros seja no universo primordial, logo não é razoável trabalhar com tais cenários, além de que, há a termodinâmica local que nos garante a existência de apenas um parâmetro de tempo.

Seja então um espaço-tempo $(\mathcal{M},\mathbf{g})$ temporalmente orientável, com dois conjuntos $\mathcal{S}$ e $\mathcal{U}$. O {\it{futuro cronológico}} $I^{+}(\mathcal{S}, \mathcal{U})$ de $\mathcal{S}$ relativo ao conjunto $\mathcal{U}$ pode ser definido como o conjunto de todos os pontos em $\mathcal{U}$ que podem ser alcançados por pontos em $\mathcal{S}$ através de curvas futuras tipo-tempo (o vetor tangente é do tipo futuro). Da mesma forma, podemos definir o {\it{passado cronológico}} $I^{-}(\mathcal{S}, \mathcal{U})$ de  $\mathcal{S}$ relativo a $\mathcal{U}$.

O {\it{futuro causal}} $J^{+}(\mathcal{S},\mathcal{U})$ de $\mathcal{S}$ relativo a $\mathcal{U}$ é o conjunto de todos os pontos de $\mathcal{U}$ que podem relacionar-se causalmente com pontos em $\mathcal{S}$. De maneira mais rigorosa, definimos o futuro causal como a união do conjunto formado pela intersecção $\mathcal{S}\cap\mathcal{U}$ e o conjunto formado por todos os pontos de $\mathcal{U}$ que podem alcançados de $\mathcal{S}$ por curvas futuras tipo-tempo ou tipo-luz.

As condições de caualidade se referem a exigência de que $(\mathcal{M}, \mathbf{g})$ não possua curvas fechada tipo-tempo. A existência de tais estruturas criaria diversos paradoxos, como, por exemplo, se um foguete viajasse numa curva desse tipo ele poderia chegar antes de sua partida. Mais precisamente, damos o nome de {\it{condição cronológica}} a esta hipótese. Uma das proposições de Hawking acerca disso, nos informa que variedades que satisfaçam à condição cronológica não devem ser compactas, pois se forem compactas, o conjunto de pontos que violam a condição cronológica é não vazio. A prova disso está em \cite{hawking1}.

A idéia principal do {\it{diagrama de Penrose-Carter}} é estudar a estrutura causal dos pontos no infinito de $(\mathcal{M}, \mathbf{g})$ iniciando com a métrica não física $\bar{g}_{ab}$, que se relaciona com a métrica física $g_{ab}$ por uma transformação conforme, isto é,
\beq\label{2.88}\nonumber
\bar{g}_{ab}=\Omega^{2}g_{ab}\quad ,
\eeq
sendo $\Omega$ o fator conforme.
Uma escolha apropriada deste fator conforme, nos possibilita ``trazer'' os pontos no infinito para uma posição finita, e então estudar sua estrutura causal. Para trazer estes pontos do infinito usamos transformações de coordenadas que envolvam funções do tipo $\tan^{-1}(x)$, que mapeiem intervalos infinitos $[-\infty,+\infty]$ em intervalos finitos, tais como $[-\pi/2,+\pi/2]$. 

A receita básica consiste em escrever a métrica usando coordenadas nulas $(v=t+r,w=t-r)$, após isso, realizar a transformação $(v,w)\rightarrow(p=\tan^{-1}v,q=\tan^{-1}\omega)$ e então identificar o fator conforme e extrair a métrica física, que representa a estrutura dos pontos no infinito. As informações da estrutura causal da métrica física colocadas em um diagrama $t\times x$ produz o diagrama de Penrose-Carter, que é o diagrama do espaço-tempo conformalmente compactificado. 

\begin{figure}[htb]
\begin{center}
\includegraphics[height=10cm,width=8cm]{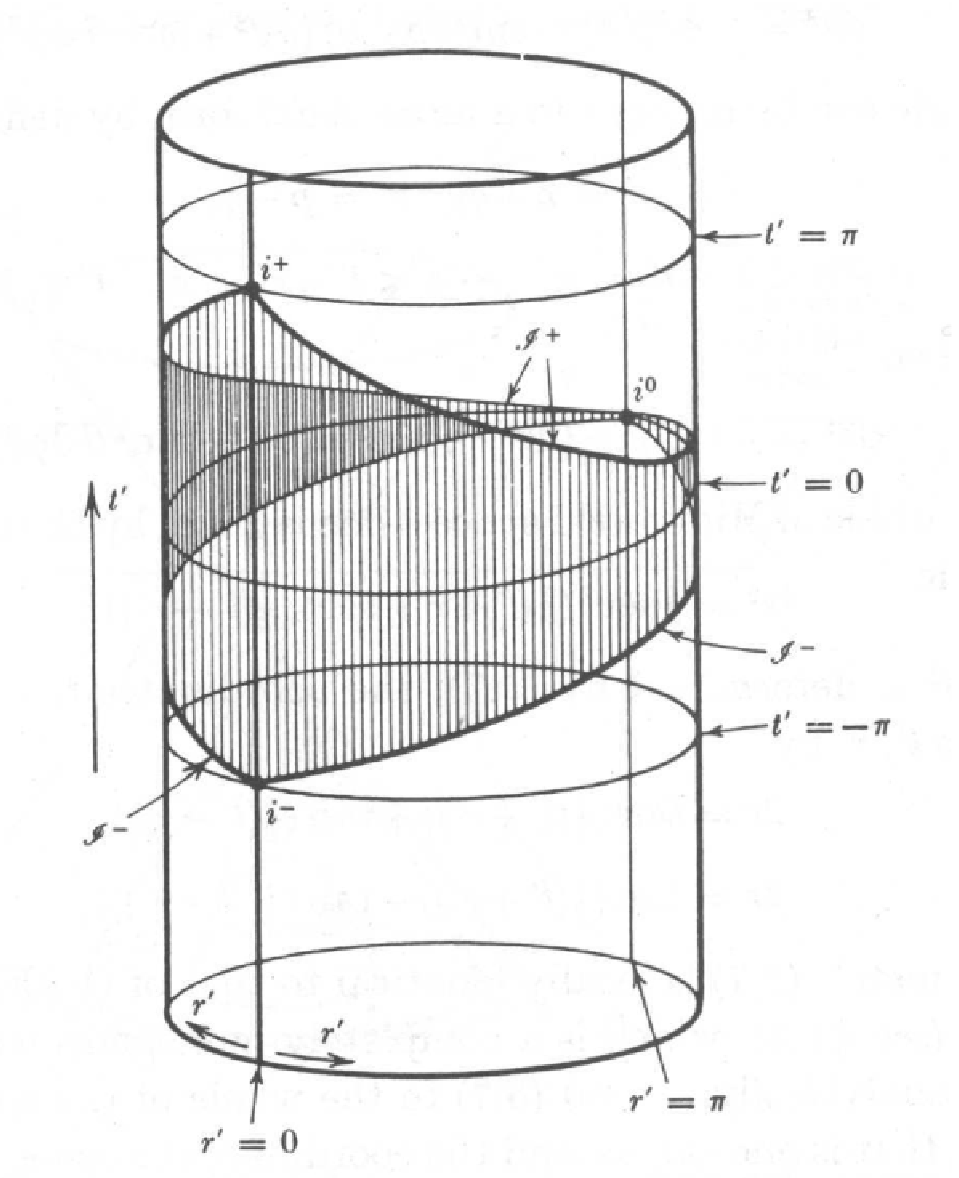}
\end{center}
\caption{Espaço-tempo de Minkowski compactificado.}\label{universo_estatico}
\end{figure}

Este processo aplicado na métrica de Minkowski
\beq\label{2.89}\nonumber
ds^{s}=-dt^{2}+dx^{2}+dy^{2}+dz^{2}\quad ,
\eeq
produz o seguinte elemento de linha
\beq\label{2.90}\nonumber
ds^{2}=\frac{1}{4}\sec^{2}(p) \sec^{2}(q) \left[4dpdq -\sin^{2}(p-q)\left(d\theta^{2}+\sin^{2}\theta d\phi^{2}\right)\right]\quad ,
\eeq
sendo o fator conforme dado por
\beq\label{2.91}\nonumber
\Omega=\frac{1}{4}\sec^{2}(p)\sec^{2}(q)\quad ,
\eeq  
e a métrica conforme $\bar{g}_{ab}$ por
\beq\label{2.92}\nonumber
d\bar{s}^{2}=dt'^{2}-dr'^{2}-\sin^{2}r'\left(d\theta^{2}+\sin^{2}\theta d\phi^{2}\right)\quad ,
\eeq
no qual $t'=p+q$ e $r'=p-q$, com domínios $-\pi<t'+r'<\pi$, $-\pi<t'-r'<\pi$ com a condição $r'\leq 0$. O elemento de linha conforme expressa a geometria dos pontos do infinfito numa região finita. No caso do espaço-tempo de Minkowski, obtemos o universo estático de Einstein como métrica conforme, cuja topologia é cilíndrica. Esta representação do espaço-tempo de Minkowski como uma região finita do universo estático de Einstein, é chamada de compactificação conforme, que pode ser representada no diagrama (\ref{universo_estatico}) retirado de \cite{hawking1}.

A fronteira dessa região representa a estrutura conforme do infinito para o espaço-tempo de Minkowski. Em termos das coordenadas $p$ e $q$, esta fronteira consiste nas superfícies nulas $p=\pi/2$ denotada por $\mathcal{T}^{+}$ e $q=-\pi/2$ chamada $\mathcal{T^{-}}$, além dos pontos $\left(p=\pi/2,q=\pi/2\right)$, $\left(p=\pi/2,q=-\pi/2\right)$ e $\left(p=-\pi/2,q=-\pi/2\right)$ denotados respectivamente por $i^{+}$, $i^{0}$ e $i^{-}$. 

As geodésicas tipo-tempo têm origem no ponto $i^{-}$ em terminam em $i^{+}$. Da mesma forma, geodésicas tipo-luz se originam na superfície $\mathcal{T}^{-}$ e terminam em $\mathcal{T}^{+}$, enquanto que as geodésicas tipo-espaço iniciam e teminam em $i^{0}$. A representação das curvas geodésicas nestes diagramas também aplica-se às curvas não geodésicas.

Os pontos $i^{+}$ e $i^{-}$ representam o infinito futuro tipo-tempo e infinito passado tipo-tempo, $\mathcal{T}^{+}$ e $\mathcal{T}^{-}$ representam o infinito futuro tipo-luz e infinito passado tipo-luz. 

\begin{figure}[h]
\begin{center}
\includegraphics[height=9cm,width=9cm]{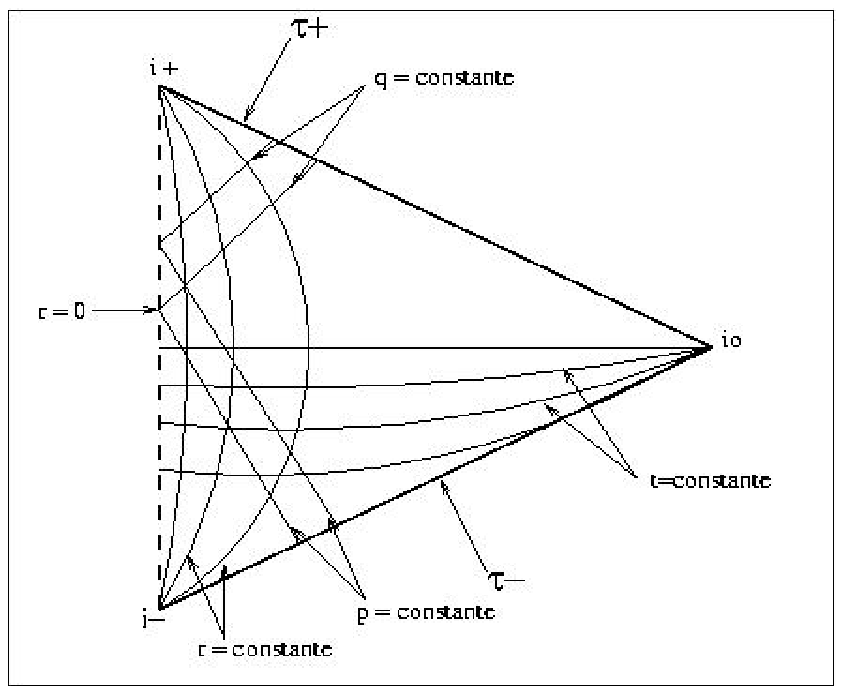}
\end{center}
\caption{Diagrama de Penrose-Carter do espaço-tempo de Minkowski.}\label{penrose_minko}
\end{figure}

O {\it{diagrama de Penrose-Carter}} é o diagrama $t\times r$ de um espaço-tempo conformalmente compactificado. Para o espaço-tempo de Minkowski o diagrama de Penrose-Carter está na figura (\ref{penrose_minko})
onde temos as curvas $r$ constante, que correspondem as histórias de esferas bidimensionais com raio constante, e as fatias tipo-tempo correspondente a $t$ constante. Geodésicas radiais nulas são representadas por retas  $p$ constante e $q$ constante fazendo angulos de $45^{o}$ e $-45^{o}$ respectivamente.  

Tendo a prescrição do diagrama de Penrose-Carter, passamos então à análise das soluções tipo buraco negro de Schwarzschild e Kerr.

\section{A solução de Schwarzschild}

A solução das equações de Einstein que descreve um espaço-tempo vazio na vizinhança de um corpo massivo esfericamente simétrico, é a solução de Schwarzschild. Esta  solução é uma boa aproximação para estudarmos o Sistema Solar, bem como qualquer sistema gravitacional que não exiba consideráveis desvios da simetria esférica. Esta solução é dada pela métrica
\beq\label{2.93}
ds^{2}=-\left(1-\frac{2m}{r}\right)dt^{2}+\left(1-\frac{2m}{r}\right)^{-1}dr^{2}+r^{2}\left(d\theta^{2}+\sin^{2}\theta d\phi^{2}\right),
\eeq
para $r>2m$, sendo $m$ uma constante de integração advinda da resolução das equações de Einstein. O valor desta constante é determinado quando consideramos o limite newtoniano da Relatividade Geral. Sendo $\Phi$ o potencial gravitacional newtoniano, então neste limite, em unidades não-relativísticas,
\beq\label{2.94}\nonumber
g_{00}\simeq 1+\frac{2\Phi}{c^{2}}=1-\frac{2G_{4}M}{c^{2}r}\quad ,
\eeq
no qual $c$ é a velocidade da luz, $G_{4}$ a constante da gravitação $4-$dimensional e $M$ a massa que dá origem ao potencial gravitacional $\Phi$. Comparando (\ref{2.94}) com a solução de Schwarzschild, determinamos a constante de integração $m$ como
\beq\label{2.95}\nonumber
m=\frac{GM}{c^{2}}\quad .
\eeq
Vemos que a solução de Schwarzschild pode ser interpretada como a descrição para o campo gravitacional de uma partícula de massa $m$ situada na origem. Esta solução é estática, no sentido de que $\partial/\partial t$ é um campo vetorial de Killing. As componentes da métrica são funções que não dependem do tempo, e não apresentam termos de rotação, pois não há termos mistos como $dtdr$. A simetria esférica  de (\ref{2.93}) aparece no fato de que $g_{00}$ e $g_{01}$ dependerem apenas da coordenada radial e não das coordenadas angulares $\theta$ e $\phi$.
 Pode-se mostrar, conforme teorema de Birkhoff, que qualquer solução das equações de Einstein para sistemas esfericamente simétricos é localmente isométrica à solução de Schwarzschild, ou seja, este teorema garante que esta solução é única.

Em geral a métrica (\ref{2.93}) descreve a geometria externa de uma estrela de raio $r_{0}$, isto é, para $r>r_{0}$. A métrica para $r<r_{0}$ depende do tensor energia-momento $T_{ab}$ no interior estelar. No caso de um completo colapso gravitacional, quando toda massa do corpo colapsa no ponto $r=0$, consideramos a solução de Schwarzschild como descrição para todos valores de $r$. Tal solução apresenta singularidades nos pontos $r=0$ e $r=2m$, ou seja, no caso do colapso gravitacional, a métrica de Schwarzschild, descreve  dois domínios da variedade $\mathcal{M}$, $0<r<2m$ ou $2m<r<\infty$.

Tomando $\mathcal{M}$ com domínio $2m<r<\infty$, é necessário determinar se $\mathcal{M}$ é extensível, isto é, descobrir se existe uma variedade $\mathcal{M'}$ com uma métrica $\mathbf{g}'$ formando o espaço-tempo $(\mathcal{M'},\mathbf{g}')$ tal que $\mathcal{M}$ é um subespaço de $\mathcal{M}'$ e que $\mathbf{g}=\mathbf{g}'$ em $\mathcal{M}$. Neste domínio isto é possível, já que o escalar de curvatura em $r=2m$ é uma função contínua, o que indica que a singularidade em $r=2m$ é apenas uma patologia do sistema de coordenadas escolhido, ou seja, deve existir um outro conjunto de coordenadas que mapeie este domínio de $\mathcal{M}$ numa imagem sem singularidades. Para efetuar a extensão maximal de $\mathcal{M}$, devemos ter todas curvas geodésicas extendidas em ambas as direções com relação ao parâmetro afim, ou devem ter terminações em uma singularidade verdadeira, ou seja, uma singularidade que não pode ser removida por nenhuma mudança de coordenadas.

Se tomarmos agora, $\mathcal{M}$ com domínio $0<r<2m$ e calcularmos o escalar de curvatura, obtido das componentes do tensor de Riemman,
\beq\label{2.96}
R^{abcd}R_{abcd}=\frac{48m^{2}}{r^{6}}\quad ,
\eeq
concluiremos que para $r\rightarrow 0$, $R^{abcd}R_{abcd}\rightarrow \infty$, o que indica a existência de uma singularidade real em $r=0$, tornando impossível a extensão de $\mathcal{M}$ além de $r=0$. O máximo que podemos fazer é extender $\mathcal{M}$ além de $r=2m$ até $r=\infty$ através do procedimento de Kruskal-Szekeres. Este procedimento consiste em reescrever (\ref{2.93}) em termos das coordenadas nulas avançadas e retardadas $(v,w)$. Descreveremos $\mathcal{M}$ com o sistema de coordenadas $(v,w,\theta,\phi)$.

Usando a condição pra uma curva geodésica ser uma geodésica tipo-luz ou nula
\beq\label{2.97}\nonumber
g_{ab}X^{a}X^{b}=0\quad ,
\eeq
as geodésicas radiais nulas do espaço-tempo descrito por (\ref{2.93}) são dadas por
\beq\label{2.98}\nonumber
\left(\frac{dt}{dr}\right)^{2}=\left(\frac{1}{1-\frac{2m}{r}}\right)^{2},
\eeq
definindo a coordenada tartaruga $r_{*}$ de Wheeler, neste caso, como
\beq\label{2.99}
r_{*}\equiv\int\frac{dr}{1-\frac{2m}{r}}=r+2m\log \left(\frac{r}{2m}-1\right)\quad .
\eeq
Então, as geodésicas radiais nulas ficam
\beq\label{2.100}\nonumber
t=r_{*}+const\quad ,\\
t=-r_{*}+const\quad .
\eeq
Definimos as coordenadas nulas $(v,w)$ por
\beq\label{2.101}\nonumber
v=t-r_{*}\quad ,\\
w=t+r_{*}\quad ,
\eeq
que consequentemente leva à definição de $r$ em termos implícitos de $v$ e $w$, já que $r_{*}=\left(w-v\right)/2$. 

Usando estas definições, a solução (\ref{2.93}) pode ser escrita como
\beq\label{2.102}
ds^{2}=-\frac{2m e^{-r/2m}}{r}e^{(w-v)/4m}dvdw+r^{2}\left(d\theta^{2}+\sin^{2}\theta d\phi^{2}\right),
\eeq
implicando em que a singularidade $r\rightarrow 2m$ corresponde a $v\rightarrow \infty$ e $\omega\rightarrow -\infty$. 

\begin{figure}[ht]
\begin{center}
\includegraphics[height=10cm,width=11cm]{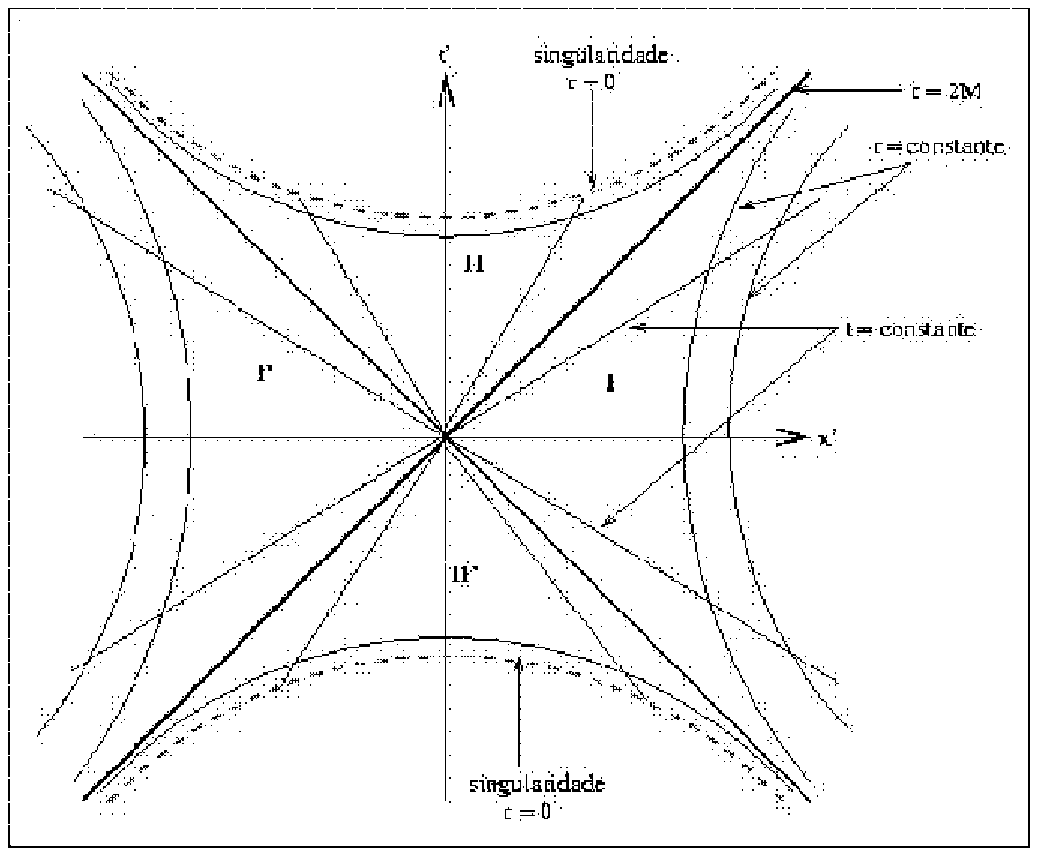}
\end{center}
\caption{Extensão de Kruskal da geometria de Schwarzschild.}\label{kruskal}
\end{figure}

Definindo novas coordenadas $V$ e $W$ como
\beq\label{2.103}\nonumber
V=-e^{-v/4m}\quad ,\\
W=e^{w/4m}\quad ,
\eeq
a  métrica (\ref{2.102}) fica
\beq\label{2.104}
ds^{2}=-\frac{32m^{3} e^{-r/2m}}{r}dVdW+r^{2}\left(d\theta^{2}+\sin^{2}\theta d\phi^{2}\right)\quad .
\eeq
Nesta forma, observamos que não há singularidade em $r=2m$, que corresponde a $V=W=0$. Para obtemos a forma de Kruskal-Szekeres da métrica de Schwarzschild fazemos a seguinte mudança de coordenadas
\beq\label{2.105}\nonumber
T=\frac{(V+W)}{2}\quad ,\\
X=\frac{(W-V)}{2}\quad ,
\eeq
em (\ref{2.104}), obtendo assim
\beq\label{2.106}
ds^{2}=\frac{32m^{3} e^{-r/2m}}{r}\left(-dT^{2}+dX^{2}\right)+r^{2}\left(d\theta^{2}+\sin^{2}\theta d\phi^{2}\right).
\eeq
A transformação de coordenadas entre as coordenadas iniciais $(t,r)$ e $(T,X)$ é dada por
\beq\label{2.107}\nonumber
X^{2}-T^{2}=\left(\frac{r}{2m}-1\right)e^{r/2m}\quad ,\\
t=4m\tanh^{-1}\left(\frac{T}{X}\right)\quad .
\eeq
A condição de que $r>0$ especifica os valores permitidos para as coordenadas $(T,X)$. Estes valores são os que obedecem a relação $X^{2}-T^{2}>-1$. Esta estrutura da extensão de Kruskal-Szekeres da geometria de Schwarzschild
é dada pelo diagrama (\ref{kruskal}), chamado de diagrama de Kruskal-Szekeres. Em tal diagrama, as geodésicas radiais nulas são retas que fazem ângulos de $45^{o}$ com os eixos coordenados. Vemos também que a singularidade verdadeira em $r=0$ corresponde aos valores $X^{2}=\pm\sqrt{\left(T^{2}-1\right)}$ e que não há singularidade quando $r=2m$. A região $I$ é correspondente a solução original de Schwarzschild em $r>2m$, que representa o campo gravitacional externo ao corpo colapsante. Esta região apresenta um comportamento assintóticamente plano, da mesma maneira que a região $I'$. Um fóton que parte da região $I$ pode ir para o infinito espacial ou atravessar a fronteira $X=T$ entrando na região $II$, mas não ir para região $I'$, o que mostra que as regiões $I$ e $I'$ são causalmente desconectadas. Não há possibilidade de um observador numa geodésica radial nula na região $II$, escapar da singularidade em $X=\sqrt{\left(T^{2}-1\right)}$, essa região é chamada de {\it{buraco negro}} e a região $II'$, que possui as mesmas propriedades que $II$ quando invertemos o tempo desta, de {\it{buraco branco}}.  

\begin{figure}[ht]
\begin{center}
\includegraphics[height=10cm,width=10cm]{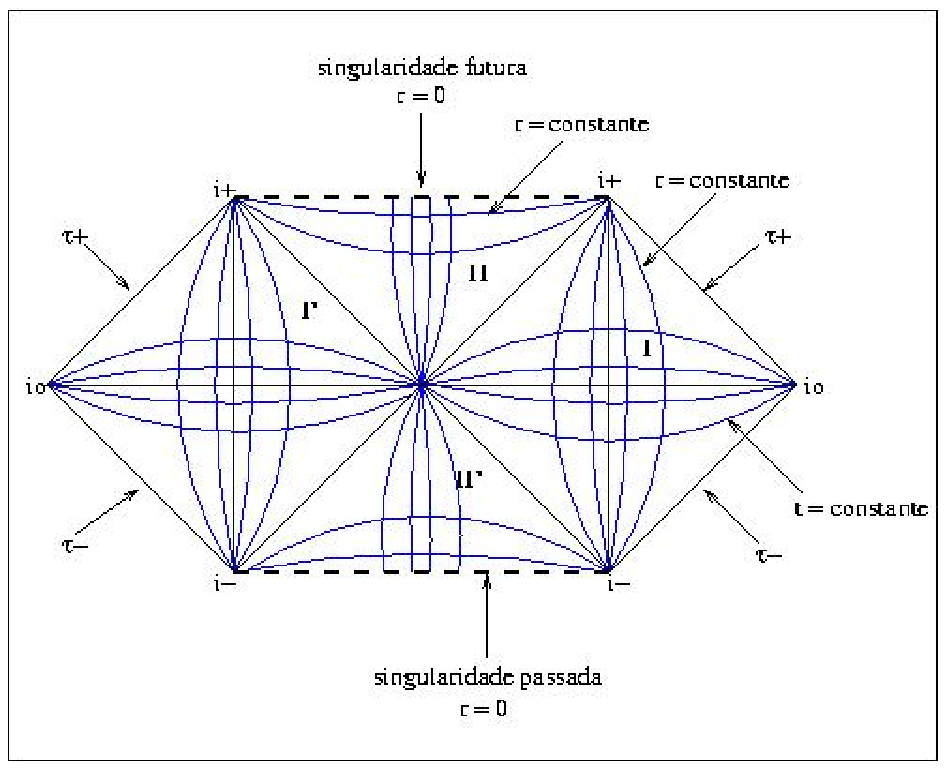}
\end{center}
\caption{Diagrama de Penrose-Carter da solução de Kruskal-Szekeres.}\label{penrose_kruskal}
\end{figure}

Para obtermos o diagrama de Penrose-Carter para a extensão de Kruskal-Szekeres da geometria de Schwarzschild se definirmos novas coordenadas nulas por 
\beq\label{2.108}\nonumber
v'=\tan^{-1}\left(\frac{V}{\sqrt{2m}}\right)\quad ,\\
w'=\tan^{-1}\left(\frac{W}{\sqrt{2m}}\right)\quad ,
\eeq 
com domínios $-\pi/2<v'<\pi/2$, $-\pi/2<w'<\pi/2$ e $-\pi<v'+w'<\pi$. Este diagrama é dado na figura (\ref{penrose_kruskal}).

Observamos neste diagrama que cada uma das regiões assintóticamente planas $I$ e $I'$ possuem infinito futuro, infinito passado e infinito tipo-luz, e que diferentemente do espaço-tempo de Minkowski, a métrica conforme é contínua mas não diferenciável em $i^{0}$.

Na extensão de Kruskal-Szekers da métrica de Schwarzschild, a superfície nula $r=2m$ é uma superfície onde cada ponto representa uma esfera bidimensional de área $16\pi m^{2}$. Se observarmos a métrica (\ref{2.26}), vemos que $g_{00}$ é sempre maior que zero quando tomamos $r>2m$, mas muda de sinal quando atravessamos o horizonte de eventos $r=2m$, isto é, $g_{00}<0$ quando $r<2m$. Isto implica que as coordenadas $t$ e $r$ invertem seus papéis, tornando o espaço-tempo para $r<2m$ não estático. O horizonte de eventos torna-se o limite estático da solução de Schwarzschild.
\section{A solução de Kerr}

Em Relatividade Geral, o campo gravitacional de um objeto em rotação e estacionário é descrito pelo espaço-tempo de Kerr \cite{kerr1}. Nas chamadas coordenadas de Boyer-Lindquist, esta solução é dada pela métrica
\beq\label{2.109}\nonumber
ds^{2}&=&-\left(1-\frac{2mr}{\Sigma}\right)dt^{2}-\frac{4amr\sin^{2}\theta}{\Sigma}dtd\theta\\
&+&\frac{\Sigma}{\Delta}dr^{2}+\Sigma d\theta^{2} +\left(r^{2}+a^{2}+\frac{2mra^{2}\sin^{2}\theta}{\Sigma}\right)\sin^{2}\theta d\phi^{2}\quad,
\eeq
 sendo $\Sigma\equiv r^{2}+a^{2}\cos^{2}\theta$ e $\Delta\equiv r^{2}-2mr+a^{2}$. A quantidade $m$ representa a massa gravitacional do corpo girante, e sendo $J$ o seu momento angular total, então $a=J/m$ representa o momento angular por unidade de massa. Os coeficientes da métrica de Kerr são independentes de $t$ e $\phi$, o que implica que esta solução é estacionária e axialmente simétrica, que são as únicas simetrias contínuas desta solução. Por simetria axial entendemos que se existe um eixo definido (neste caso o eixo z em coordenadas cartesianas, ou $\theta=0$), tal que a solução é invariante por rotações em torno deste eixo, então a solução é axialmente simétrica. Existem duas simetrias discretas para este espaço-tempo, a reflexão em $t$ e $\phi$, que não ocorrem separadamente mas em conjunto, isto é, esta solução admite apenas a inversão simultânea\beq\label{2.110}\nonumber
t\rightarrow -t, \hspace{0.3cm} \phi\rightarrow -\phi\quad .
\eeq
Este tipo de simetria representa o fato do espaço-tempo de Kerr ser a solução para um corpo em rotação, já que se invertemos o spin e tomarmos o tempo negativo é equivalente a direção positiva do tempo com spin positivo. 

Temos ainda a segunda simetria discreta
\beq\label{2.1101}\nonumber
t\rightarrow -t, \hspace{0.3cm}a\rightarrow -a\quad, 
\eeq
 que sugere que $a$ especifica a direção do spin do corpo girante. A solução (\ref{2.109}) é assinto\-ticamente plana, isto é, para $r\rightarrow \infty$, obtemos a métrica de Minkowski. Quando tomamos $a=0$, não só os termos relacionados a rotação vão a zero, mas a métrica se reduz a solução de Schwarzschild.

A solução de Kerr admite uma singularidade espaço-temporal em forma de anel. Esta singularidade não removível aparece quando $\Sigma=0$, a partir do cálculo da quantidade $R_{abcd}R^{abcd}$. Então se
\beq\label{2.111}\nonumber
\Sigma=r^{2}+a^{2}\cos^{2}\theta=0\quad, 
\eeq
devemos ter $r=\cos\theta=0$, que determina uma singularidade com estrutura de um anel de raio $a$ no plano equatorial $z=0$, já que em coordenadas cartesianas $z=r\cos\theta$. O espaço-tempo de Kerr admite duas superfícies $S_{+}$ e $S_{-}$ onde o desvio para o vermelho é infinito, tais superfícies representam limites estacionários, como no caso de Schwarzschild, são superfícies onde as coordenadas tipo-tempo trocam de papel com as coordenadas tipo-espaço. Ainda admite duas superfícies nulas, que definem dois horizontes de eventos.

As superfícies com desvio para o vermelho infinito são obtidas quando $g_{00}$ é nulo. Disto obtemos os limites estacionários
\beq\label{2.112}
r_{S_{\pm}}=m\pm\sqrt{m^{2}-a^{2}\cos^{2}\theta}\quad .
\eeq
No lmite de Schwarzschild $a\rightarrow 0$, a superfície $S_{+}$ se reduz ao horizonte de eventos $r=2m$ e $S_{-}$ a $r=0$. Estas superfícies definidas por (\ref{2.112}) são axialmente simétricas, com $S_{+}$ possuindo um raio $r=2m$ no equador, e assumindo $a^{2}<m^{2}$, um raio $m+\sqrt{m^{2}-a^{2}}$ nos pólos. A superfície $S_{-}$ está completamente contida por $S_{+}$. A existência dessas superfícies indica a existência de superfícies nulas que definem horizontes de eventos. Para encontrar tais horizontes, devemos procurar por superfícies $r=constante$ onde a componente $g^{11}$ se anule. Então se
\beq\label{2.113}\nonumber
g^{11}=-\frac{\Delta}{\Sigma^{2}}=-\frac{r^{2}-2mr+a^{2}}{r^{2}+a^{2}\cos^{2}\theta}\quad ,
\eeq
para que ele se anule, devemos ter
\beq\label{2.114}\nonumber
\Delta=r^{2}-2mr+a^{2}=0,
\eeq
o que implica em dois horizontes de eventos, sempre com $a^{2}<m^{2}$
\beq\label{2.115}\nonumber
r=r_{\pm}=m\pm\sqrt{m^{2}-a^{2}}.
\eeq

A métrica de Kerr apresenta dois horizontes de eventos e duas superfícies com desvio para o vermelho infinito. Esta configuração define três regiões em que esta métrica é regular. A região I cujo domínio é $r_{+}<r<+\infty$, a região II compreendida no intervalo $r_{-}<r<r_{+}$ e a região III definida por $0<r<r_{-}$.
No limite $a=0$ os horizontes de eventos se reduzem a $r=2m$ e $r=0$. De fato, na solução de Schwarzschild o horizonte de eventos e a superfície de infinito desvio para o vermelho coincidem. A região entre o limite estacionário $S_{+}$ e o horizonte de eventos $r=r_{+}$ é chamada de {\it{ergosfera}}.

Destas considerações, vemos que para $a<m$ a solução de Kerr descreve um buraco negro com parâmetro de rotação diferente de zero. No caso em que $a>m$, não há a formação do buraco negro, já que não há o aparecimento de um horizonte de eventos mas apenas a formação de uma singularidade em $r=0$, que é visível a todos observadores externos. Tal singularidade, pela ausência de um horizonte de eventos, é chamada de {\it{singularidade nua}}. Segundo o {\it{hipótese da censura cósmica}}, tais obejtos não devem existir na natureza, mas tal hipótese nunca foi provada de maneira geral. Por outro lado, no caso de buracos negros em rotação, não parece haver nenhum mecanismo dinâmico que faça com que uma estrela com $a<m$ vá para um estado em que $a>m$.

\begin{figure}[ht]
\begin{center}
\includegraphics[height=10cm,width=10cm]{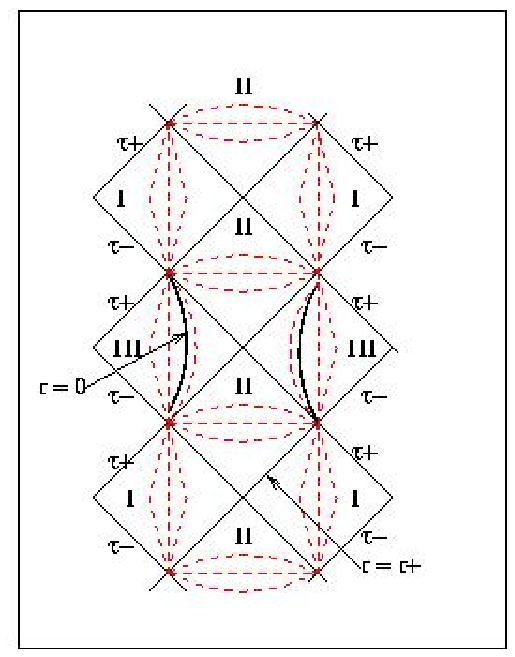}
\end{center}
\caption{Estrutura conforme da métrica de Kerr ao longo do eixo de simetria para $a<m$.}\label{kerr2}
\end{figure}

A extensão maximal da métrica de Kerr ($a<m$) é obtida utilizando-se as coordenadas de Eddington-Finkelstein avaçadas e retardadas
\beq\label{2.116}\nonumber
du_{\pm}&=&dt\pm\frac{r^{2}+a^{2}}{\Delta}dr,\\\nonumber
d\phi_{\pm}&=&d\phi\pm\frac{a}{\Delta}dr.
\eeq
Na figura (\ref{kerr2}) temos a estrutura conforme ao longo do eixo de simetria. As regiões I determinadas pelo intervalo $r_{+}<r<+\infty$ são estacionárias e assintóticamente planas. As regiões II $r_{-}<r<r_{+}$ não são estacionárias e apresentam superfícies aprisionadas fechadas. As regiões III $-\infty<r<r_{-}$ contêm a singularidade em forma de anel do tipo-tempo, que pode ser evitada. Esta região contém curvas fechadas tipo-tempo. Tais curvas violam o princípio da causalidade, não sendo fisicamente consideradas. Não há violação da causalidade nas regiões I e II. 

Segundo Penrose \cite{penrose2}, é possível realizar uma espécie de extração de energia de um buraco negro que possui uma ergosfera. Dada uma partícula se movendo numa trajetória geodésica a partir do infinito, com energia
$E_{0}\equiv -p_{0}^{a}K_{a}>0$ ao longo desta trajetória, com $p_{0}^{a}=mv_{0}^{a}$ sendo o momento da partícula, $v_{0}^{a}$ são as componentes da quadri-velocidade da partícula e $K_{a}$ são as componentes do vetor de Killing. Então suponhamos que tal partícula se divida em dois fragmentos com momentos $p_{1}^{a}$ e $p_{2}^{a}$, onde $p_{0}^{a}=p_{1}^{a}$. Sendo $K_{a}$ um vetor do tipo-espaço é possível escolher $p_{1}^{a}$ um vetor tipo-tempo futuro direcionado, tal que $E_{1}\equiv -p_{1}^{a}K_{a}<0$. Então $E_{2}\equiv -p_{2}^{a}K_{a}$ será maior que $E_{0}$. Isto indica que o segundo fragmento de energia $E_{2}$ pode escapar para o infinito onde terá mais energia que a partícula original. Esta quantidade de energia excedente foi extraída do buraco negro.
\section{Perturbações de Buracos Negros}

O estudo das perturbações em soluções tipo buraco negro em Relatividade Geral é um ramo de pesquisa que nos permite a análise de inúmeros temas, tais como a radiação gerada por pequenos corpos caindo no buraco negro, o perfil de ondas gravitacionais muito tempo após a formação do buraco, espalhamento e absorção de ondas por buracos negros, a estabilidade das soluções tipo buraco negro e vários outros temas. 

Esta seção é devotada ao estudo de perturbações lineares em torno da solução de Schwarzschild e Kerr, sobretudo as perturbações escalares. Os campos físicos considerados como perturbações são tomados como fracos, isto é, o efeito de seu tensor energia-momento na métrica de fundo do buraco negro pode ser desprezado. Em síntese, serão consideradas apenas perturbações lineares dessas soluções. 

Os primeiros estudos teóricos acerca desse tema derivam do trabalho de Regge Whee\-ler da década de 50 \cite{regge1}. Eles se propuseram a estudar a estabilidade da singularidade de Schwarzschild frente a pequenos desvios de sua simetria esférica. Para  a obtenção das equações lineares que governam estas pequenas perturbações na simetria esférica, foi necessário estudar a equação de Eintein linearizada sobre o espaço-tempo descrito por tal métrica. Esta equação é obtida escrevendo a métrica perturbada $\bar{g}_{\mu\nu}$ como a soma de dois termos: um representando a métrica não perturbada de fundo, ${}^{(\circ)}{g}_{\mu\nu}$, e outro representando a perturbação dado por $h_{\mu\nu}$. Disto então, podemos escrever \cite{nollert}
\beq\label{2.117}
\bar{g}_{\mu\nu}={}^{(\circ)}{g}_{\mu\nu}+h_{\mu\nu}\quad , 
\eeq
onde consideramos $|h_{\mu\nu}|<<1$. Usando (\ref{2.117}) nas equações de Einstein obtemos as equações que governam $h_{\mu\nu}$ em primeira ordem. 

As componentes $\bar{\Gamma}^{\kappa}_{\mu\nu}$ da conexão métrica perturbada são dadas por, considerando no máximo termos de primeira ordem em $h_{\mu\nu}$,
\beq\label{2.118}
\bar{\Gamma}^{\kappa}_{\mu\nu}={}^{(\circ)}\Gamma^{\kappa}_{\mu\nu}+\delta\Gamma^{\kappa}_{\mu\nu}\quad ,
\eeq
 sendo
\beq\label{2.119}
\delta\Gamma^{\kappa}_{\mu\nu}=\frac{1}{2}{}^{(\circ)}g^{\kappa\alpha}\left[h_{\alpha\nu ;\mu}+h_{\alpha\mu ;\nu}-h_{\mu\nu ;\alpha}\right]\quad .
\eeq
Com isto, podemos escrever as componentes do tensor de Ricci perturbado como
\beq\label{2.120}
\delta R_{\mu\nu}=\delta\Gamma^{\alpha}_{\mu\alpha ;\nu}-\delta\Gamma^{\alpha}_{\mu\nu ;\alpha}\quad .
\eeq 

No vácuo, as equações de perturbação reduzem-se a
\beq\label{2.121}
\delta R_{\mu\nu}=0\quad ,
\eeq
que representam dez equações diferenciais parcias das componentes lineares de $h_{\mu\nu}$. Como é mostrado em \cite{nollert}, este conjunto de equações pode ser reduzido, se fizermos a separação de variáveis
\beq\label{2.122}
h_{\mu\nu}=\sum^{\infty}_{L=0}\sum^{L}_{M=-L}\sum^{10}_{n=1}C^{n}_{LM}(t,r)\left(Y^{n}_{LM}\right)_{\mu\nu}(\theta,\phi)\quad ,
\eeq
sendo $C^{n}_{LM}(t,r)$ os coeficientes da expansão e $\left(Y^{n}_{LM}\right)_{\mu\nu}(\theta,\phi)$ os harmônicos esféricos tensorias. Seguindo isto, foi possível mostrar que as perturbações se dividem em duas classes de paridade, para cada valor de $L$. São denominadas perturbações polares, as que se transformam com $(-1)^{L}$, perturbações axiais as que se transformam como $(-1)^{L+1}$. Se considerarmos como campos perturbativos fracos o campo escalar o campo eletromagnético e o campo gravitacinal nessa prescrição, obtemos  uma equação diferencial tipo barreira de potencial \cite{chandra}\cite{zerilli1}\cite{regge1}\cite{nollert}, dada por
\beq\label{2.123}
\frac{\partial^{2}}{\partial t^{2}}\Psi(x,t)-\frac{\partial^{2}}{\partial x^{2}}\Psi(x,t)+V(x)\Psi(x,t)=0\quad ,
\eeq
sendo $\Psi(x,t)$ o campo perturbativo, $x\equiv r+2M\ln(r/2M-1)$ a coordenada tartaruga e $V(x)$ o potencial efetivo dado por
\beq\label{2.124}
V(x)=\left(1-\frac{2M}{r}\right)\left[\frac{L(L+1)}{r^{2}}+\frac{2M\sigma}{r^{3}}\right]\quad ,
\eeq
no qual $\sigma$ assume os valores $+1$, $0$ e $-3$, correspondendo, respectivamente, a perturbação por campo escalar, eletromagnético e gravitacional. Temos então que as perturbações li\-neares na métrica de Schwarzschild são descritas por uma equação de onda unidimensional com um potencial real.

Apesar da equação (\ref{2.124}) sintetizar as perturbações lineares da solução de Schwarzschild, vamos verificar com mais detalhes a perturbação gerada por um campo escalar definido na vizinhança de um  buraco negro esfericamente simétrico, e em seguida estudaremos a perturbação pelo mesmo campo num buraco negro de Kerr.

\begin{figure}[ht]
\begin{center}
\includegraphics[height=8cm,width=11cm]{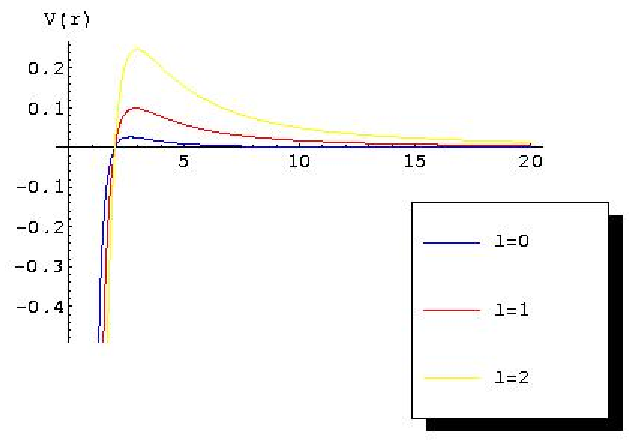}
\end{center}
\caption{Potencial escalar efetivo com ponto máximo $r=3M$ para $r_{*}=0$}\label{potencial_escalar}
\end{figure}

Um campo escalar sem massa $\Phi$ tem sua evolução determinada pela equação de Klein-Gordon,
\beq\label{2.125}
\Box\Phi\equiv\frac{1}{\sqrt{-g}}\frac{\partial}{\partial x^{\mu}}\left[\sqrt{-g}g^{\mu\nu}\frac{\partial}{\partial x^{\nu}}\Phi\right]\quad ,
\eeq
sendo $g$ o determinate da métrica $g_{\mu\nu}$, cujas componentes seguem da solução de Schwarzschild (\ref{2.93}). Usando o fato da solução ser esfericamente simétrica podemos realizar a seguinte separação de variáveis em $\Phi$
\beq\label{2.126}
\Phi_{Lm}=\frac{\Psi_{L}(r,t)}{r}Y_{Lm}(\theta,\phi)\quad .
\eeq
Com isso obtemos duas equações diferenciais,
\beq\label{2.126}
\frac{\partial}{\partial\theta}\left[\sin\theta\frac{\partial}{\partial\theta}Y_{lm}\right]&+&\frac{\partial^{2}}{\partial\phi^{2}}Y_{Lm}=L(L+1)\sin^{2}\theta Y_{Lm}\quad ,\\\label{2.127}
-\frac{\partial^{2}}{\partial t^{2}}\Psi (r,t)&+&\left(1-\frac{2M}{r}\right)\frac{\partial}{\partial r}\left[\left(1-\frac{2M}{r}\right)\frac{\partial}{\partial r}\Psi (r,t)\right]\\\nonumber
&=&\left(1-\frac{2M}{r}\right)\left[\frac{L(L+1)}{r^{2}}+\frac{2M}{r^{3}}\right]\Psi(r,t)\quad .
\eeq
A primeira equação tem como solução os harmônicos esféricos $Y_{Lm}$. A equação redial-temporal (\ref{2.127}) pode ser simplificada através da coordenada tartaruga (\ref{2.99})
que implica na relação
\beq\label{2.129}
\frac{\partial}{\partial r_{*}}=\left(1-\frac{2M}{r}\right)\frac{\partial}{\partial r}\quad ,
\eeq
que usada em (\ref{2.127}) dá como resultado 
\beq\label{2.130}
\left[\frac{\partial^{2}}{\partial r_{*}^{2}}-\frac{\partial^{2}}{\partial t^{2}}-V_{L}(r)\right]\Psi_{L}(r,t)=0\quad .
\eeq
Se considerarmos uma dependência temporal harmônica, tal como $\Psi_{L}(r,t)=\hat{\Psi}_{L}(r,t)e^{-i\omega t}$, reduzimos (\ref{2.130}) em uma equação diferencial ordinária
\beq\label{2.131}
\left[\frac{d^{2}}{dr_{*}^{2}}+\omega^{2}-V_{L}(r)\right]\hat{\Psi}_{L}(r)=0,
\eeq
cujo potencial efetivo é dado por
\beq\label{2.132}
V_{L}(r)=\left(1-\frac{2M}{r}\right)\left[\frac{L(L+1)}{r^{2}}+\frac{2M}{r^{3}}\right].
\eeq
Este potencial corresponde a uma barreira de potencial , cujo máximo é $r=3M$ (figura(\ref{potencial_escalar})). Com isso podemos tratar vários dos problemas de perturbações de buracos negros com métodos da espalhamento da mecânica quântica.

Podemos verificar que quando $r_{*}\rightarrow \infty$, que corresponde ao infinito espacial, e quando $r_{*}\rightarrow -\infty$, correspondendo ao horizonte de eventos, o potencial efetivo (\ref{2.132}) vai a zero, ou seja
\beq\label{2.133}\nonumber
V_{L}(r_{*}\rightarrow\pm\infty)\rightarrow 0,
\eeq
então o comportamento assintótico das soluções de (\ref{2.131}) é da forma
\beq\label{2.134}
\hat{\Psi}_{L}(r,\omega)\sim e^{\pm i\omega r},
\eeq
tanto no horizonte de eventos($r_{*}\rightarrow -\infty$), quanto no infinito($r_{*}\rightarrow +\infty$).

Dado este perfil assintótico das soluções (\ref{2.134}), podemos construir soluções para a equação de onda a partir das condições de contorno. Uma classe de soluções de parti\-cular interesse é aquela que satisfaz as condições de contorno chamadas de {\it{causais}}. Ela correspon\-de a situação em que não haja ondas emergindo do horizonte de eventos do buraco negro, isto é
\beq\label{2.135}\nonumber
\hat{\Psi}_{L}^{in}(r_{*})\rightarrow -\infty)&\sim& e^{-i\omega r_{*}}\quad ,\\ 
\hat{\Psi}_{L}^{in}(r_{*}\rightarrow +\infty)&\sim& A_{out}(\omega)e^{i\omega r_{*}}+A_{in}(\omega)e^{-i\omega r_{*}}\quad ,
\eeq
supondo que $\omega$ seja um número real e positivo. Como em \cite{frolov1} chamaremos esta classe de soluçõe de modos {\it{in}}, e seu complexo conjugado de modos {\it{out}}. Da teoria das equações diferenciais \cite{butkov}, sabemos que duas soluções linearmente independentes de (\ref{2.131}) têm o wronskiano constante em $r_{*}$. Calculando o wronskiano dos modos {\it{in}} e {\it{out}}  no horizonte de eventos e no infinito, temos que 
\beq\label{2.136}
|A_{in}|^{2}=|A_{out}|^{2}+1\quad ,
\eeq
de maneira similar
\beq\label{2.137}
|T|^{2}+|R|^{2}=1,
\eeq
sendo $T=1/A_{in}$ e $R=A_{out}/A_{in}$, os coeficientes de transmissão de reflexão respectivamente. Então parte da onda incidente é absovida pelo buraco negro e outra parte refletida de volta ao infinito.

Um segundo par de soluções básicas, os modos {\it{up}} e modos {\it{down}}, podem ser definidos de maneira análoga. Os modos {\it{up}} correspondem a ondas puramente emergentes no infinito, que são dados pelas seguintes condições de contorno
\beq\label{2.138}\nonumber
\hat{\Psi}_{L}^{up}(r_{*}\rightarrow -\infty)&\sim& B_{out}(\omega)e^{i\omega r_{*}}+B_{in}(\omega)e^{-i\omega r_{*}}\\
\hat{\Psi}_{L}^{up}(r_{*}\rightarrow +\infty)&\sim& e^{+i\omega r_{*}}.
\eeq
O modo {\it{down}} é o complexo conjugado do modo {\it{up}}. Usando novamente o fato do wronskiano referente a estas soluções ser constante, temos que 
\beq\label{2.139}
B_{out}(\omega)=A_{in}(\omega),\\
B_{in}=-\bar{A}_{out}(\omega)=-{A}_{out}(-\omega),
\eeq
com a barra denotando a operação de conjugação complexa. Quaisquer duas soluções mencionadas({\it{in}}, {\it{out}}, {\it{up}}, {\it{down}}) podem ser escolhidas como soluções básicas de um dado problema.

Podemos estudar as perturbações lineares na solução de Kerr da mesma forma que foi feito para o caso esfericamente simétrico, mas o caso com rotação se apresenta mais intrincado devido a simetria axial da solução. A evolução de campos escalares nesta geometria foi estudada por Brill \cite{brill1}. Por algum tempo acreditou-se que não seria possível separar a equação de onda nessa geometria a não ser para campos de spin zero. Mas no trabalho \cite{teuk1}, Teukolsky mostrou, usando o formalismo de Newman-Penrose \cite{chandra}, que as perturbações escalares, eletromagnéticas e gravitacionais de um buraco negro de Kerr podem ser descritas por uma única equação, chamada de equação mestra de Teukolsky, que no caso de ausência de fontes é dada por
\beq\label{2.140}\nonumber
&&\left[\frac{(r^{2}+a^{2})^{2}}{\Delta}-a^{2}\sin^{2}\theta\right]\frac{\partial^{2}\Psi}{\partial t^{2}}+\frac{4Mar}{\Delta}\frac{\partial^{2}\Psi}{\partial t\partial\phi}+\left[\frac{a^{2}}{\Delta}-\sin^{-2}\theta\right]\frac{\partial^{2}\Psi}{\partial\phi^{2}}\\\nonumber
&-&\Delta^{-s}\frac{\partial}{\partial r}\left(\Delta^{s+1}\frac{\partial\Psi}{\partial r}\right)-\sin^{-1}\theta\frac{\partial}{\partial\theta}\left(\sin\theta\frac{\partial\Psi}{\partial\theta}\right)-2s\left[\frac{a(r-M)}{\Delta}\right]\frac{\partial\Psi}{\partial\phi}\\
&+&\frac{i\cos\theta}{\sin^{2}\theta}\frac{\partial\Psi}{\partial\phi}-2s\left[\frac{M(r^{2}-a^{2})}{\Delta}-r-ia\cos\theta\right]\frac{\partial\Psi}{\partial t}+(s^{2}cot^{2}\theta-s)\Psi=0 ,
\eeq 
na qual $s$ assume os valores $0,\pm 1, \pm 2$, correspondendo respectivamente a perturbações escalares, eletromagnéticas e gravitacionais. 

Usando o {\it{Ansatz}} de Teukolsky
\beq\label{2.141}
\Psi_{Lm}=R_{Lm}(r, \omega)S_{Lm}(\theta, \phi)e^{-i\omega t}\quad ,
\eeq
a equação (\ref{2.140}) pode ser separada em duas equações diferenciais acopladas
\beq\label{2.142}
&&\Delta^{-s}\frac{d}{dr}\left(\Delta^{s+1}\frac{dR}{dr}\right)+\left[\frac{K^{2}-2is(r-M)K}{\Delta}+4is\omega r-\lambda\right]R=0,\\\label{2.143}
&&\frac{1}{\sin\theta}\frac{d}{d\theta}\left(\sin\theta\frac{dS}{d\theta}\right)+\left[a^{2}\omega^{2}\cos^{2}\theta-\frac{m^{2}}{\sin^{2}\theta}-2a\omega s\cos\theta\right]S\\\nonumber
&&-\left[\frac{2\cos\theta ms}{\sin^{2}\theta}-s^{2}\cot^{2}\theta + E-s^{2}\right]S=0\quad ,
\eeq
em que
\beq\label{2.144}\nonumber
K\equiv (r^{2}+a^{2})\omega -am,
\lambda\equiv E-s(s+1)+(a\omega)^{2}-2am\omega\quad ,
\eeq
e $E=E(a,\omega)$ é a constante de separação. Pelo fato dessas equações serem acopladas não podemos resolvê-las diretamente, pois as funções $R$ e $S$ dependem da frequência $\omega$. A equação (\ref{2.143}) forma um problema de Sturm-Liouville para cada $(a\omega)^{2}$ complexo e $m$ inteiro positivo, cujas auto-funções são os harmônicos esferoidais $S_{L|m|}(\theta\phi)$ com autovalores $E(L,|m|,\omega)$. Estas autofunções formam um conjunto completo no caso de $\omega$ ser real \cite{siedel}.

Usando a seguinte transformação \cite{frolov1}
\beq\label{2.145}
\chi=\Delta^{s/2}\sqrt{r^{2}+a^{2}}R\quad ,
\eeq
 na equação radial (\ref{2.142}), esta pode ser reescrita como
\beq\label{2.146}
\left[\frac{d^{2}}{dr_{*}^{2}}+V(r_{*};\omega,s,\lambda)\right]\chi=0\quad ,
\eeq
em que a coordenada tartaruga neste caso é dada por
\beq\label{2.147}
\frac{d}{dr_{*}}=\frac{\Delta}{r^{2}+a^{2}}\frac{d}{dr}\quad ,
\eeq
e o potencial efetivo é
\beq\label{2.148}
V(r_{*};\omega,s,\lambda)=\frac{K^{2}-2is(r-M)K+\Delta(4ir\omega s-\lambda)}{\left(r^{2}+a^{2}\right)^{2}}-G^{2}-\frac{dG}{dr_{*}}\quad ,
\eeq
onde
\beq\label{2.149}\nonumber
G=\frac{s(r-M)}{r^{2}+a^{2}}+\frac{r\Delta}{\left(r^{2}+a^{2}\right)^{2}}\quad .
\eeq
O problema das perturbações em torno da solução de Kerr se reduz, da mesma forma que no caso sem rotação, a um problema de penetração de uma barreira de potencial.

O comportamento assintótico do potencial (\ref{2.148}) leva ao  comportamento
\beq\label{2.150}\nonumber
\chi(r\rightarrow r_{+})&\sim& \Delta^{\pm s/2}e^{\pm \bar{\omega}r_{*}}\quad ,\\
\chi(r\rightarrow +\infty)&\sim& r^{\pm s}e^{\mp \bar{\omega}r_{*}}\quad .
\eeq
Para $R$ teremos
\begin{equation}\label{2.151}
R(r\rightarrow r_{+})\sim\left\{
\begin{array}{ll}
e^{i\bar{\omega}r_{*}}\\
\Delta^{-s}e^{-i\bar{\omega}r_{*}}
\end{array}
\right\}\quad ,
\end{equation}

\begin{equation}\label{2.152}
R(r\rightarrow +\infty)\sim\left\{
\begin{array}{ll}
\frac{e^{+i\omega r_{*}}}{r^{-(2s+1)}}\\
\frac{e^{-i\omega r_{*}}}{r}
\end{array}
\right\}\quad ,
\end{equation}
sendo $\bar{\omega}$ definido como $\bar{\omega}=\omega-M\Omega_{H}$, com $\Omega_{H}$ representando a velocidade angular do buraco negro.

Nesta seção apresentamos as princiapis idéias sobre perturbações em buracos negros em especial para os buracos negros de Schwarzschild e Kerr. Na próxima seção definiremos modos quase-normais de um buraco negro, e o detalhamento da evolução de campos na métrica de Kerr.

\section{Modos Quasi-Normais(MQN)}

Segundo Chandrasekhar \cite{chandra}, a descrição da evolução de uma pequena perturbação inicial de um buraco negro pode, em princípio, ser determinada pela expressão dessa perturbação em termos de uma superposição de modos normais de vibração. No estágio final, entretanto, espera-se que  esta perturbação inicial decaia de maneira que expresse as principais caracterísiticas do buraco negro pertur\-bado, independentemente da natureza da perturbação inicial. Espera-se, então, que os estágios finais da perturbação, o buraco negro oscile com frêquências e taxas de amortecimento características unicamente dele próprio \cite{frolov1}. Disto, definimos MQN como soluções das equações de perturbação cujas frequências sejam complexas, que descrevam ondas que apenas imergem no horizonte de eventos e ondas puramente emergindo no infinito.

As condições de contorno para que se obtenha os MQN são
\beq\label{2.153}
\Psi(r_{*}\rightarrow \pm\infty)\sim e^{\pm i\omega r_{*}}\quad ,
\eeq
cujas frequências são números complexos
\beq\label{2.154}
\omega=Re(\omega)+Im(\omega).
\eeq

A questão da estabilidade das soluções tipo buraco negro define o sinal da parte imaginária das frequências (\ref{2.154}). A perturbação inicial deve cair de maneira amortecida, dada a definição de MQN, com o tempo, isto é, quando $t\rightarrow \infty$ devemos ter $e^{-i\omega t}=e^{-iRe(\omega)t}e^{Im(\omega)t}\rightarrow 0$, que será possível apenas se $Im(\omega)<0$. 

Os modos ou frequências quasi-normais, são as frequências $\omega$ determinadas pelas soluções de uma equação do tipo (\ref{2.123})(\ref{2.146}) juntamente com as condições de contorno (\ref{2.153}). Devemos considerar apenas ondas puramente emergentes no infinito e puramente imergentes no horizonte de eventos, em métricas que sejam assintóticamente planas ou assintóticamente de-Sitter, uma vez que é esta estrutura assintótica que determina estas condições. Segundo \cite{leaver1}, devido a semelhança entre MQN e ressonâncias de espalhamento, estes modos podem ser identificados como pólos de uma apropriada função de Green. 

Dada uma equação de onda do tipo (\ref{2.123}) com potenciais $V\geq 0$, podemos usar a trasformação de Laplace para representar as soluções que tenham frequências quasi-normais \cite{kokotas1}. A trasformação de Laplace $\bar{\Psi}(s,r_{*})$, com $s>0$ e real, da solução $\Psi(t,r_{*})$ é dada pelo operador
\beq\label{2.155}
\mathcal{L}[\Psi(t,r_{*}),s]=\bar{\Psi}(s,r_{*})=\int^{\infty}_{0}e^{-st}\Psi(t,r_{*})\quad ,
\eeq
aplicando este operador em (\ref{2.123}) obtemos a seguinte equação diferencial ordinária
\beq\label{2.156}
s^{2}\bar{\Psi}-\bar{\Psi}^{''}+V\bar{\Psi}=s\bar{\Psi}(0,r_{*})+\frac{\partial}{\partial t}\bar{\Psi}(0,r_{*})\quad .
\eeq
A parte inomogênea dessa equação é determinada pelas condições de contorno, denotaremos esta parte por
\beq\label{2.157}
h(s,r_{*})=s\bar{\Psi}(0,r_{*})+\frac{\partial}{\partial t}\bar{\Psi}(0,r_{*})\quad .
\eeq

Podemos apresentar a solução formal da equação (\ref{2.156}) através do método das funções de Green \cite{butkov}, que consiste em representar a solução $\bar{\Psi}$ em termos da função de Green $\hat{G}(r_{*},r_{*}^{'};s)$ e do termo inomogêneo (\ref{2.157}), tal como
\beq\label{2.158}
\bar{\Psi}=\int \hat{G}(r_{*},r_{*}^{'};s)h(s,r_{*}^{'})dr_{*}^{'}\quad .
\eeq

Todas funções de Green podem ser construídas a partir de duas soluções linearmente independentes da equação (\ref{2.156}) quando (\ref{2.157}) for zero. Sejam essas duas soluções $f_{-}(s,r_{*})$ e $f_{+}(s,r_{*})$, então a função de Green correspondente será
\begin{equation}\label{2.159}
\hat{G}(r_{*},r_{*}^{'};s)=\frac{1}{W(s)}\left\{
\begin{array}{ll}
f_{-}(s,r_{*}^{'})f_{+}(s,r_{*}) \hspace{0.4cm} r_{*}^{'}<r_{*}\\
f_{-}(s,r_{*})f_{+}(s,r_{*}^{'}) \hspace{0.4cm} r_{*}^{'}>r_{*}\
\end{array}
\right\}\quad ,
\end{equation}
sendo $W(s)$ o wronskiano, que nesse caso é dado por
\beq\label{2.160}
W(s)=f_{-}(s,r_{*})\frac{\partial}{\partial r_{*}}f_{+}(s,r_{*})-f_{+}(s,r_{*})\frac{\partial}{\partial r_{*}}f_{-}(s,r_{*})\quad .
\eeq
O comportamento assintótico dos potenciais $V(r_{*})$ são utilizados na seleção das funções $f_{-}(s,r_{*})$, $f_{+}(s,r_{*})$. 

Da representação em termos das funções de Green da transformada de Laplace de $\Psi(t,r_{*})$, podemos obter, usando a fórmula de inversão complexa de Bromwich, formalmente $\Psi(t,r_{*})$ como
\beq\label{2.161}
\Psi(t,r_{*})=\frac{1}{2\pi i}\int \bar{\Psi}(t,r_{*})e^{st}ds\quad .
\eeq
Tomamos a continuação analítica dessa integral em $s$, cuja integração é feita ao longo da reta $s=j$ do plano complexo. O número $j$ é real e escolhido de modo que a reta $s=j$ fique à direita de todas singularidades de $\bar{\Psi}(t,r_{*})$, e completada com um semi-círculo $C$ de raio $R$. Usando o teorema dos resíduos neste contorno, temos que
\beq\label{2.162}
\oint \bar{\Psi}(t,r_{*})e^{st}ds=2\pi i\sum_{s_{i}}Res(s_{i})=-\Psi_{QN}\quad ,
\eeq
sendo $Res(s_{i})$ o resíduo correspondente ao pólo $s_{i}$. Temos aisnda que 
\beq\label{2.163}
\oint \bar{\Psi}(t,r_{*})e^{st}ds=\Psi+\Psi_{C}+\Psi_{t}\quad ,
\eeq
ou seja, de maneira geral a onda pode ser escrita como a soma de três termos
\beq\label{2.164}
\Psi=\Psi_{QN}+\Psi_{C}+\Psi_{cont}\quad ,
\eeq
sendo o primeiro termo a contribuição quasi-normal da representação da onda. Por (\ref{2.162}), vemos que esta contribuição provém dos pólos na função $\bar{\Psi}$, que são originados pelos zeros do Wronskiano. Além disso, temos o termo $\Psi_{cont}$ referente a integração pelo contorno usado para escapar da singularidade de ramificação, e o termo $\Psi_{C}$ oriundo da integração pelo arco $C$.

A condição para que o Wrosnkiano se anule é que as funções $f_{-}(s,r_{*})$ e $f_{+}(s,r_{*})$ sejam linearmente dependentes, isto é,
\beq\label{2.165}
f_{+}(s,r_{*})= q(s_{i})f_{-}(s,r_{*})\quad ,
\eeq
sendo $q(s_{i})$ uma constante que depende do modo quasi-normal.

Na seção anterior mencionamos que quaisquer duas das classes de soluções {\it{in}}, {\it{out}}, {\it{up}}, {\it{down}}, podem ser escolhidas como soluções básicas de um problema envolvendo perturbações no buraco negro de Schwarzschild. Para a caracterização de MQN, os modos {\it{in}} (\ref{2.135}) e modos {\it{up}} (\ref{2.138}) são os tipos adequados. O Wronskiano construído a partir desses modos é \beq\label{2.166}
W_{[in,up]}=\hat{\Psi}_{L}^{in}(r_{*})\frac{d}{dr_{*}}\hat{\Psi}_{L}^{up}(r_{*})-\hat{\Psi}_{L}^{up}(r_{*})\frac{d}{dr_{*}}\hat{\Psi}_{L}^{in}(r_{*})=2i\omega A_{in}(\omega)=2i\omega B_{out}(\omega).
\eeq

\begin{figure}[ht]
\begin{center}
\includegraphics[height=10cm,width=14cm]{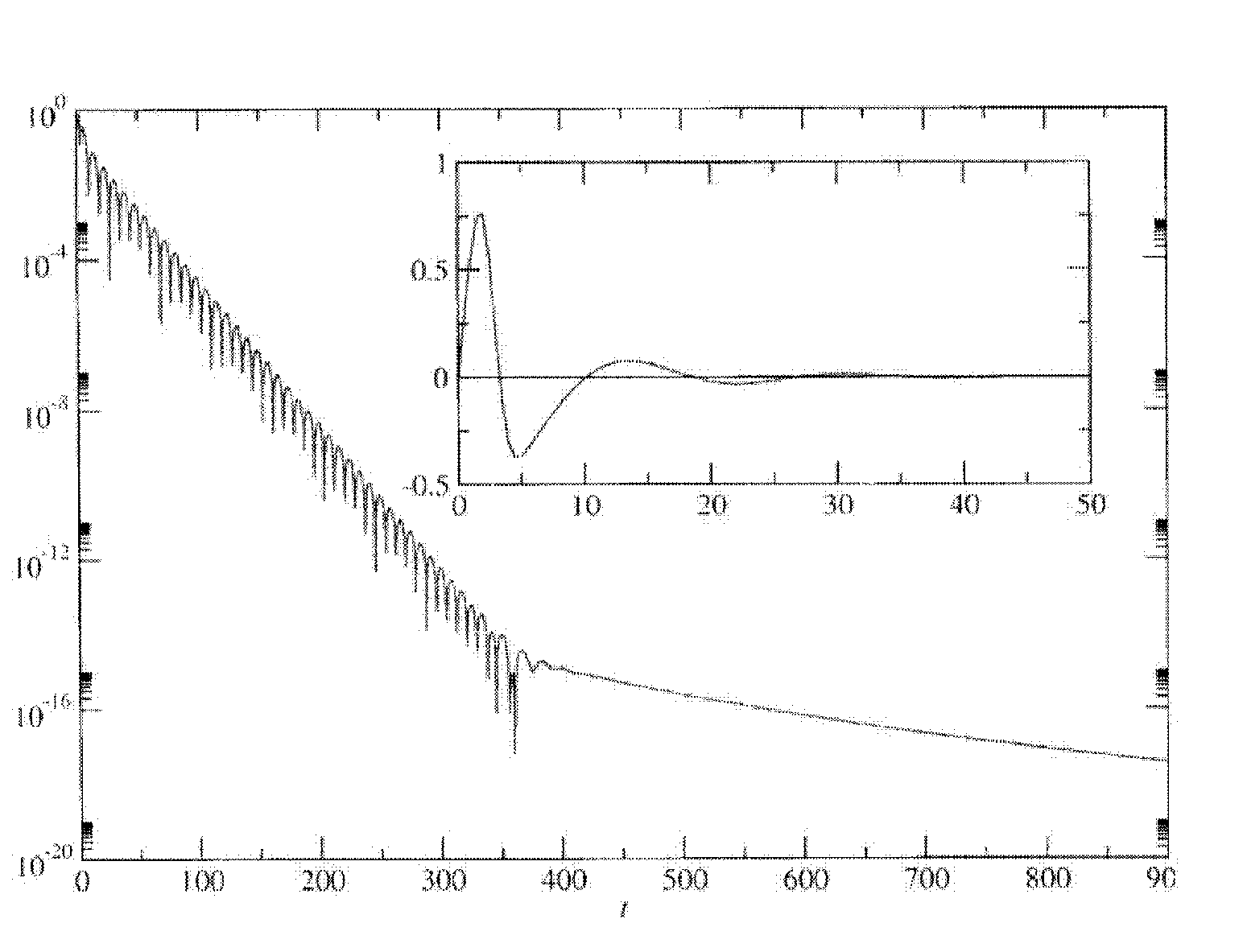}
\end{center}
\caption{Perfil da evolução temporal de uma perturbação inicial da métrica de  Schwarzschild em escala loagrítmica no quadro maior e linear no menor.}\label{perturbacao_inicial}
\end{figure}

Para grandes valores de $r$ o potencial (\ref{2.132}) decai fracamente. Como consequência, o decaimeto de uma perturbação inicial não terá o comportamento de uma exponecial decrescente. No decaimento de um campo escalar massivo no espaço-tempo de  Schwarzschild, observa-se que para grandes valores de $r$, o campo decai como uma lei de potência do tipo \cite{price1},
\beq\label{2.167}
\Psi(r,t)\sim t^{-(2L+P+1)}\quad ,
\eeq
sendo $P=1$, se o campo for estático inicialmente e $P=2$ se não. A figura (\ref{perturbacao_inicial}) mostra a evolução temporal de uma perturbação inicial em um buraco negro de  Schwarzschild.

Uma das primeiras técnicas para a resolução do problema de encontrar os MQN, é o método WKB proposto por Schutz e Will \cite{schutz1}. Este método é baseado no procedimento padrão WKB usado para o estudo de espalhamentos em potenciais barreira. Eles mostraram que as frequências complexas podem ser estimadas a partir da relação
\beq\label{2.268}
\left[M\omega_{n}\right]^{2}=V_{L}(r=r_{max})-i\left(n+\frac{1}{2}\right)\left[-2\frac{d^{2}}{dr^{2}_{*}}V_{L}(r=r_{max})\right]^{1/2}\quad ,
\eeq
sendo $r_{max}$ o máximo da barreira de potencial. O modo quasi-normal fundamental $n=0$, $L=1$ é $M\omega\approx (0.37-0.09)$. Se considerarmos um buraco negro esfericamente simétrico, cuja massa seja $M=10M_{\odot}$, este modo fundamental corresponde a uma frequência de 1.2kHz, com um tempo de amortecimento de 0.55ms. Na tabela (\ref{mqs}) tem-se as primeiras quatro frequências quasi-normais.
\begin{table}[h]
\begin{center}
\begin{tabular}{|l|l|l|l|}
\hline
n & L=2 & L=3 & L=4\\
\hline
0 & 0.37367 - 0.8896i & 0.59944 -0.09270i & 0.80918 - 0.9416i\\
\hline
1 & 0.34671 - 0.27391i & 0.58264 -0.28130i & 0.79663 - 0.28443i\\
\hline
2 & 0.30105 - 0.47828i & 0.55168 - 0.47909i & 0.77271 - 0.47991i\\
\hline
3 & 0.25150 - 0.70514i & 0.51196 - 0.69034i & 0.73984 - 0.68392i\\
\hline
\end{tabular}\caption{Primeiras frequências quasi-nomais do buraco negro de Schwarzschild.}\label{mqs}
\end{center}
\end{table}
Observamos nesta tabela que aparte imaginária das frequências aumenta muito rapidamente, o que indica que modos com  $n$ grande não contribuem de maneira significante para o sinal gravitacional. A parte real se torna constante para modos cada vez maiores. O perfil assintótico das frequências quasi-normais foi obtido por Nollert \cite{nollert1} dado pela expressão
\beq\label{2.269}
M\omega_{n}\approx 0.0437+\frac{\gamma_{1}}{\sqrt{2n+1}}+\cdots -i\left[-\frac{1}{8}(2n+1)+\frac{\gamma_{1}}{\sqrt{2n+1}}\cdots\right]\quad ,
\eeq
com $\gamma_{1}$ assumindo os valores 0.343, 0.7545 e 2.81, para $L=2$, 3 e 6, respectivamente, em que $n\rightarrow +\infty$. Para grande valores de $L$, a distribuição das frequências quasi-normais é dada por \cite{iyer1}
\beq\label{2.270}
3\sqrt{3}M\omega_{n}\approx L+\frac{1}{2}-i\left(n+\frac{1}{2}\right)\quad .
\eeq

No caso de buracos negros em rotação, para cada valor de multipolo $L$ existem $2L+1$ modos distintos, que se aproximam dos valores obtidos para a métrica de Schwarzschild quando $a\rightarrow 0$. Estes modos correspondem a diferentes valores de $m$, onde $-L\leq m\leq L$. O trabalho pioneiro no cálculo dessas frequências para o caso com rotação é o de Detweiler \cite{detweiler1}. Um dos seus principais resultados é que para $a\rightarrow M$ e $L=m$, a parte imaginária dos MQN fica praticamente constante e a parte real cresce monotonicamente. Para $L=-m$ a parte imaginária tende a zero e a parte real para $-m/2$. Isto sugere que um grande número de modos quasi-normais possuem um amortecimento que aumenta vagarosamente quando $a\rightarrow M$. Segundo \cite{leaver2}, estes modos devem coalecer num único modo sem amortecimento para $a=M$. Quando parâmetro de rotação é igual a massa  temos o buraco negro de Kerr extremo, e é possível mostrar que existe uma sequência infinita de frequências resonantes com mesmo limite, neste caso.
\section{O espalhamento super-radiante}

Seja o espalhamento de uma onda numa barreira de potencial gerada por um buraco negro, em geral espera-se que parte da onda incidente penetre na barreira de potencial e seja absorvida pelo buraco negro e outra seja espalhada para o infinito. Como consequência, teremos a amplitude da onda espalhada menor ou igual que a amplitude da onda incidente. Considerando o caso de buracos negros em rotação isto não é necessariamente verdade. Uma onda incidente num  buraco negro de Kerr pode ter sua amplitude amplificada num experimento de espalhameto no potencial desse buraco negro \cite{starobinski1} \cite{starobinski2}. Esta energia adicional, que aparece na amplitude espalhada no infintito é devido a absorção da energia de rotação do buraco negro.

Consideremos então, o espalhamento de um campo escalar pelo potencial (\ref{2.148}). O comportamento assintótico para tal campo, neste caso, é dado por
\beq\label{2.271}\nonumber
\chi^{in}(r_{*}\rightarrow r_{+})&\sim& e^{-i\bar{\omega}r_{*}}\quad ,\\
\chi^{in}(r_{*}\rightarrow +\infty)&\sim&  A_{out}e^{i{\omega}r_{*}} +A_{in}e^{-i\omega r_{*}}\quad .
\eeq
Usando o complexo conjugado de $\chi^{in}$, que chamaremos de $\chi^{out}$, podemos calcular o Wrons\-kiano, sendo $\chi^{in}$ e $\chi^{out}$ as soluções escolhidas para (\ref{2.147}). Então, usando a condição de que o Wronskiano seja constante, temos que
\beq\label{2.272}
\left(1-\frac{m\Omega_{H}}{\omega}\right)|T|^{2}=1-|R|^{2}\quad ,
\eeq
os coeficientes de trasmissão $T$ e reflexão $R$ serão discutido mais adiante no capítulo 4, onde estudaremos este fenômeno para o caso em que tenhamos uma dimensão extra. Desta equação fica claro que teremos o fenômeno de amplificação da onda espalhada(super-radiância $R>1$) quando 
\beq\label{2.273}
\omega<m\Omega_{H}=\frac{ma}{2Mr_{+}}\quad ,
\eeq 
apenas para valores positivo de $m$.

Podemos observar esta extração de energia do buraco negro através das condições de contorno (\ref{2.151}). Se $\bar{\omega}$ se torna negativo, a solução com o termo $e^{-i\bar{\omega}r_{*}}$ será vista por um observador no infinito como uma onda emergente do horizonte de eventos. De acordo com \cite{starobinski1}, o máximo de amplificação da onda incidente é de $0.3\%$ para uma onda escalar, $4.4\%$ para o campo eletromagnético e $138\%$ para ondas gravitacionais. Voltaremos ao assunto sobre a super-radiância no capítulo 4.


\chapter{Será o nosso Universo uma brana?}

\section{Motiva\c c\~oes para os modelos de mundo brana}

A teoria de supercordas apresenta-se at\'e o momento com grandes possibilidades de realizar a unifica\c c\~ao das quatro intera\c c\~oes fundamentais numa \'unica descri\c c\~ao coerente. Uma de suas caracter\'{\i}sticas mais surpreendentes \'e a necessidade do aparecimento de dimens\~oes espaciais extras al\'em das tr\^es usuais. Esta id\'eia j\'a havia sido utilizada na d\'ecada de $1920$ por Kaluza e Klein, no contexto da unifica\c c\~ao da gravita\c c\~ao, cuja descri\c c\~ao \'e dada pela Relatividade Geral, com Eletromagnetismo de Maxwell numa \'unica teoria atrav\'es do uso de uma dimens\~ao espacial extra compactificada.

Existem verdadeiramente, cinco teorias de supercordas livres de anomalias: tipo I, tipo IIA, tipo IIB, SO($32$) e heter\'otica $E_{8}\times E_{8}
$. O fato \'e que todas elas para serem quânticamente consistentes, necessitam em geral dez dimens\~oes espaciais. \'E nesse contexto que Horava e Witten conjecturaram, usando o fato das dualidades entre as diferentes teorias de supercordas, que as cinco teorias livres de anomalias mais a supergravidade em onze dimens\~oes s\~ao de fato diferentes aspectos de uma \'unica teoria, denominada teoria M \cite{witten1}. Esta teoria seria verdadeiramente a prescri\c c\~ao \'unica para todas as intera\c c\~aoes fundamentais. \'E nesse contexto que surgem os modelos que pretendem encontrar algumas propriedades n\~ao perturbativas da conjectura de Horava-Witten, tais como os modelos de mundo brana.

Nessa busca por uma descri\c c\~ao n\~ao perturbativa da teoria M descobriu-se que outros objetos de dimens\~ao maior que a das cordas, que aparecem devido as condi\c c\~oes de contorno das equa\c c\~oes de movimento na folha mundo, chamados de $p-$branas, no qual $p$ se refere a dimensionalidade da brana, realizam papel fundamental na teoria. Considerando o regime de baixo acoplamento na intera\c c\~ao entre cordas, as $p-$branas n\~ao aparecem na teoria perturbativa, sendo, portanto, um aspecto exato da descri\c c\~ao. De particular import\^ancia dentre as $p-$branas, est\~ao as $D-$branas, que s\~ao as estruturas onde as cordas abertas se prendem. Falando de maneira simples, cordas abertas, que descrevem o setor n\~ao gravitacional da teoria, tem suas pontas presas nas branas, enquanto que as cordas fechadas, que descrevem o setor gravitacional, podem mover-se livremente em todas dimens\~oes. A esse conjunto de dimens\~oes a que as cordas fechadas tem acesso chamamos de {\it{bulk}}. Classicamente, este cen\'ario \'e realizado atrav\'es da localiza\c c\~ao da mat\'eria e campos radiativos na brana, enquanto as ondas gravitacionais podem se propagar em todo bulk.

Na conjectura de Horava-Witten, os campos de calibre do modelo padr\~ao s\~ao confinados em duas $(1+9)-$branas localizadas nos pontos fixos de um orbifold $S_{1}/Z_{2}$.\footnote{Espa\c co obtido pela identifica\c c\~ao $x\rightarrow -x$ com pontos fixos, isto \'e, tal como na reta real o n\'umero zero \'e identificado com ele mesmo portanto \'e um ponto fixo do orbifold.} As $6$ dimens\~oes extras nas branas s\~ao compactificadas num raio muito pequeno, pr\'oximo da escala de Planck, para que este cen\'ario esteja em acordo com os experimentos.

Podemos tomar esta descri\c c\~ao de Horava-Witten como uma teoria efetiva em $5$ dimens\~oes, onde consideramos os campos de calibre definidos apenas nas $(3+1)-$branas, ou simplesmente $3-$branas, localizadas nos pontos fixos de um orbifold num bulk $5-$dimensional. Esta dimens\~ao espacial adicional pode ser maior que a escala de Planck, o que torna o estudo desses modelos bastante atraente, pois se espera na pr\'oxima gera\c c\~ao de aceleradores de part\'{\i}culas tal como o LHC ({\it{Large Hadron Collider}}), chegar at\'e a faixa de energia onde se podem encontrar ind\'{\i}cios de dimens\~oes espaciais extras previstas por estes modelos de mundo brana.

Considerar a descri\c c\~ao de Horava-Witten, com uma teoria efetiva em $5-$dimens\~oes \'e o ponto de partida para o modelo de Randall-Sundrum de duas $3-$branas definidas num  bulk $5-$dimensional \cite{rs1} (RS1). Quando fazemos o raio do orbifold tender ao infinito, obtemos o modelo de Randall-Sundrum com uma \'unica $3-$brana num bulk $5-$dimensional (RS2), cuja dimens\~ao extra \'e infinita. Mas estas n\~ao s\~ao as \'unicas tentivas de aplicar as id\'eias advindas da conjectura da teoria M. Antes da tentativa de Randall-Sundrum, Arkani-Hamed-Dimopoulos-Dvali (ADD) \cite{add} tiveram a id\'eia de diminuir a escala de Planck $M_{p}\equiv M_{4}$ atrav\'es do volume da dimens\~ao espacial extra compacta, considerando que a gravidade se ``espalhe'' por este volume. Ent\~ao a escala da Planck em quatro dimens\~oes n\~ao seria verdadeiramente a escala fundamental, mas apenas uma escala efetiva.

Para compreender as id\'eias do modelo ADD consideremos
a  a\c c\~ao gravitacional de Einstein-Hilbert em $(4+d)$-dimens\~oes \cite{maartens1},
\be
\label{1.1}
S_{g}=\frac{1}{16\pi G_{4+d}}\int d^{4}x d^{d}y \sqrt{-g_{4+d}}\left[R_{4+d}-2\Lambda_{4+d}\right]\quad ,
\ee
na qual $G_{4+d}$ \'e a constante da gravita\c c\~ao universal, $g_{4+d}$ o determinante da m\'etrica, $\Lambda_{4+d}$ a constante cosmol\'ogica e  $R_{4+d}$ o escalar de Ricci, todas quantidades escritas em  $(4+d)$ dimens\~oes. E o sistema de coordenadas que cobre esta variedade $(4+d)-$dimensional \'e $X^{A}=(x^{\mu},..,y^{d})$, sendo as letras gregas os \'{\i}ndices $4-$dimensionais tais como $\mu$ e as letras latinas min\'unsculas representam o sistema de coordenadas das dimens\~oes extras como $d$ e as latinas mai\'usculas representam \'{\i}ndices que percorrem todas as dimens\~oes, tal conven\c c\~ao dever\'a ser usada ao longo de todo este texto, al\'em do sistema de unidade geom\'etrico no qual $c=\hbar=1$.
 
Aplicando o princ\'{\i}pio de mínima ação em (\ref{1.1}), obtemos as equa\c c\~oes de campo de Einstein,
\be
\label{1.2}
G_{AB}^{4+d}\equiv R_{AB}^{4+d} g_{AB}^{4+d}-\frac{1}{2}g_{AB}^{4+d}R^{4+d}=-\Lambda_{4+d} g_{AB}^{4+d}+\kappa^{2}_{4+d} T_{AB}^{4+d}\quad ,
\ee
sendo  $T_{AB}^{4+d}$ o tensor energia momento, e $\kappa^{2}_{4+d}$ a constante de acoplamento da intera\c c\~ao gravitacional, cuja rela\c c\~ao com a constante da gravita\c c\~ao universal e a massa de Planck $M_{4+d}^{2+d}$ em $4+d$ dimens\~oes \'e
\be
\label{1.3}
\kappa^{2}_{4+d}=8\pi G_{4+d}=\frac{8\pi}{M_{4+d}^{2+d}}\quad .
\ee

No limite de campo fraco a equa\c c\~ao(\ref{1.2}) se reduz à equa\c c\~ao de Poisson $(4+d)-$dimensional, cuja solu\c c\~ao \'e o potencial gravitacional

\be
V_{4+d}=\frac{\kappa^{2}_{4+d}}{r^{1+d}}\quad ,
\ee
sendo $r$ a coordenada radial em $(1+d)$ dimens\~oes. Nesta express\~ao fica claro que se n\~ao tivermos nenhuma dimens\~ao espacial extra obteremos o potencial gravitacional usual, mas do contr\'ario a lei da gravita\c c\~ao universal deve levar corre\c c\~oes. Podemos entender isso da seguinte forma: se a escala de comprimento das dimens\~oes extras for $L$, ent\~ao para dist\^ancias onde $r<L$, isto \'e para dist\^ancias onde possamos medir estas dimens\~oes extras, o potencial gravitacional deve ser corrigido e assumir a forma $V=r^{-(1+d)}$, mas se observarmos a regi\~ao onde $r>>L$, ent\~ao estas dimens\~oes adicionais n\~ao poder\~ao contribuir para o potencial gravitacional, logo para este caso o potencial gravitacional \'e simplesmente $V=r^{-1}$. 

Uma outra possibilidade \'e considerarmos $r=L$ nas $d$ dimens\~oes extras, assim modi\-ficando o potencial gravitacional para
\be
\label{1.4}
V=\frac{\kappa^{2}_{4+d} L^{-d}}{r}=\frac{8\pi}{L^{d}M^{2+d}_{4+d}r}\quad , 
\ee
ou seja 
\be
\label{1.5}
M^{2}_{4}=M_{4+d}^{2+d}L^{d}\quad ,
\ee
a escala de Planck $4-$dimensional se torna uma constante de acoplamento efetiva, descre\-vendo a gravidade apenas em regi\~oes muito maiores que a escala das dimens\~oes extras. 

Este tipo de hip\'otese de um universo com dimens\~oes espaciais extras seria impossi\-bilitado de se medir dentro da precis\~ao atual dos experimentos, mas pode sugerir uma solu\c c\~ao para o problema da hierarquia, que se refere à grande diferen\c ca entre a escala de acoplamento da for\c ca eletrofraca $M_{ew}$ e a escala de acoplamento da gravita\c c\~ao $M_{4}$, j\'a que se usarmos $M_{4+d}$ como constante de acoplamento fundamental, e fazendo-a da ordem da escala eletrofraca $10^{3}GeV$ e $M_{4}=10^{19}GeV$ podemos inferir sobre a escala de comprimento $L^{d}$ das dimens\~oes extras 
\be
\label{1.6}
L=\frac{(M_{4})^{\frac{2}{d}}}{M_{4+d}^{1+\frac{2}{d}}}=10^{\frac{32}{d}}(TeV)^{-1}\quad ,
\ee
mas como $1cm=10^{17}(TeV)^{-1}$ , ent\~ao

\be
\label{1.7}
L=10^{\frac{32}{d}-17}cm\quad .
\ee
Nesta express\~ao se tivermos $d=1$, implica na viola\c c\~ao da gravita\c c\~ao de Newton em dist\^ancias da ordem do Sitema Solar, mas se considerarmos duas dimens\~oes espaciais extras,
\be
\label{1.8}
L=10^{-1}mm .
\ee

Os experimentos atuais que procuram desvios na lei da gravita\c c\~ao universal  em escalas submilim\'etricas, permitem viola\c c\~oes na regi\~ao abaixo de (\ref{1.8}). Ent\~ao se considerarmos a escala de Planck como uma constante de acoplamento efetiva obteremos uma solu\c c\~ao para o problema da hierarquia.

No modelo ADD, como visto pela express\~ao (\ref{1.7}), devemos ter mais que uma dimens\~ao espacial extra para que tenhamos compatibilidade entre a teoria e os experimentos, al\'em disso estas dimens\~oes extras s\~ao planas. Ao contr\'ario, nos modelos de Randall-Sundrum , as dimens\~oes extras s\~ao curvadas ou {\it{warped}} com o bulk sendo uma por\c c\~ao de um espa\c co-tempo anti-De Sitter $5-$dimensional($AdS_{5}$). Como na conjectura de Horava-Witten, as branas de Randall-Sundrum possuem simetria $Z_{2}$ (simetria de espelho), e possuem um termo escalar chamado de tens\~ao que desempenha o papel de conter a influ\^encia da constante cosmol\'ogica negativa do bulk na brana. Nos modelos RS, as tr\^es dimens\~oes espaciais constatadas experimentalmente s\~ao protegidas das dimens\~oes extras no regime baixas de energias, n\~ao pelo fato destas serem compactificadas, mas sim por estas serem curvadas, ou seja em RS podemos ter dimens\~oes extras grandes.

Os modelos RS s\~ao modelos fenomenol\'ogicos que refletem no m\'{\i}nimo algumas carac\-ter\'{\i}sticas poss\'{\i}veis da conjectura da teoria M. No modelo RS com duas branas (RS2) \'e poss\'{\i}vel o desenvolvimento de id\'eias acerca do princ\'{\i}pio hologr\'afico, que aparecem na conjectura da teoria M. Falando de maneira n\~ao formal, a holografia sugere que a din\^amica da gravita\c c\~ao num espa\c co-tempo de $d-$dimens\~oes pode ser determinado pelo conhecimento dos campos definidos na fronteira $d-1$ deste espa\c co-tempo. A correspond\^encia ADS/CFT \'e um exemplo da holografia, no qual temos a din\^amica cl\'assica da gravita\c c\~ao em um espa\c co-tempo $AdS_{d}$ equivalente a din\^amica qu\^antica da teoria de campos conforme na fronteira $d-1$. O modelo RS2 com um bulk $AdS_{5}$ satisfaz esta correspond\^encia no regime perturbativo de baixa ordem \cite{duff}.

\section{Solu\c c\~ao de Randall-Sundrum}

O primeiro modelo RS \cite{rs1} foi proposto com o intuito de resolver o problema da hierarquia entre as constantes de acoplamento da intera\c c\~ao eletrofraca e da gravitacional por meio de uma \'unica dimens\~ao espacial extra compacta entre duas branas, cuja localiza\c c\~ao destas se d\'a nos pontos fixos do orbifold $S_{1}/Z_{2}$.

Fa\c camos agora neste modelo de duas branas (RS2) a dimens\~ao extra compacta com dom\'{\i}nio $-\pi r_{c}<y<\pi r_{c}$, sendo $r_{c}$ o raio de compactifica\c c\~ao. A nossa brana (nosso universo) está localizada em $y=\pi r_{c}$ a segunda brana, a qual n\~ao temos acesso, em $y=0$.

 Consideraremos as letras gregas representando os \'{\i}ndices das quantidades $4-$dimensionais ($\mu, \nu=0...3$) e as letras latinas mai\'usculas os índices $5-$dimensionais($M, N=0...4$). Dadas estas condi\c c\~oes sup\~oe-se a seguinte a\c c\~ao para este modelo, no qual o bulk n\~ao possui mat\'eria,
\beq
\label{1.9}\nonumber
S&=&\int d^{4}x \int dy\sqrt{-g_{5}}\left[-2\Lambda +R_{5}\right]M^{3}_{5} +\\
&& \int d^{4}x\sqrt{-g_{4}^{vis}}\left[L_{vis}-\lambda_{vis}\right]+ \int d^{4}x\sqrt{-g^{esc}_{4}}\left[L_{esc}-\lambda_{esc}\right],
\eeq
na qual as quantidades $5-$dimensionais s\~ao a constante cosmol\'ogica $\Lambda$, a escala de Planck fundamental, o escalar de Ricci $R_{5}$, o determinate da m\'etrica  $g_{5}$, e  as quantidades $4-$dimensionais $\lambda_{vis}$, $\lambda_{esc}$, as tens\~oes na brana vis\'{\i}vel e escondida respectivamente. Al\'em disso, $L_{vis}$ e $L_{esc}$ representam a lagrangiana da mat\'eria presente em cada brana.

As branas s\~ao hipersuperf\'{\i}cies singulares segundo o formalismo de Israel \cite{israel}. Do ponto de vista $5-$dimensional, as lagrangeanas da mat\'eria confinada em cada brana podem ser expressas com fun\c c\~oes delta de Dirac. Logo podemos escrever
\beq
\label{1.10}\nonumber
L_{vis}-\lambda_{vis}&=&\sqrt{-g_{4}^{vis}}\lambda_{vis}\delta(y-\pi)\quad ,\\
L_{esc}-\lambda_{esc}&=&\sqrt{-g_{4}^{esc}}\lambda_{esc}\delta(y)\quad.
\eeq
Desta forma, variando a a\c c\~ao (\ref{1.9}) obteremos
\beq
\label{1.11}\nonumber
\delta S&=&\int d^4 x dy {\delta(\sqrt{-g_{5}})[-2\Lambda+ R_{5}]M^{3}_{5}+\sqrt{-g_{5}}M^{3}_{5}[\delta (R_{5})]}\\
&+&\int d^4 x\left[\delta(\sqrt{-g_{4}^{esc}})\lambda_{esc}\delta(y)\right] + \int  d^4 x[\delta(\sqrt{-g_{4}^{vis}})\lambda_{vis}\delta(y-\pi r_{c})]\quad ,
\eeq
onde
\beq\label{1.12}\nonumber
\delta \sqrt{-g_{5}}&=&-\frac{1}{2}\sqrt{-g_{5}}g^{MN}\delta g_{MN}\quad ,\\\nonumber
\delta \sqrt{-g_{4}^{vis}}&=&-\frac{1}{2}\sqrt{-g_{4}^{vis}}g^{\mu\nu}_{vis}\delta g_{\mu\nu}^{vis}\quad ,\\\nonumber
\delta \sqrt{-g_{4}^{esc}}&=&-\frac{1}{2}\sqrt{-g_{4}^{esc}}g^{\mu\nu}_{esc}\delta g_{\mu\nu}^{esc}\quad ,\\
\delta R&=& -R^{MN}\delta g_{MN}\quad .
\eeq
Na \'ultima identidade usamos a equa\c c\~ao de Palatini para anular $\delta R_{MN}$. Substituindo (\ref{1.12}) em (\ref{1.11}), obtemos
\beq\label{1.13}\nonumber
\delta S&=&\int d^{4}x dy\left[-\frac{1}{2}\sqrt{-g_{5}}g^{MN}\delta g_{MN}M_{5}^{3}[R_{5}-2\Lambda]+\sqrt{-g_{5}}M^{3}_{5}[-R^{MN}\delta g_{MN}]\right]\\\nonumber
&+&\int d^4 x\left[-\frac{1}{2}\sqrt{-g_{4}^{vis}}g^{\mu\nu}_{vis}\delta g_{\mu\nu}^{vis}\lambda_{vis}\delta(y-\pi r_{c})]\right]\\
&+&\int d^4 x\left[-\frac{1}{2}\sqrt{-g_{4}^{esc}}g^{\mu\nu}_{esc}\delta g_{\mu\nu}^{esc}\lambda_{esc}\delta(y)\right]\quad . 
\eeq

Usando a seguinte rela\c c\~ao entre a m\'etrica $5-$dimensional e a m\'etrica induzida em cada brana,
\beq\label{1.14}\nonumber
\delta g_{\mu\nu}^{vis}&=&\delta g_{MN}\delta^{M}_{\mu}\delta^{N}_{\nu}\quad ,\\
\delta g_{\mu\nu}^{esc}&=&\delta g_{MN}\delta^{M}_{\mu}\delta^{N}_{\nu}\quad  ,
\eeq
e aplicando o princ\'{\i}pio de mínima ação $\delta S=0$, obtemos as equa\c c\~oes de campo
\beq\label{1.15}\nonumber
\sqrt{-g_{5}}\left[R^{MN}-\frac{1}{2}g^{MN}R_{5}\right]&=&-\frac{1}{2M^{3}_{5}}\left[ \sqrt{-g_{5}}\Lambda M^{3}_{5} g^{MN}+ \sqrt{-g_{4}^{vis}}g_{vis}^{\mu\nu}\delta^{M}_{\mu}\delta^{N}_{\nu}\lambda_{vis}\delta(y-\pi r_{c})\right]\\
&-&\frac{1}{2M^{3}_{5}}\sqrt{-g_{4}^{esc}}g_{esc}^{\mu\nu}\delta^{M}_{\mu}\delta^{N}_{\nu}\lambda_{esc}\delta (y)\quad .
\eeq

O {\it{Ansatz}} de Randall e Sundrum para resolver estas equa\c c\~oes de campo \'e um elemento de linha n\~ao fator\'avel $5-$dimensional, onde \'e introduzindo o chamado fator de {\it{warp}} que multiplica o termo $4-$dimensional,
\be\label{1.16}
ds^{2}=e^{2\sigma(y)}\eta_{\mu\nu}dx^{\mu}dx^{\nu} + dy^{2}\quad ,
\ee
no qual $\eta_{\mu\nu}$ \'e a m\'etrica de Minkowski em 4 dimens\~oes, note que  tomamos o raio de compactifica\c c\~ao $r_{c}$ igual a 1. Ent\~ao segundo esta proposta podemos calcular as equa\c c\~oes de campo e encontrar as propriedades da fun\c c\~ao $\sigma(y)$.

As \'unicas componentes n\~ao nulas da conex\~ao para a m\'etrica (\ref{1.16}), com $i, j=1,2,3$ (neste caso índices repetidos n\~ao devem ser somados), s\~ao
\beq\label{1.17}\nonumber
\Gamma^{0}_{0 y}=\Gamma^{0}_{y 0}=\Gamma^{0}_{0 y}=\Gamma^{i}_{i y}=\Gamma^{i}_{i y}=\frac{d\sigma(y)}{dy}\quad ,\\
\Gamma^{y}_{0 0}=-\Gamma^{y}_{i i}=e^{2\sigma(y)}\frac{d\sigma(y)}{dy}\quad .
\eeq
Usando estes resultados, \'e poss\'{\i}vel encontrar as componentes n\~ao nulas do tensor de Ricci,
\beq\label{1.18}
R_{yy}&=&-\frac{\partial}{\partial y}\left(\Gamma_{y0}^{0}+3\Gamma^{i}_{yi}\right)-\left(\Gamma^{0}_{0y}\Gamma^{0}_{0y}+3\Gamma^{i}_{yi}\Gamma^{i}_{yi}\right)\\\nonumber
&=&-4\left(\frac{d^{2}}{dy^{2}}\sigma(y)+\left(\frac{d}{dy}\sigma(y)\right)^{2}\right)\quad ,\\
R_{ii}&=&\frac{\partial}{\partial y}\Gamma^{y}_{ii} + 2\Gamma^{y}_{ii}\Gamma^{0}_{0y}=-e^{2\sigma(y)}\left(\frac{d^{2}}{dy^{2}}\sigma(y)+ 4 \left(\frac{d}{dy}\sigma(y)\right)^{2}\right)\quad ,\\
R_{00}&=&\frac{\partial}{\partial y}\Gamma^{y}_{00}+2\Gamma^{y}_{00}\Gamma^{0}_{y0}=e^{2\sigma(y)}\left(\frac{d^{2}}{dy^{2}\sigma(y)}+4\left(\frac{d}{dy}\sigma(y)\right)^{2}\right)\quad .
\eeq
O escalar de curvatura $5-$dimensional ser\'a
\be\label{1.19}
R=-8\frac{d^{2}}{dy^{2}}\sigma(y)-20\left(\frac{d}{dy}\sigma(y)\right)^{2}\quad .
\ee

Substituindo os resultados (\ref{1.18}-\ref{1.19}) nas equa\c c\~oes de campo (\ref{1.15}), obtemos duas equa\c c\~oes independentes, sendo $G_{MN}$ o tensor de Einstein $5-$dimensional, 
\beq\label{1.20}
G_{yy}&=&-\Lambda g_{yy}\Rightarrow 6\left(\frac{d}{dy}\sigma(y)\right)^{2}=-\Lambda\quad ,\\\nonumber
G_{00}&=&-G_{ii}=-\left[3\frac{d^{2}}{dy^{2}}\sigma(y)+6\left(\frac{d}{dy}\sigma(y)\right)\right]\\\label{1.21}
&=&\frac{1}{2M^{3}_{5}}\left(2M^{3}_{5}\Lambda+\delta(y-\pi)\lambda_{vis}+\delta(y)\lambda_{esc}\right)\quad .
\eeq

A equa\c c\~ao (\ref{1.20}) pode ser integrada diretamente. Considerando a simetria de orbifold, temos
\be\label{1.22}
\frac{d}{dy}\sigma(y)=\sqrt{-\frac{\Lambda}{6}}\Rightarrow \sigma(y)=\pm|y|\sqrt{-\frac{\Lambda}{6}}\quad .
\ee
Substituindo o resultado de (\ref{1.20}) em (\ref{1.21}),
\be\label{1.23}
\frac{d^{2}}{dy^{2}}\sigma(y)=-\frac{1}{6M_{5}^{3}}\left[\lambda_{vis}\delta(y-\pi)+\lambda_{esc}\delta(y)\right]\quad .
\ee

Derivando a equa\c c\~ao (\ref{1.22}) duas vezes,
\beq\label{1.24}\nonumber
\frac{d}{dy}\sigma(y)&=&\pm\frac{d}{dy}|y|\sqrt{-\frac{\Lambda}{6}}=\pm\left[\Theta(y)-\Theta(y-\pi)\right]\sqrt{-\frac{\Lambda}{6}}\quad ,\\
\frac{d^{2}}{dy^{2}}\sigma(y)&=&\pm 2\left[\delta(y)-\delta(y-\pi)\right]\sqrt{-\frac{\Lambda}{6}}\quad .
\eeq

Comparando (\ref{1.24}) com (\ref{1.23}), chegamos a conclus\~ao  que  a proposta de Randall-Sundrum será solu\c c\~ao das equa\c c\~oes de campo se tivermos
\be\label{1.25}
-\lambda_{vis}=\lambda_{esc}=\pm 12M_{5}^{3}\sqrt{-\frac{\Lambda}{6}}\quad ,
\ee
definindo,
\be\label{1.26}
\kappa=\frac{1}{l}=\sqrt{-\frac{\Lambda}{6}}\quad ,
\ee
onde $l$ \'e a escala de curvatura do bulk. Portanto temos, para (\ref{1.25})
\be\label{1.27}
-\lambda_{vis}=\lambda_{esc}=\pm 12M_{5}^{3}\kappa\quad .
\ee
Note que \'e imprescind\'{\i}vel que o bulk seja AdS para que haja um {\it{fine tuning}} entre os termos de tens\~ao das branas e a constante cosmol\'ogica no bulk. Ent\~ao a solu\c c\~ao de Randall-Sundrum ser\'a
\be\label{1.28}
ds^{2}=e^{\pm 2k|y|}\eta_{\mu\nu}dx^{\mu}dx^{\nu}+dy^{2}\quad .
\ee

\section{Gera\c c\~ao da hierarquia}

Ap\'os termos certeza de que a proposta de Randall-Sundrum \'e uma solu\c c\~ao $5-$dimensional das equa\c c\~oes de Einstein no v\'acuo, devemos estudar como o termo exponencial gera a hierarquia entre a escala de Planck efetiva $M_{4}$ e a escala de Planck fundamental $M_{5}$. Para isto, consideremos flutua\c c\~oes gravitacionais sem massa  em torno da solu\c c\~ao (\ref{1.28}),
\be\label{1.29}
\hat{g}_{\mu\nu}=\eta_{\mu\nu}+{h_{\mu\nu}}\quad ,
\ee
no qual $\hat{g}_{\mu\nu}$ \'e a m\'etrica $4-$dimensional perturbada e $|h_{\mu\nu}|<<1$ representa a pequena perturba\c c\~ao na m\'etrica de Minkowski $\eta_{\mu\nu}$. 

Estamos interessados em perturba\c c\~oes de modo zero, ou seja, os gr\'avitons {\it{f\'{\i}sicos}} advindos da decomposi\c c\~ao de Kaluza-Klein de $\hat{g_{\mu\nu}}$, isto implica que o modo zero deve ser independente de $y$.
Tomando apenas o termo da a\c c\~ao (\ref{1.9}) que representa a curvatura
\be\label{1.30}
S=\int d^{4}x \int dy\sqrt{e^{\pm 8|y|\kappa}[-\hat{g}]}M_{5}^{3}e^{\pm 2\kappa |y|}\hat{R}\quad ,
\ee
na qual $\hat{g}$ e $\hat{R}$ s\~ao quantidades escritas com (\ref{1.29}). Ent\~ao integrando em $y$ temos a a\c c\~ao efetiva em 4 dimens\~oes,
\be\label{1.31}
S_{4}=\pm\int d^{4}x M^{3}_{5} \sqrt{-\hat{g}}\hat{R}\frac{1}{\kappa}\left[e^{\pm 2\kappa \pi r_{c}}-1\right]\quad ,
\ee
disto podemos indentificar a massa de Planck efetiva $M_{4}$ como
\be\label{1.32}
M^{2}_{4}=\pm \frac{M_{5}^{3}}{\kappa}\left[e^{\pm 2\kappa \pi r_{c}}-1\right]\quad .
\ee
Se tomarmos o sinal negativo na equa\c c\~ao acima, vemos que a rela\c c\~ao entre as duas escalas de massa depende fracamente do raio de compactifica\c c\~ao no limite de $\kappa r_{c}$ grande, possibilitando a gera\c c\~ao de hierarquia exponencial entre essas duas quantidades apesar do pequeno efeito da exponencial na determina\c c\~ao da escala de Planck. Dado que $g_{\mu\nu}^{vis}=e^{-2\pi \kappa r_{c}}\hat{g_{MN}\delta^{M}_{\nu}\delta^{N}_{\nu}}$, \'e poss\'{\i}vel mostrar que qualquer massa $m_{0}$ medida por um observador na brana vis\'{\i}vel corresponde a uma massa f\'{\i}sica $m=e^{-2\pi \kappa r_{c}} m_{0}$ na teoria fundamental em mais dimens\~oes apenas com uma \'unica dimens\~ao extra, contrastando com o modelo ADD, que necessita de pelo menos duas dimens\~oes espaciais extras. Esse cen\'ario \'e chamado modelo de Randall-Sundum de duas branas, onde a brana em que vivemos possui tens\~ao negativa, o que leva a uma intera\c c\~ao gravitacional repulsiva, o que n\~ao nos parece razo\'avel. Podemos superar esta dificuldade abandonando a dimens\~ao extra compacta fazendo $r_{c}\rightarrow \infty$ e escolhendo o sinal positivo para a tens\~ao na brana vis\'{\i}vel (\ref{1.27}). Isto foi feito por Randall e Sundrum no trabalho seguinte \cite{rs2}, mostrando que a escala de Planck continua finita apesar de n\~ao haver nenhuma compactifica\c c\~ao na dimens\~ao extra. Uma consequ\^encia disto \'e a exist\^encia de um estado ligado do gr\'aviton sem massa $4-$dimensional localizado na brana vis\'{\i}vel, fazendo com que recobremos a lei da gravita\c c\~ao $1/r^{2}$ observada em escalas de baixas energias.

Vamos considerar, com o prop\'osito de mostrar a localiza\c c\~ao deste modo zero na brana, a solu\c c\~ao (\ref{1.28}) perturbada com o sinal negativo \cite{tanaka}
\be\label{1.32}
ds^{2}=\left[e^{-2\pi\kappa r_{c}}\eta_{\mu\nu}+h^{\mu\nu}\right]dx^{\mu}dx^{\nu}+dy^{2}\quad ,
\ee
sendo $h_{\mu\nu}=f(x^{\mu},y)$ e $\hat{g}^{\mu\nu}=\left[e^{-2\pi\kappa r_{c}}\eta_{\mu\nu}+h^{\mu\nu}\right]$.

Calculemos ent\~ao todas quantidades necess\'aria para a constru\c c\~ao das equa\c c\~oes de campo para a perturba\c c\~ao $h_{\mu\nu}$.
Os termos não nulos da conex\~ao s\~ao
\beq\label{1.33}\nonumber
\Gamma^{\mu}_{y\nu}&=&\frac{1}{2}\hat{g}^{\mu\kappa}\frac{\partial}{\partial y}\hat{g}_{\nu\kappa}\quad ,\\
\Gamma^{y}_{\nu\alpha}&=&-\frac{1}{2}\hat{g}^{yy}\frac{\partial}{\partial y}\hat{g}_{\nu\alpha}\quad ,
\eeq
e as componentes do tensor de Ricci $(R^{(4)}_{\mu\nu}$ denota a parte puramente $4-$dimensional)
\beq\label{1.34}\nonumber
R_{\nu\alpha}&=&R^{(4)}_{{\mu\nu}}+\frac{\partial}{\partial y}\Gamma^{y}_{\nu\alpha}+ \Gamma^{y}_{\nu\alpha}\Gamma^{\mu}_{y\mu}-\Gamma^{\beta}_{y\nu}\Gamma^{y}_{\beta\kappa}-\Gamma^{y}_{\nu\mu}\Gamma^{\mu}_{y\kappa}\quad ,\\
R_{yy}&=&-\left[\frac{\partial}{\partial y}\Gamma^{\mu}_{\mu y}+  \Gamma^{\beta}_{y\mu}\Gamma^{\mu}_{y\beta}\right]\quad .
\eeq
Usando o calibre de Randall-Sundrum
\be\label{1.35}\nonumber
\frac{\partial}{\partial x^{\mu}}h^{\mu\nu}=h^{\mu}_{\mu}=0\quad ,
\ee
temos explicitamente
\beq\label{1.36}\nonumber
\Gamma^{y}_{\nu\alpha}&=&-\frac{1}{2}\frac{\partial}{\partial y}\left[e^{-2\kappa|y|}\eta_{\nu\alpha}+h_{\nu\alpha}\right]=\kappa\frac{d}{dy}|y|e^{-2\kappa|y|}\eta_{\nu\alpha}-\frac{\partial}{\partial y}h_{\nu\alpha}\quad ,\\\nonumber
\Gamma^{\mu}_{\nu y}&=& \frac{1}{2}\frac{\partial}{\partial y}\left[h_{\nu\kappa}+e^{-2\kappa|y|}\eta_{\nu\kappa}\right]\left[\eta^{\mu\kappa}-e^{2\kappa|y|}h^{\mu\kappa}\right]\frac{e^{2\kappa|y|}}{2}\\
&=&\kappa\frac{d}{dy}|y|e^{2\kappa |y|}\eta_{\nu\kappa}h^{\mu\kappa}-\kappa\frac{d}{dy}|y|\delta^{\mu}_{\nu}+\eta^{\mu\kappa}\frac{\partial}{\partial y}h_{\kappa\nu}\quad .
\eeq
Tomando a derivada primeira em rela\c c\~ao a $y$
\beq\label{1.37}\nonumber
\frac{\partial}{\partial y}\Gamma^{y}_{\nu\alpha}&=&-\left[2\kappa^{2}\left(\frac{d}{dy}|y|\right)^{2} e^{-2\kappa|y|}\eta_{\nu\alpha}+ \frac{1}{2}\frac{\partial^{2}}{\partial y^{2}}h_{\nu\alpha}-\kappa e^{-2\kappa |y|}\eta_{\nu\alpha}\frac{d^{2}}{dy^{2}}|y|\right]\quad ,\\\nonumber
\frac{\partial}{\partial y}\Gamma^{\mu}_{y\nu}&=&2\kappa e^{2\kappa |y|}\frac{d}{dy}|y|\frac{\partial}{\partial y}h^{\mu}_{\nu}+\frac{1}{2}e^{2\kappa |y|}\frac{\partial^{2}}{\partial y^{2}}h^{\mu}_{\nu}+\kappa e^{2\kappa |y|}h^{\mu}_{\nu}\frac{d^{2}}{dy^{2}}|y|-\kappa\delta^{\mu}_{\nu}\frac{d^{2}}{dy^{2}}|y|\\
&+&2\kappa^{2} e^{2\kappa |y|}h^{\mu}_{\nu}\left(\frac{d}{dy}|y|\right)^{2}\quad .
\eeq
Al\'em disso, com
\beq\label{1.38}\nonumber
\Gamma^{\beta}_{\mu y}\Gamma^{\mu}_{y\beta}&=&4\kappa^{2}\left(\frac{d}{dy}|y|\right)^{2}\quad ,\\\nonumber
\Gamma^{y}_{\nu\kappa}\Gamma^{\mu}_{\mu y}&=&-4\kappa^{2}e^{-2\kappa |y|}\eta_{\nu\kappa}\left(\frac{d}{dy}|y|\right)^{2}+2\kappa \frac{\partial}{\partial y}h_{\nu\kappa}\frac{d}{dy}|y|\quad ,\\
\Gamma^{y}_{\nu y}\Gamma^{y}_{\beta\kappa}&=&\kappa^{2}\left[h_{\nu\kappa}-e^{-2\kappa|y|}\eta_{\nu\kappa}\right]\left(\frac{d}{dy}|y|\right)^{2}+\kappa\frac{\partial}{\partial y}h_{\nu\kappa}\frac{d}{dy}|y|\quad ,
\eeq
\'e poss\'{\i}vel calcular explicitamente (\ref{1.34})
\beq\label{1.39}\nonumber
R_{yy}&=& -4\kappa\left[\kappa\left(\frac{d}{dy}|y|\right)^{2}-\frac{d^{2}}{dy^{2}}|y|\right]\quad ,\\
R_{\nu\alpha}&=&rR_{\nu\alpha}^{(4)}+\kappa e^{-2\kappa |y|}\eta_{\nu\alpha}\left(\frac{d}{dy}|y|\right)^{2}+\kappa\frac{\partial}{\partial y}h_{\nu\alpha}\frac{d}{dy}|y|+\kappa^{2}h_{\nu\alpha}\left(\frac{d}{dy}|y|\right)^{2}\quad ,
\eeq
o escalar de curvatura 
\be\label{1.40}
R=\hat{g}^{\nu\alpha}R_{\nu\alpha}=\hat{g}^{\nu\alpha}R_{\nu\alpha}^{(4)}-16\kappa^{2}\left(\frac{d}{dy}|y|\right)^{2}+ 4\kappa\frac{d^{2}}{dy^{2}}|y|\quad .
\ee

Usando as identidades 
\beq\label{1.41}\nonumber
\left(\frac{d}{dy}|y|\right)^{2}&=&1\quad ,\\\nonumber
\frac{d^{2}}{dy^{2}}|y|&=&2\delta(y)\quad ,
\eeq
podemos reescrever o tensor de Einstein como
\be\label{1.42}
G_{\nu\alpha}=G_{\nu\alpha}^{(4)}+8\kappa^{2}h_{\nu\alpha}+6\kappa^{2}e^{-2\kappa |y|}\eta_{\nu\alpha}-\frac{1}{2}\frac{\partial ^{2}}{\partial y^{2}}h_{\nu\alpha}-8\kappa h_{\nu\alpha}\delta(y)-6\kappa\eta_{\nu\alpha}e^{-2\kappa |y|}\delta(y)\quad .
\ee
Igualando este resultado a (\ref{1.25}) na brana, temos
\beq\label{1.42}\nonumber
G_{\mu\nu}^{(4)}-\frac{1}{2}\frac{\partial^{2}}{\partial y^{2}}h_{\mu\nu}-2\kappa h_{\mu\nu}+2\kappa^{2} h_{\mu\nu}=0\quad ,
\eeq
e no calibre de Randall-Sundrum
\be\label{1.43}\nonumber
G_{\mu\nu}^{(4)}=\frac{1}{2}e^{2\kappa |y|}\frac{\partial^{2}}{\partial x^{c}\partial x_{c}}h_{\mu\nu}\quad  .
\ee
Disto obtemos as equa\c c\~oes de campo para as perturba\c c\~oes $h_{\mu\nu}(x^{\mu},y)$
\be\label{1.44}
\left[\frac{1}{2}e^{2\kappa |y|}\frac{\partial^{2}}{\partial x^{c}\partial x_{c}}-\frac{1}{2}\frac{\partial^{2}}{\partial y^{2}}+2\kappa^{2}-2\kappa\delta(y)\right]h_{\mu\nu}=0\quad .
\ee

Para obtermos os modos de $h_{\mu\nu}$ devemos utilizar a decomposi\c c\~ao de Kaluza-Klein para estas flutua\c c\~oes $5-$dimensionais, que correspondem a um número infinito de modos do campo $h_{\mu\nu}(x^{\mu})$ vistos por um observador confinado na brana, que correspondem aos modos de Fourier dos campos em 5 dimens\~oes $h_{\mu\nu}(x^{\mu},y)$. Isto s\'o \'e poss\'{\i}vel se a dimens\~ao extra for peri\'odica e compactada com raio $2\pi r_{c}$, o que \'e o caso do modelo de Randall-Sundrum com duas branas, mas n\~ao do modelo com dimens\~ao extra infinita. Para este \'ultimo em vez de uma s\'erie de Fourier obtemos uma transformada de Fourier e em vez de um espectro discretizado infinito para os modos $h_{\mu\nu}(x^{\mu})$ teremos um espectro cont\'{\i}nuo. As perturbações gravitacionais em modelos de mundo brana foram estudadas de forma pormenorizada em \cite{abdalla1}. Para verificar  se o modo zero das flutua\c c\~oes est\'a confinado na brana, para que a lei da gravita\c c\~ao de Newton seja recobrada, fa\c camos a seguinte separa\c c\~ao de vari\'aveis 
\be\label{1.45}\nonumber
h_{\mu\nu}(x^{\mu},y)=e^{imx}\Omega_{\mu\nu}(y),
\ee
substituindo na equa\c c\~ao de campo (\ref{1.44}), obtemos 
\be\label{1.46}
\left[\frac{\partial^{2}}{\partial y^{2}}+4\kappa\delta(y)+m^{2}e^{2\kappa |y|}-4\kappa^{2}\right]\Omega_{\mu\nu}(y)=0,
\ee
Para o modo zero, isto \'e, para $m=0$, a solu\c c\~ao \'e
\be\label{1.47}
\Omega^{0}=e^{-2\kappa |y|},
\ee
que \'e algo que procur\'avamos para conseguir o confinamente deste modo, j\'a que o alcance do gr\'aviton de modo zero, segundo esta solu\c c\~ao \'e $1/2\kappa=l/2$, onde $l$ e a escala de curvatura do bulk, e mede a penetra\c c\~ao dos gr\'avitons na dimens\~ao extra. Esta escala \'e tipicamente pequena, possibilitando o confinamento do modo zero. A solu\c c\~ao geral para a equa\c c\~ao de campo (\ref{1.46}) \'e uma combina\c c\~ao linear de funções de Bessel do primeiro tipo $J_{n}(x)$ de segundo tipo $Y_{n}(x)$,
\be\label{1.48}
\Omega_{\mu\nu}=A_{\mu\nu} J_{2} (my)+B_{\mu\nu} Y_{2}(my),
\ee
na qual $A_{\mu\nu} $ e $B_{\mu\nu}$ s\~ao constantes a serem determinadas pelas condi\c c\~oes de contorno. No modelo de Randall-Sundrum com dimesão extra infinita, as condi\c c\~oes de contorno levam ao espectro cont\'{\i}nuo dos modos de Kaluza-Klein, enquanto no modelo de duas branas com dimens\~ao extra compacta leva a discretização desses modos.

Acerca das idéias dos modelos de Randall-Sundrum, podemos dizer que sempre podemos identidficar uma dessas hipersuperfícies como o nosso universo e outra como uma contraparte {\it{escondida}}. No primeiro desses cenários, a nossa brana possui tensão negativa e \'e possível gerar completamente a hierarquia com apenas uma dimensão extra compacti\-ficada com um raio $r_{c}\leq 0.1mm$, enquanto que no segundo cen\'ario fazemos a brana escondida tender ao infinito, possibilitando o aparecimento de uma dimensão extra infinita, que não gera completamente a hierarquia mas apresenta o modo zero de Kaluza-Klein completamente localizado na brana física com tensão positiva.
\section{Formalismo Shiromizu-Maeda-Sasaki}

Esta seção se baseia no trabalho de  Shiromizu, Maeda e Sasaki \cite{maeda} acerca de uma apre\-sentação covariante para as quantidades físicas descritas nos modelos de mundo brana. A idéia fundamental é projetar todas as quantidades geom\'etricas $5-$dimensionais em 4 dimensões usando a equação de Gauss e a equação de Codazzi.

Primeiramente, seguindo o trabalho de W. Israel \cite{israel}, buscaremos as relações de Gauss e Codazzi. Para isso consideremos o bulk como sendo  uma variedade $M$ com métrica $g_{MN}$ de assinatura $(+----...)$ em $D$ dimensões, uma hipersuperfície $V$ $(D-1)$dimensional, que será identificado como uma brana, com métrica $q_{\mu\nu}$ também de assinatura $(+----...)$, imersa em $M$.

Consideremos uma curva $\lambda\in M$ parametrizada por $t$, e um campo vetorial ${\bf{A}}$ definido nessa curva, então sua derivada absoluta nessa curva será
\be\label{1.49}
\left(\frac{\partial}{\partial t}{\bf{A}}\right)^{M}\equiv \frac{\partial}{\partial t}A^{M}+A^{N}\Gamma^{M}_{NP}\frac{d}{dt}x^{P}\quad .
\ee
Sendo ${\bf{n}}$ o vetor normal a hipersuperfície $V$,
\beq\label{1.58}
{\bf{n}}\cdot{\bf{n}}=\varepsilon({\bf{n}})\quad ,
\eeq
onde $\varepsilon({\bf{n}})=+1$ quando ${\bf{n}}$ for do tipo espaço ($V$ tipo tempo) e $\varepsilon({\bf{n}})=-1$ para ${\bf{n}}$ do tipo tempo ($V$ tipo espaço).

Utilizando a base de vetores ${\bf e}_{(\alpha)}$ do espaço tangente $T(V)$ de $V$ podemos escrever o desclocamento infinitesimal em $V$ como
\be\label{1.51}
d{\bf{s}}={\bf e}_{(\alpha)}d\xi^{\alpha}\quad ,
\ee
sendo $\{\xi^{\alpha}\}$ o sistema de coordenadas da hipersuperfície.

A derivada covariante de um campo vetorial $A^{\alpha}$ não pertencente a $T(V)$, intrínseca a $V$, é 
\be\label{1.52}
A_{\alpha ;\beta}={\bf e}_{(\alpha)}\cdot \frac{\partial}{\partial \xi^{\beta}}{\bf{A}}=\frac{\partial}{\partial \xi^{\beta}}-A^{\delta}\Gamma_{\delta,\alpha\beta}\quad ,
\ee
sendo 
\be\label{1.53}
\Gamma_{\delta,\alpha\beta}={\bf e}_{(\delta)}\frac{\partial}{\partial \xi^{\beta}}{\bf{e}_{(\alpha)}}\quad .
\ee
Assim, a derivada covariante intrínseca e o tensor de Riemann $R^{\delta}_{\alpha\beta\gamma}$ associado, não dependem do bulk onde $V$ está mergulhada. Estas quantidades são invariantes frente a mudança de $M$ que preserva a métrica $q_{\mu\nu}$ de $V$.

As propriedades geométricas não intrínsecas aparecem quando perguntamos como $V$ se apresenta para um observador em $M$. Este tipo de quantidades são medidas através da variação do vetor unitário (\ref{1.58}) em relação ao sistema de coordenadas $\{\xi^{\alpha}\}$, ou seja devemos calcular 
\be\label{1.54}
\frac{\partial}{\partial \xi^{\alpha}}{\bf{n}}=K^{\beta}_{\alpha}{\bf{e}_{(\beta)}}\quad ,
\ee 
no qual $K_{\alpha\beta}$ é definido como a {\it{curvatura extrínseca}} de $V$ com relação a $M$. Multiplicando ambos lados desta equação por ${\bf{e}_{(\lambda)}}$
\be\label{1.55}
K^{\alpha}_{\beta}{\bf{e}_{(\alpha)}}{\bf{e}_{(\lambda)}}={\bf{e}_{(\lambda)}}\cdot \frac{\partial}{\partial\xi^{\beta}}{\bf{n}}\quad ,
\ee
e usando o fato de que $q_{\alpha\lambda}={\bf{e}_{(\alpha)}}\cdot {\bf{e}_{(\lambda)}}$, temos
\be\label{1.56}
K_{\beta\lambda}={\bf{e}_{(\lambda)}}\cdot\frac{\partial}{\partial\xi^{\beta}}{\bf{n}}\quad ,
\ee
mas
\be\label{1.56}
\frac{\partial}{\partial \xi^{\alpha}}\left({\bf{n}}\cdot{\bf{e}_{(\beta)}}\right)=0\Rightarrow {\bf{e}_{(\beta)}}\cdot\frac{\partial}{\partial\xi^{\alpha}}{\bf{n}}=-{\bf{n}}\cdot\frac{\partial}{\partial \xi^{\alpha}}{\bf{e}_{(\beta)}}\quad ,
\ee
pois ${\bf{n}}\bot{\bf{e}_{(\beta)}}$, logo
\be\label{1.57}
K_{\alpha\beta}=-{\bf{n}}\cdot\frac{\partial}{\partial \xi^{\alpha}}{\bf{e}_{(\beta)}}=-{\bf{n}}\cdot\frac{\partial}{\partial \xi^{\beta}}{\bf{e}_{(\alpha)}}=K_{\beta\alpha}\quad .
\ee

Podemos obter a equação de Gauss-Weingarten usando este resultado.
Para isso multipliquemos toda equação por ${\bf{n}}$, e usando (\ref{1.58}) 
\be\label{1.59}
-\varepsilon({\bf{n}})\frac{\partial}{\partial \xi^{\alpha}}={\bf{n}}K_{\alpha\beta}\quad ,
\ee 
multiplicando ambos lados por $\varepsilon({\bf{n}})$ obtemos
\be\label{1.60}
-\frac{\partial}{\partial \xi^{\alpha}}{\bf{e}_{(\beta)}}=\varepsilon({\bf{n}})K_{\alpha\beta}{\bf{n}}\quad ,
\ee
usando o resultado 
\be\label{1.61}
-\frac{\partial}{\partial \xi^{\alpha}}{\bf{e}_{(\beta)}}={\bf{e}_{(\delta)}}\Gamma^{\delta}_{\alpha\beta}\quad ,
\ee
chegamos a equação de Gauss-Weingarten
\be\label{1.62}
\frac{\partial}{\partial \xi^{\beta}}{\bf{e}_{\alpha}}=-\varepsilon({\bf{n}})K_{\alpha\beta}{\bf{n}}+{\bf{e}_{\delta}}\Gamma^{\delta}_{\alpha\beta}\quad .
\ee
 Para encontrarmos as equações de Gauss e Codazzi, começamos operando com $\partial/\partial \xi^{\alpha}$ em (\ref{1.62}) usando (\ref{1.54}) e (\ref{1.57}),
\be\label{1.63}
\frac{\partial}{\partial\varepsilon^{\gamma}}\left(\frac{\partial}{\partial \xi^{\beta}}{\bf{e}_{(\alpha)}}\right)={\bf{n}}\frac{\partial}{\partial \xi^{\gamma}}K_{\alpha\beta}+K_{\alpha\beta}K^{\delta}_{\gamma}{\bf{e}_{(\delta)}} + {\bf{e}_{(\delta)}}\frac{\partial}{\partial\xi^{\gamma}}\Gamma^{\delta}_{\alpha\beta}+\Gamma^{\delta}_{\alpha\beta}K_{\delta\gamma}{\bf{n}}+\Gamma^{\lambda}_{\alpha\beta}\Gamma^{\delta}_{\lambda\gamma}{\bf{e}_{\delta}}\quad .
\ee
Se considerarmos o campo vetorial não tangente ${\bf{A}}(u,v)$ definido em uma superfície $x^{M}=x^{M}(u,v) \in M$, podemos calcular a relação de comutação de Ricci, que define o tensor de curvatura $ R^{M}_{QNP}A^{Q}$
\be\label{1.64}
\left[\left(\frac{\partial}{\partial u}\frac{\partial}{\partial v}-\frac{\partial}{\partial v}\frac{\partial}{\partial u}\right){\bf{A}}\right]^{M}=\left[\frac{\partial}{\partial u},\frac{\partial}{\partial v}\right]{\bf{A}}^{M}=R^{M}_{QNP}A^{Q}\frac{\partial x^{N}}{\partial u}\frac{\partial x^{P}}{\partial v}\quad , 
\ee
que pode ser calculada usando (\ref{1.62}) e (\ref{1.63}) com $\varepsilon(\bf{n})=-1$. Assim,
\beq\label{1.65}\nonumber
&\left[\frac{\partial}{\partial\xi^{\gamma}},\frac{\partial}{\partial\xi^{\beta}}\right]{\bf{e}_{(\alpha)}}={\bf{n}}\left(\frac{\partial}{\partial \xi^{\gamma}}K_{\alpha\beta}-\frac{\partial}{\partial \xi^{\beta}}K_{\alpha\gamma}+\Gamma^{\delta}_{\alpha\beta}K_{\gamma\delta}-\Gamma^{\delta}_{\alpha\gamma}K_{\delta\beta}\right)+\\
&{\bf{e}_{(\delta)}}\left(K_{\alpha\beta}K_{\gamma}^{\delta}-K_{\alpha\delta}K^{\delta}_{\beta}+\frac{\partial}{\partial\xi^{\gamma}}\Gamma^{\delta}_{\alpha\beta}-\frac{\partial}{\partial\xi^{\beta}}\Gamma^{\delta}_{\alpha\gamma}+\Gamma^{\lambda}_{\alpha\beta}\Gamma^{\delta}_{\lambda\gamma}-\Gamma^{\lambda}_{\alpha\gamma}\Gamma^{\delta}_{\lambda\beta}\right)\quad .
\eeq
Podemos agora representar explicitamente o tensor de curvatura extrínseco $R^{M}_{NQP}$ em função das quantidades intrínsecas a $V$, então usando (\ref{1.64}),
\be\label{1.66}
\left[\frac{\partial}{\partial\xi^{\gamma}},\frac{\partial}{\partial\xi^{\beta}}\right]e^{M}_{(\alpha)}=R^{M}_{QLR} e^{Q}_{(\alpha)}e^{L}_{(\gamma)}e^{R}_{(\beta)}\quad ,
\ee
comparando este resultado com (\ref{1.65}) chegamos a equação de Gauss, que corresponde a parte não normal
\beq\label{1.67}
\nonumber
R_{MQLR} e^{M}_{(\delta)} e^{Q}_{(\alpha)}e^{L}_{(\gamma)}e^{R}_{(\beta)}=K_{\alpha\beta}K_{\delta\gamma}-K_{\alpha\gamma}K_{\delta\beta}+\left(\frac{\partial}{\partial\xi^{\gamma}}\Gamma_{\delta\alpha\beta}-\frac{\partial}{\partial\xi^{\beta}}\Gamma_{\delta\alpha\gamma}+\Gamma^{\lambda}_{\alpha\beta}\Gamma_{\delta\lambda\gamma}-\Gamma^{\lambda}_{\alpha\gamma}\Gamma_{\delta\lambda\beta}\right)\quad ,
\eeq
ora, o termo entre parênteses nada mais é do que o tensor de curvatura da hipersuperfície $V$, $R_{\delta\alpha\gamma\beta}$, logo a equação de Gauss pode ser escrita como
\be\label{1.68}
R_{MQLR} e^{M}_{(\delta)} e^{Q}_{(\alpha)}e^{L}_{(\gamma)}e^{R}_{(\beta)}=K_{\alpha\beta}K_{\delta\gamma}-K_{\alpha\gamma}K_{\delta\beta}+R_{\delta\alpha\gamma\beta}\quad . 
\ee
Da parte normal, obtemos a equação de Codazzi
\be\label{1.69}
R_{MQLR}n^{M}e^{Q}_{(\alpha)}e^{L}_{(\gamma)}e^{R}_{(\beta)}=K_{\alpha\gamma ; \beta}-K_{\alpha\beta ; \gamma}\quad .
\ee
A métrica $q_{\alpha\beta}$ induzida na hipersuperfície é relacionada com a métrica $g_{MN}$ de $M$ por, 
\be\label{1.70}
q_{\alpha\beta}=g_{MN}e^{M}_{(\alpha)}e^{N}_{(\beta)}-n_{M}n_{N}\quad , 
\ee
e usando as relações
\beq\label{1.69}\nonumber
e^{M}_{\delta}&=& e^{M}e_{\delta}= q^{M}_{\delta}\quad ,\\\nonumber
e^{Q}_{\alpha}&=& e^{Q}e_{\alpha}= q^{M}_{\alpha}\quad ,
\eeq
podemos reescrever as equações de Gauss e Codazzi, respectivamente, como
\be\label{1.70}
R^{\alpha}_{\beta\gamma\delta}=R^{M}_{NPQ}q^{\alpha}_{M}q^{N}_{\beta}q^{P}_{\gamma}q^{Q}_{\delta} + K^{\alpha}_{\gamma}K_{\beta\delta}-K^{\alpha}_{\delta}K_{\beta\gamma}\quad ,
\ee
\be\label{1.71}
K^{\beta}_{\alpha;\beta}-K_{;\alpha}=R_{PQ}n^{Q}q^{P}_{\alpha}\quad ,
\ee
sendo $K=K^{\alpha}_\alpha$ o traço da curvatura extrínseca. 

Vamos fazer um estudo das quantidades geométricas envolvidas o mais geral possível sem nos atermos, por enquanto, a modelos de universo tipo brana. Não faremos distinção entre os índices da hipersuperfície $(V,q_{\mu\nu})$ e os do bulk $(M,g_{AB})$ tomando ambos representados por letras gregas. 

Da expressão (\ref{1.70}) segue que
\beq\label{1.72}\nonumber
q_{\mu\nu}&=&g_{\mu\nu}-n_{\mu}n_{\nu}\quad ,\\\nonumber
q^{\mu\nu}&=&g^{\mu\nu}-n^{\mu}n^{\nu}\quad ,\\
q^{\mu}_{\nu}&=&\delta^{\mu}_{\nu}-n^{\mu}n_{\nu}\quad ,
\eeq
além disso
\beq\label{1.73}\nonumber
n_{\mu}n^{\mu}=1\quad ,\\\nonumber
q_{\mu\nu}n^{\mu}=0\quad ,\\
g_{\mu\nu}n^{\mu}n^{\nu}=1\quad .
\eeq

Desta forma reescrevemos as equações de Gauss e Codazzi, respectivamente, como
\beq\label{1.74}
{}^{(4)}R^{\alpha}_{\beta\gamma\delta}={}^{(5)}R^{\mu}_{\nu\rho\delta}q^{\alpha}_{\mu}q^{\nu}_{\beta}q^{\rho}_{\gamma}q^{\sigma}_{\delta}+K^{\alpha}_{\gamma}K_{\beta\delta}-K^{\alpha}_{\delta}-K_{\beta\gamma}\quad ,\\
K^{\nu}_{\mu;\nu}-K_{;\mu}={}^{(5)}R_{\rho\sigma}n^{\sigma}q^{\rho}_{\mu}\quad .
\eeq

Precisamos das equações de Einstein para relacionar estas quantidades geométricas com as fontes de energia-momento, mas para fazer isso necessitamos do tensor de Ricci, que aparece quando contraímos $\alpha$ e $\gamma$ na  equação de Gauss usando $q^{\gamma}_{\alpha}$ 
\beq\label{1.75}\nonumber
{}^{(4)}R^{\alpha}_{\beta\gamma\delta}q^{\gamma}_{\alpha}={}^{(5)}R^{\mu}_{\nu\rho\delta}q^{\gamma}_{\alpha}q^{\alpha}_{\mu}q^{\nu}_{\beta}q^{\rho}_{\gamma}q^{\sigma}_{\delta}+K^{\alpha}_{\gamma}K_{\beta\delta}q^{\gamma}_{\alpha}-q^{\gamma}_{\alpha}K^{\alpha}_{\delta}-K_{\beta\gamma}\quad ,
\eeq
mas 
\beq\label{1.76}\nonumber
K=q^{\gamma}_{\alpha}K^{\alpha}_{\gamma}\quad ,\\\nonumber
q^{\alpha}_{\mu}=q^{\alpha\lambda}q_{\lambda\mu}\quad ,
\eeq
então, usando a s identidades (\ref{1.70}) temos
\beq\label{1.77}\nonumber
{}^{(4)}R_{\beta\delta}&=&{}^{(5)}R^{\mu}_{\nu\rho\sigma}q^{\gamma}_{\alpha}q^{\alpha\lambda}\left(g_{\lambda\mu}-n_{\lambda}n_{\nu} \right)q^{\nu}_{\beta}q^{\rho}_{\gamma}q^{\sigma}_{\delta}+KK_{\beta\delta}-K^{\alpha}_{\delta}K_{\beta\alpha}\\\nonumber
&=&{}^{(5)}R_{\lambda\nu\rho\sigma}q^{\gamma\lambda}q^{\nu}_{\beta}q^{\rho}_{\gamma}q^{\sigma}_{\delta}-{}^{(5)}R^{\mu}_{\nu\rho\sigma}q^{\gamma}_{\alpha}q^{\alpha\lambda}n_{\lambda}n_{\mu}q^{\nu}_{\beta}q^{\rho}_{\gamma}q^{\sigma}_{\delta}+KK_{\beta\delta}-K^{\alpha}_{\delta}K_{\beta\alpha}\quad ,
\eeq
usando novamente (\ref{1.70})
\beq
&=&{}^{(5)}R_{\lambda\nu\rho\sigma}\left(g^{\gamma\lambda}-n^{\gamma}n^{\lambda}\right)\left(\delta^{\rho}_{\gamma}-n^{\rho}n_{\gamma}\right)q^{\nu}_{\gamma}q^{\sigma}_{\delta}+KK_{\beta\delta}-K^{\alpha}_{\delta}K_{\beta\alpha}\quad ,\\\nonumber
&=&{}^{(5)}R_{\lambda\nu\rho\sigma}g^{\lambda\rho}q^{\nu}_{\beta}q^{\sigma}_{\delta}-{}^{(5)}R_{\lambda\nu\rho\sigma}g^{\gamma\lambda}n^{\rho}n_{\gamma}q^{nu}_{\beta}q^{\sigma}_{\delta}+{}^{(5)}R_{\lambda\nu\rho\sigma}n^{\sigma}n^{\lambda}n^{\rho}n_{\gamma}q^{\nu}_{\beta}q^{\sigma}_{\delta}\\\nonumber
&-&{}^{(5)}R_{\lambda\nu\rho\sigma}n^{\sigma}n^{\lambda}\delta^{\rho}_{\gamma}q^{\nu}_{\beta}q^{\sigma}_{\delta}+KK_{\beta\delta}-K^{\alpha}_{\delta}K_{\beta\alpha}\quad ,
\eeq
finalmente,
\beq\label{1.78}
{}^{(4)}R_{\beta\delta}={}^{(5)}R_{\nu\sigma}q^{\nu}_{\beta}q^{\sigma}_{\delta}-{}^{(5)}R^{\gamma}_{\nu\rho\sigma}n^{\rho}n_{\gamma}q^{\nu}_{\beta}q^{\sigma}_{\delta}+KK_{\beta\delta}-K^{\alpha}_{\delta}K_{\beta\alpha}\quad .
\eeq

Calculando o escalar de Ricci $4-$dimensional ${}^{(4)}R=q^{\mu\nu}R_{\mu\nu}$,
\beq\label{1.79}\nonumber
{}^{(4)}R_{\beta\delta}q^{\beta\delta}&=&{}^{(5)}R_{\nu\sigma}q^{\nu}_{\beta}q^{\sigma}_{\delta}q^{\beta\delta}-{}^{(5)}R^{\gamma}_{\nu\rho\sigma}n^{\rho}n_{\gamma}q^{\nu}_{\beta}q^{\sigma}_{\delta}q^{\beta\delta}+q^{\beta\delta}KK_{\beta\delta}-K^{\alpha}_{\delta}K_{\beta\alpha}q^{\beta\delta}\quad ,\\
{}^{(4)}R&=&{}^{(5)}R_{\nu\sigma}q^{\nu\sigma}-{}^{(5)}R^{\gamma}_{\nu\rho\sigma}n^{\rho}n_{\gamma}q^{\nu}_{\beta}q^{\sigma}_{\delta}q^{\beta\delta}-K^{\alpha\beta}K_{\alpha\beta}+K^{2}.
\eeq

Usando (\ref{1.78}) e (\ref{1.79}) podemos construir o tensor de Einstein $4-$dimensional ${}^{(4)}G_{\beta\delta}$ 
\beq\label{1.80}\nonumber
{}^{(4)}G_{\beta\delta}&=&{}^{(4)}R_{\beta\delta}-\frac{1}{2}q_{\beta\delta}{}^{(4)}R,\\\nonumber
&=&{}^{(5)}R_{\nu\sigma}q^{\nu}_{\beta}q^{\sigma}_{\delta}-{}^{(5)}R^{\gamma}_{\nu\rho\sigma}n^{\rho}n_{\gamma}q^{\nu}_{\beta}q^{\sigma}_{\delta}+KK_{\beta\sigma}-K^{\alpha}_{\delta}K_{\beta\delta}\\\nonumber
&-&\frac{1}{2}q_{\beta\delta}\left[{}^{(5)}R_{\tau\theta}q^{\tau\theta}-{}^{(5)}R^{\gamma}_{\tau\rho\theta}q^{\beta\delta}n_{\gamma}n^{\rho}q^{\tau}_{\beta}q^{\theta}_{\delta}-K^{2}-K^{\alpha\beta}K_{\alpha\beta}\right]\quad ,
\eeq
definindo 
\be\label{1.81}
\widetilde{E}_{\beta\delta}\equiv {}^{(5)}R^{\gamma}_{\nu\rho\sigma}n^{\rho}n_{\gamma}q^{\nu}_{\beta}q^{\sigma}_{\delta}\quad ,
\ee
então
\beq\label{1.82}\nonumber
{}^{(4)}G_{\beta\delta}&=&\left[{}^{(5)}R_{\nu\sigma}-\frac{1}{2}q_{\nu\sigma}{}^{(5)}R_{\tau\theta}\left(g^{\tau\theta}-n^{\tau}n^{\theta}\right)\right]q^{\nu}_{\beta}q^{\sigma}_{\delta}-\widetilde{E}_{\beta\delta}\\
&+&KK_{\alpha\beta}-K^{\alpha}_{\delta}K_{\beta\alpha}
-\frac{1}{2}q_{\beta\delta}\left(K^{2}-K^{\alpha\beta}K_{\alpha\beta}\right)+\frac{1}{2}{}^{(5)}R^{\gamma}_{\tau\rho\theta}n_{\gamma}n^{\rho}q^{\tau}_{\beta}q^{\theta}_{\delta}q^{\beta\delta}q_{\beta\delta}\quad ,
\eeq

Utilizando as relações (\ref{1.73}), obtemos a seguinte relação
\be\label{1.83}\nonumber
\frac{1}{2}{}^{(5)}R_{\tau\theta}g_{\nu\sigma}n^{\tau}n^{\sigma}q^{\nu}_{\beta}q^{\sigma}_{\delta}=\frac{1}{2}{}^{(5)}R_{\tau\theta}q_{\beta\delta}n^{\tau}n^{\theta}\quad ,
\ee
que substituída na equação (\ref{1.82}) dá como resultado 
\beq\label{1.84}\nonumber
{}^{(4)}G_{\beta\delta}&=&{}^{(5)}G_{\nu\sigma}q^{\nu}_{\beta}q^{\sigma}_{\delta}-\widetilde{E}_{\beta\delta}+KK_{\beta\delta}-K^{\alpha}_{\delta}K_{\beta\delta}-\frac{1}{2}q_{\beta\delta}\left(K^{2}-K^{\alpha\beta}K_{\alpha\beta}\right)\\\nonumber
&+&\frac{1}{2}q_{\beta\delta}\left[\left(g^{\tau\theta}-n^{\tau}n^{\theta}\right){}^{(5)}R_{\lambda\tau\rho\theta}n^{\lambda}n^{\rho}+R_{\tau\theta}n^{\tau}n^{\theta}\right]\quad.
\eeq

Se usarmos a simetria do tensor de Riemman
\beq\label{1.85}\nonumber
{}^{(5)}R_{\lambda\tau\rho\theta}=-{}^{(5)}R_{\lambda\tau\theta\rho}&=&-{}^{(5)}R_{\theta\rho\tau\lambda}={}^{(5)}R_{\theta\rho\lambda\tau}\\\nonumber
g^{\lambda\rho}n^{\tau}n^{\theta}{}^{(5)}R_{\lambda\tau\rho\theta}&=&g^{\tau\theta}n^{\lambda}n^{\rho}{}^{(5)}R_{\lambda\tau\rho\theta}\quad ,
\eeq
então
\be\label{1.86}
{}^{(5)}R_{\lambda\tau\rho\theta}n^{\lambda}n^{\theta}n^{\tau}n^{\rho}=0\quad ,
\ee

Estas relações nos permitem escrever o tensor de Einstein como
\beq\label{1.87}
{}^{(4)}G_{\beta\delta}&=&{}^{(5)}G_{\nu\sigma}q^{\nu}_{\beta}q^{\sigma}_{\delta}-\widetilde{E}_{\beta\delta}+KK_{\beta\delta}-K^{\alpha}_{\delta}K_{\beta\alpha}\\
&-&\frac{1}{2}q_{\beta\delta}\left(K^{2}-K^{\alpha\beta}K_{\alpha\beta}\right)+q_{\alpha\beta}{}^{(5)}R_{\tau\theta}n^{\tau}n^{\theta}\quad ,
\eeq
com
\be\label{1.88}
\widetilde{E}_{\beta\delta}={}^{(5)}R^{\gamma}_{\nu\rho\sigma}n^{\rho}n_{\gamma}q^{\nu}_{\beta}q^{\sigma}_{\delta}\quad .
\ee
Esta última expressão pode ser escrita em função do tensor de Weyl $5-$dimensional se usarmos a decomposição do tensor de Riemann no tensor de Ricci, escalar de Ricci e tensor de Weyl, tal como em \cite{weinberg}
\beq\label{1.89}\nonumber
{}^{(5)}R_{\gamma\nu\rho\sigma}&=&\frac{1}{D-2}\left[g_{\gamma\rho}{}^{(5)}R_{\nu\sigma}-g_{\gamma\sigma}{}^{(5)}R_{\nu\rho} -g_{\nu\rho}{}^{(5)}R_{\gamma\sigma}+g_{\nu\sigma}{}^{(5)}R_{\gamma\rho}\right]\\\nonumber
&-&\frac{{}^{(5)}R}{(D-1)(D-2)}\left[g_{\gamma\rho}g_{\nu\sigma}-g_{\gamma\sigma}g_{\nu\rho}\right] +{}^{(5)}C_{\gamma\nu\rho\sigma}\quad ,
\eeq 
no qual $D$ é o número total de dimensões, em nosso caso consideraremos $D=5$. Deste modo podemos reescrever(\ref{1.88}),
\beq\label{1.90}\nonumber
\widetilde{E}_{\beta\delta}&=&\frac{1}{3}\left[g_{\gamma\rho}{}^{(5)}R_{\nu\sigma}-g_{\gamma\sigma}{}^{(5)}R_{\nu\rho}-g_{\nu\rho}{}^{(5)}R_{\gamma\sigma}+g_{\nu\sigma}{}^{(5)}R_{\gamma\rho}\right]n^{\rho}n^{\gamma}q^{\nu}_{\beta}q^{\sigma}_{\delta}\\\nonumber
&-&\frac{1}{12}\left[g_{\gamma\rho}g_{\nu\sigma}-g_{\gamma\sigma}g_{\nu\rho}\right]{}^{(5)}Rn^{\rho}n^{\sigma}q^{\nu}_{\beta}q^{\sigma}_{\delta}+{}^{(5)}C_{\gamma\nu\rho\sigma}n^{\rho}n^{\gamma}q^{\nu}_{\beta}q^{\sigma}_{\delta}\quad ,\\\nonumber
&=&\frac{1}{3}\left[{}^{(5)}R_{\nu\sigma}q^{\nu}_{\beta}q^{\sigma}_{\delta}+g_{\nu\sigma}n^{\rho}n^{\gamma}q^{\nu}_{\beta}q^{\sigma}_{\delta}{}^{(5)}R_{\gamma\rho}\right]\\\nonumber
&-&\frac{1}{12}g_{\nu\sigma}q^{\nu}_{\beta}q^{\sigma}_{\delta}{}^{(5)}R+ {}^{(5)}C_{\gamma\nu\rho\sigma}n^{\rho}n^{\gamma}q^{\nu}_{\beta}q^{\sigma}_{\delta}\quad .
\eeq

Consideremos agora a equação de Einstein $5-$dimensional
\be\label{1.91}
{}^{(5)}R_{\alpha\beta}-\frac{1}{2}g_{\alpha\beta}{}^{(5)}R=\kappa_{5}^{2}T_{\alpha\beta}\quad ,
\ee
de onde segue o escalar de Ricci a partir da contração dos índices $\alpha$ e $\beta$
\be\label{1.92}
{}^{(5)}R=-\frac{2}{3}\kappa_{5}^{2}T^{\alpha}_{\alpha}\quad .
\ee
Assim sendo, obtemos
\be\label{1.93}
{}^{(5)}R_{\nu\sigma}=\frac{\kappa_{5}^{2}}{3}\left[3T_{\nu\sigma}-\kappa_{5}^{2}g_{\nu\sigma}T^{\alpha}_{\alpha}\right]\quad .
\ee
Definindo
\be\label{1.94}
E_{\beta\delta}\equiv {}^{(5)}C_{\gamma\nu\rho\sigma}n^{\rho}n^{\gamma}q^{\nu}_{\beta}q^{\sigma}_{\delta}\quad .
\ee
Substituindo os resultados (\ref{1.92})-(\ref{1.94}) na expressão para $\widetilde{E}_{\beta\delta}$ chegamos a
\beq\label{1.95}\nonumber
\widetilde{E}_{\beta\delta}&=&{}^{(5)}C_{\gamma\nu\rho\sigma}n^{\rho}n^{\gamma}q^{\nu}_{\beta}q^{\sigma}_{\delta}+\frac{\kappa_{5}^{2}}{3}T_{\nu\sigma}q^{\nu}_{\beta}q^{\sigma}_{\delta}\\
&-&\frac{\kappa_{5}^{2}}{6}q_{\beta\delta}T^{\alpha}_{\alpha}+\frac{\kappa_{5}^{2}}{3}T_{\gamma\rho}n^{\rho}n^{\gamma}q_{\beta\delta}\quad .
\eeq
Depois disto, a equação de Einstein (\ref{1.82}) pode ser escrita como
\beq\label{1.96}\nonumber
{}^{(5)}G_{\beta\delta}&=&\frac{2}{3}\kappa_{5}^{2}\left[T_{\nu\sigma}q^{\nu}_{\beta}q^{\sigma}_{\delta}+\left(T_{\gamma\rho}n^{\gamma}n^{\rho}-\frac{1}{4}T^{\alpha}_{\alpha}\right)\right]\\
&+&KK_{\beta\delta}-K^{\alpha}_{\delta}K_{\beta\alpha}-\frac{1}{2}q_{\beta\delta}\left(K^{2}-K^{\alpha\beta}K_{\alpha\beta}\right)-E_{\beta\delta}\quad .
\eeq
Este é um dos resultados que procurávamos, o outro vem da equação de Codazzi (\ref{1.71})
\beq\label{1.97}\nonumber
D_{\alpha}K^{\alpha}_{\nu}-D_{\mu}K&=&\frac{K_{5}^{2}}{3}\left(3T_{\nu\rho}-T^{\alpha}_{\alpha}g_{\nu\rho}\right)n^{\nu}n^{\rho}\\\nonumber
&=&\kappa_{5}^{2}T_{\nu\rho}n^{\nu}q^{\rho}_{\mu}-\frac{\kappa^{2}_{5}}{3}T_{\alpha}^{\alpha}g_{\nu\rho}n^{\nu}q^{\rho}_{\mu}\quad ,
\eeq
mas como $g_{\nu\rho}n^{\nu}q^{\rho}_{\mu}=0$\quad ,
teremos
\be\label{1.99}
D_{\alpha}K^{\alpha}_{\mu}-D_{\mu}K=\kappa_{5}^{2}T_{\nu\rho}n^{\nu}q^{\rho}_{\mu}\quad .
\ee

Os resultados (\ref{1.96}) e (\ref{1.99}) foram obtidos de forma geral, levando em conta apenas as equações de Gauss, Codazzi e Einstein. Para usarmos estas relações obtidas no estudo de modelos de mundo brana devemos escrever o tensor energia momento $5-$dimensional na forma
\be\label{1.100}
T_{\mu\nu}=-\Lambda_{5}g_{\mu\nu}+\delta(y)\left[\tau_{\mu\nu}-\lambda q_{\mu\nu}\right]\quad .
\ee
O primeiro termo é a contribuição ao tensor energia-momento da constante cosmológica do bulk, o segundo termo é todo confinado na brana, formado pela energia de vácuo ou tensão da brana $\lambda$ e pelo tensor energia momento $\tau_{\mu\nu}$ da matéria e campos confinados. Esta contribução da brana pode ser totalmente determinada se soubermos quais são os campos e partículas presentes na brana. Disto temos que $\tau_{\mu\nu}n^{\nu}=0$, pois não pode haver fluxo na direção $y$ de partículas e campos que são confinados.

Há uma forma muito útil de relacionar a curvatura extrínseca $K_{\mu\nu}$ ao tensor energia-momento da brana. Tal forma foi obtida por W.Israel \cite{israel} e é conhecida como condições de junção de Israel. Derivaremos a seguir tais vínculos.

Seja então uma hipersuperfície $(V,q_{\mu\nu})$ que divide a variedade $(M,g_{\mu\nu})$ em dois subes\-paços com extremidades $M^{+}$ e $M^{-}$. Suponhamos que $V$ seja uma camada superficial, então a curvatura extrínseca medida em uma das extremidades deve ser diferente da medida no outro, isto é,
\be\label{1.101}
K_{\mu\nu}^{+}-K_{\mu\nu}^{-}=\gamma_{\mu\nu}\quad ,
\ee
sendo em geral $\gamma_{\mu\nu}$ é um tensor não nulo, que segundo a idéia de Israel é devido a existência de uma distribuição de energia finita sobre a hipersuperfície $V$. Esta descontinuidade na curvatura extrínseca pode então em princípio ser expressa em função deste tensor energia momento definido em $(V,q_{\mu\nu})$. Usaremos o chamado sistema de coordenas normais definidas pelo conjunto $x^{\mu}=\{x^{\alpha},y\}$, onde $y$ representa a distância geodésica de um ponto em $V$ até um ponto em $M$. Seja então uma quantidade finita $y=\epsilon$ que é e localização da extremidade  $V^{-}$ e $V^{+}$ localizada em $y=0$. Com isso o vetor normal a $V$ pode ser expresso por 
\beq\label{1.102}\nonumber
n_{\alpha}&=&\delta^{y}_{\alpha}\\
K^{+}_{\mu\nu}-K^{-}_{\mu\nu}&=&K_{\mu\nu}^{\pm}=\frac{1}{2}\left[\frac{\partial}{\partial y}g_{\mu\nu}\right]^{\pm}\quad .
\eeq
Da equação de Codazzi (\ref{1.71}), temos
\be\label{1.103}\nonumber
R_{yQLR}e^{Q}_{(\alpha)}e^{L}_{(\gamma)}e^{R}_{(\beta)}=K_{\alpha\gamma;\beta}-K_{\alpha\beta;\gamma}\quad ,
\ee
tomando a componente normal desta relação,
\be\label{1.104}
R_{yQyR}e^{Q}_{(\alpha)}e^{R}_{(\beta)}=\frac{\partial}{\partial y}K_{\alpha\beta}\quad .
\ee

Tomando o traço do tensor de Riemann na equação de Gauss (\ref{1.68}) temos
\be\label{1.105}
R^{M}_{QMR}e^{Q}_{(\alpha)e^{R}_{(\beta)}}=K_{\alpha\beta}K^{M}_{M}-K_{\beta}^{M}K_{\alpha M}+R^{M}_{\alpha M \beta}\quad .
\ee

Dos resultados (\ref{1.104}) e (\ref{1.105}) chegamos a
\beq\label{1.106}
R_{QR}e^{Q}_{(\alpha)}e^{R}_{(\beta)}=-\frac{\partial}{\partial y}K_{\alpha\beta}+K_{\alpha\beta}K^{M}_{M}-K_{\beta}^{M}K_{\alpha M}+R^{M}_{\alpha M \beta}\equiv -\frac{\partial}{\partial y}K_{\alpha\beta}+H_{\alpha\beta}\quad .
\eeq
Mas da equação de Einstein
\be\label{1.107}\nonumber
R_{QR}=\kappa^{2}_{5}\left(T_{QR}-\frac{1}{3}g_{QR}T^{M}_{M}\right)\quad ,
\ee
vemos que ela pode ser integrada usando (\ref{1.106}) no intervalo $[0,\epsilon]$
\beq\label{1.108}\nonumber
\int^{\epsilon}_{0}R_{QR}e^{Q}_{(\alpha)}e^{R}_{(\beta)}dy=\gamma_{\alpha\beta}+H_{\alpha\beta}\epsilon\quad ,
\eeq
tomando o limite $\epsilon\rightarrow 0$
\be\label{1.109}\nonumber
\lim_{\epsilon\rightarrow 0}\left[\kappa^{2}_{5}\int^{\epsilon}_{0}\left(T_{QR}-\frac{1}{3}g_{QR}T^{M}_{M}\right)e^{Q}_{(\alpha)}e^{R}_{(\beta)}\right]=\gamma_{\alpha\beta}\quad .
\ee

Definindo o tensor energia momento da hipersuperfície $V$ por 
\be\label{1.110}\nonumber
S_{\alpha\beta}\equiv\lim_{\epsilon\rightarrow 0}\left[\int^{\epsilon}_{0}T_{QR}e^{Q}_{(\alpha)}e^{R}_{(\beta)}\right]\quad ,
\ee
podemos relacionar a descontinuidade $\gamma_{\alpha\beta}$ da curvatura extrínseca com o conteúdo de energia da brana tal como
\be\label{1.111}
\gamma_{\alpha\beta}=\kappa_{5}^{2}\left[S_{\alpha\beta}-\frac{1}{3}q_{\alpha\beta}S^{\rho}_{\rho}\right]\quad ,
\ee
que são as equações de junção de Israel. A prescrição consiste em considerar o tensor métrico contínuo ao longo da brana, mas com derivadas descontínuas. Essa descontinuidade na derivada indica a existência de uma distribução de energia na hipersuperfície. Então se integrarmos esta parte descontínua nas equações de Gauss e Codazzi termos explicitamente a relação, da mesma forma que em (\ref{1.111}), que identifica a descontinuidade na curvatura da brana com seu conteúdo de energia.  

Usando a simetria espelho $Z_{2}$ podemos calcular tanto a curvatura extrínseca em $V^{+}$ como em $V^{-}$, portanto podemos omitir tais índices. Subtituindo as equações de junção (\ref{1.111}) em (\ref{1.96}) com \be\label{112}\nonumber
S_{\mu\nu}=-\lambda q_{\mu\nu}+\tau_{\mu\nu}\quad ,
\ee
chegamos nas equações para a interação gravitacional na brana
\beq\label{1.113}\nonumber
{}^{(4)}G_{\mu\nu}&=&\frac{\kappa^{2}_{5}}{2}\left(\Lambda+\frac{\kappa^{2}_{5}}{6}\lambda^{2}\right)q_{\mu\nu}+8\pi G_{4}\tau_{\mu\nu}\\
&+&\frac{\kappa^{4}_{5}}{4}\left(\frac{1}{3}\tau\tau_{\mu\nu}+\frac{1}{4}q_{\mu\nu}\tau^{\alpha\beta}\tau_{\alpha\beta}-\tau_{\mu\alpha}\tau^{\alpha}_{\nu}-\frac{1}{6}q_{\mu\nu}\tau^{2}\right)-E_{\mu\nu}\quad ,
\eeq
com 
\be\label{1.114}\nonumber
E_{\mu\nu}={}^{(5)}C^{\alpha}_{\beta\rho\sigma}n_{\alpha}n^{\rho}q^{\beta}_{\mu}q^{\rho}_{\nu}\quad .
\ee
Observemos que as equações de Einstein usuais são obtidas se fizermos $\kappa_{5}\rightarrow 0$ mantendo $G_{4}$ constante, e que o lado esquerdo da equação (\ref{1.113}) contém termos quadráticos do tensor energia momento dos campos e matéria na brana. Estes termos são de grande importância na época em que o universo era jovem, onde as energias envolvidas eram da ordem da escala de Planck e, portanto, seus efeitos não podem ser desprezados. Além disso, temos ainda o termo $E_{\mu\nu}$, que é a parte {\it{elétrica}} do tensor de Weyl. Este termo carrega a informação sobre o campo gravitacional fora da brana. Pode-se mostrar que este termo é nulo quando o bulk é um espaço-tempo puramente AdS. Em geral, para os outros tipos de bulk a parte elétrica do tensor de Weyl será diferente de zero. Segundo \cite{maeda}, podemos relacionar tal quantidade com o conteúdo de matéria confinado na brana. 

Tomando a equação de Codazzi (\ref{1.99}) e a expressão (\ref{1.111}), chegaremos a equação de conseravção para a matéria na brana
\be\label{1.115}\nonumber
D_{\nu}K^{\nu}_{\mu}-D_{\mu}K\propto D_{\nu}\tau^{\nu}_{\mu}=0\quad ,
\ee 
além disso, das identidades de Bianchi,
\be\label{1.116}\nonumber
D^{\mu}{}^{(4)}G_{\mu\nu}=0\quad ,
\ee
implicam na relação entre $E_{\mu\nu}$ e $\tau_{\mu\nu}$
\be\label{1.117}\nonumber
D^{\mu}E_{\mu\nu}=\frac{\kappa_{5}^{4}}{4}\left[\tau^{\alpha\beta}\left(D_{\nu}\tau_{\alpha\beta}-D_{\beta}\tau_{\nu\alpha}\right)+\frac{1}{3}D^{\mu}\tau \left(\tau_{\mu\nu}-q_{\mu\nu}\tau\right)\right]\quad .
\ee
O termo $E_{\mu\nu}$ não é possível de ser determinado sem conhecermos em detalhes a geome\-tria do bulk espaço-temporal, mas sua divergência é conhecida a partir da configuração de matéria da brana. Para um observador confinado, este termo aparece como um efeito {\it{não-local}} ou {\it{fantasma}}, já que não pode ser determinado pelas informações puramente $4-$dimensionais. Esta expressão mostra qualitativamente como as variações $4-$dimensionais no conteúdo de matéria e energia confinados podem ser fontes para os modos de Kaluza-Klein. 

A equação para o campo gravitacional induzido (\ref{1.113}) apresenta duas importantes modificações das equações de Einstein $4-$dimensionais padrão. Estas correções surgem a partir dos efeitos da dimensão extra. A primeira delas  é o aparecimento de correções  no regime de altas energias, representado pelo segundo termo do lado direito de (\ref{1.113}) que é proporcional a termos quadráticos no tensor energia-momento da matéria e campos confinados. Tal termo é desprezível se a tensão da brana ou energia de vácuo $\lambda$ for muito maior que a densidade de energia $\rho$ contida na brana, mas domina o regime em que  $\lambda\ll \rho$. O segundo fator modificador é a projeção do tensor de Weyl na brana representado por $E_{\mu\nu}$. Este fator carrega os efeitos dos grávitons $5-$dimensionais. Por construção, $E_{\mu\nu}$ não tem traço e possui 9 componentes independentes que são reduzidas a 5 devido a equação (\ref{1.117}).

\section{Campos escalares no Bulk}

Pode-se mostrar que a solu\c c\~ao (\ref{1.48}) não se aplica apenas ao setor gravitacional. No trabalho \cite{walter}, os autores mostraram que se obtem a mesma equa\c c\~ao diferencial para o a parte do campo escalar que é função da dimensão extra. A proposta é estender a decomposição de Kaluza-Klein para os campos no bulk não gravitacionais. 
Consideremos a ação para um campo escalar no bulk, cuja dimensão extra é compactificada com os pontos fixos de orbifold sendo a posição das branas ,
\be\label{1.5.1}
S=\frac{1}{2}\int d^{4}x \int_{-\pi}^{\pi} dy\sqrt{-g_{5}}g^{AB}\left[\frac{\partial}{\partial x^{A}}\Phi \frac{\partial}{\partial x^{B}}\Phi-m^{2}\Phi^{2}\right]\quad ,
\ee
onde $m$ é da ordem de $M_{5}$. Explicitamente
\be\label{1.5.2}
S=\frac{1}{2}\int d^{4}x \int_{-\pi}^{\pi} dy\sqrt{-g_{5}}[g^{\mu\nu}\frac{\partial}{\partial x^{\mu}}\Phi \frac{\partial}{\partial x^{\nu}}\Phi + g^{yy}\frac{\partial}{\partial x^{\mu}}\Phi \frac{\partial}{\partial x^{\nu}}\Phi-m^{2}\Phi]\quad ,
\ee
substituindo a solução de Randall-Sundrum (\ref{1.28}), com
\be\label{1.5.3}\nonumber
\sqrt{-g_{5}}=e^{-4\sigma(y)}
\ee
temos
\be\label{1.5.4}
S=\frac{1}{2}\int d^{4}x \int_{\pi}^{\pi} dy r_{c}\left[e^{-2\sigma(y)}\eta^{\mu\nu}\frac{\partial}{\partial x^{\mu}}\Phi \frac{\partial}{\partial x^{\nu}}\Phi -\frac{1}{r_{c}^{2}}e^{-4\sigma(y)}\frac{\partial}{\partial x^{\mu}}\Phi \frac{\partial}{\partial x^{\nu}}\Phi -m^{2}e^{-4\sigma(y)}\Phi^{2}\right]\quad ,
\ee
mas
\beq\label{1.5.5}\nonumber
e^{-4\sigma(y)}\Phi\frac{\partial}{\partial y}\Phi&=&\frac{\partial}{\partial y}\left[e^{-4\sigma(y)}\Phi\frac{\partial}{\partial y}\Phi\right]
-\Phi\frac{\partial}{\partial y}\left[e^{-4\sigma(y)}\frac{\partial}{\partial y}\Phi\right]\\
\int^{\pi}_{-\pi}e^{-4\sigma(y)}\frac{\partial}{\partial y}\Phi\frac{\partial}{\partial y}\Phi dy&=&\left[e^{-4\sigma(y)}\Phi\frac{\partial}{\partial y}\Phi\right]^{\pi}_{-\pi}-\int^{\pi}_{-\pi}\Phi\frac{\partial}{\partial y}\left[e^{-4\sigma(y)}\frac{\partial}{\partial y}\phi\right]\quad ,
\eeq
e usando as condições de Israel
\be\label{1.5.6}\nonumber
\left[\frac{\partial}{\partial y}\Phi\right]_{y=\pi}=\left[\frac{\partial}{\partial y}\Phi\right]_{y=-\pi}\quad ,
\ee
temos que
\be\label{1.5.7}\nonumber
\int^{\pi}_{-\pi}e^{-4\sigma(y)}\left[\frac{\partial}{\partial y}\Phi\frac{\partial}{\partial y}\Phi\right] dy=-\int^{\pi}_{-\pi}\Phi\left[e^{-4\sigma(y)}\frac{\partial}{\partial y}\Phi\right] dy\quad ,
\ee
então podemos reescrever a ação (\ref{1.5.4}) como
\be\label{1.5.8}
S=\frac{1}{2}\int d^{4}x \int^{\pi}_{-\pi}r_{c}dy\left[e^{-2\sigma(y)}\eta^{\mu\nu}\frac{\partial}{\partial x^{\mu}}\Phi\frac{\partial}{\partial x^{\nu}}\Phi-\frac{1}{r_{c}^{2}}\Phi\frac{\partial}{\partial y}\left[e^{-4\sigma(y)}\frac{\partial}{\partial y}\Phi\right]-m^{2}e^{-4\sigma(y)}\Phi^{2}\right]\quad .
\ee

Para realizarmos a decomposição de Kaluza-Klein escrevemos $\Phi(x^{\mu},y)$ como a seguinte soma de modos
\be\label{1.5.9}
\Phi(x^{\mu},y)=\sum_{n}\psi_{n}(x^{\mu})\frac{\phi_{n}(y)}{\sqrt{r_{c}}}\quad ,
\ee
e substituindo na ação (\ref{1.5.8}) temos
considerando $\phi_{n}(y)$ com a normalização 
\be\label{1.5.11}
\int^{\pi}_{-\pi}dy e^{-2\sigma(y)}\phi_{n}\phi_{m}=\delta_{nm}\quad ,
\ee
a  equação de movimento para $\phi_{n}(y)$
\be\label{1.5.10}
-\frac{\phi_{n}}{r_{c}^{2}}\frac{d}{dy}\left[e^{-4\sigma(y)}\frac{d\phi_{n}}{dy}\right]+m^{2}e^{-4\sigma(y)}\phi_{n}^{2}=m_{n}^{2}e^{-2\sigma(y)}\phi_{n}^{2}\quad ,
\ee
disto (\ref{1.5.8}) pode ser escrita na forma
\be\label{1.5.11}
S=\frac{1}{2}\sum_{n}\int d^{4}x\left[\eta^{\mu\nu}\frac{\partial}{\partial x_{\mu}}\psi_{m}\frac{\partial}{\partial x_{\nu}}\psi_{n}-m_{n}^{2}\phi_{n}^{2}\right]\quad .
\ee
Observamos nesta última expressão que da mesma forma como ocorre compactificação de Kaluza-Klein, o campo escalar $\Phi(x^{\mu},y)$ se apresenta para um observador $4-$dimensional como um campo que apresenta uma parte totalmente $4-$dimensional mais um termo totalmente dependente da dimensão extra compactificada. A magnitude desse termo é dada por uma torre infinita de modos representados pelas {\it{massas}} $m_{n}^{2}$. Este resultado está em pleno acordo com o obtido por Randall-Sundrum para o setor perturbativo gravitacional. Fazendo a  mudança de variáveis
\beq\label{1.5.11}\nonumber
z_{n}\equiv\frac{m_{n}}{\kappa}e^{\sigma(y)}\quad ,\\\nonumber
f_{n}=e^{-2\sigma(y)}\phi_{n}\quad ,
\eeq
em (\ref{1.5.10}) chegamos a seguinte equação diferencial
\be\label{1.5.12}
z_{n}^{2}\frac{d^{2}}{dz_{n}^{2}}f_{n}+z_{n}\frac{d}{dz_{n}}f_{n}+f_{n}\left[z_{n}^{2}-\left(4+\frac{m^{2}}{\kappa^{2}}\right)\right]=0\quad ,
\ee
 cuja solução são as funções de Bessel de ordem $\upsilon=\sqrt{4+m^{2}/\kappa^{2}}$,
\be\label{1.5.13}
\phi_{n}(y)=e^{2\sigma(y)}\left[A_{n\upsilon}J_{\upsilon}\left(\frac{m_{n}}{\kappa}e^{2\sigma(y)}\right)+B_{n\upsilon}Y_{\upsilon}\left(\frac{m_{n}}{\kappa}e^{2\sigma(y)}\right)\right]\quad ,
\ee
da mesma forma que encontrada por Randall-Sundrum. Então podemos concluir que da mesma forma que no setor perturbativo gravitacional, o setor perturbativo escalar gera a mesma equação diferencial que governa a parte proveniente da dimensão extra compacta do campo escalar no bulk. Mostraremos no último capítulo que a equação (\ref{1.5.12}) também é obtida no caso em que consideramos a métrica induzida na brana sendo o espaço-tempo de Schwarzschild ou Kerr.

\section{É possível modelar buracos negros no mundo brana?}

O fato de o campo gravitacional ter acesso as dimensãoes extras, traz problemas referentes ao confinamento da matéria na brana e consequentemente ao aparecimento de condições favoráveis a formação de objetos compactos, que eventualmente levariam a um colapso gravitacional e a formação de buracos negros. Além disso, há o aparecimento de fenômenos não locais, vistos por um observador confinado, referente ao termo $E_{\mu\nu}$. Ainda há a dificuldade de implementar a continuidade da métrica induzida e da curvatura extrínseca através da brana.

Sendo a região exterior a uma estrela ou buraco negro, uma região de vácuo, então as seguintes equações são satisfeitas
\beq\label{1.6.1}\nonumber
R_{\mu\nu}&=&-E_{\mu\nu},\\\nonumber
R^{\mu}_{\mu}&=&E^{\mu}_{\nu}=0\quad ,\\
D^{\mu}E_{\mu\nu}&=&0\quad .
\eeq
Como o termo de Weyl $E_{\mu\nu}$ carrega correções em altas energias aos modos de Kaluza-Klein, o processo de colapso gravitacional será em geral diferente do processo tratado em Relatividade Geral $4-$dimensional.

O candidato a buraco negro no mundo brana proposto por Chamblin, Hawking e Reall \cite{hawking3} é o mais {\it{natural}}. O termo de Weyl é nulo nessa construção, que consiste em assumir que a métrica induzida na brana é a métrica de Schwarzschild, a qual se expande no bulk, através de um empilhamento de métricas induzidas através do bulk extra dimensional. Denotando a métrica induzida por $\hat{g}_{\mu\nu}$, temos que a métrica $5-$dimensional é dada por
\beq\label{1.6.2}
{}^{(5)}ds^{2}=e^{-2\kappa |y|}\hat{g}_{\mu\nu}dx^{\mu}dx^{\nu}+dy^{2}\quad ,
\eeq
sendo
\beq\label{1.6.3}
\hat{g}_{\mu\nu}=e^{2\kappa |y|}g_{\mu\nu}=-\left(1-\frac{2GM}{r}\right)dt^{2}+\left(1-\frac{2GM}{r}\right)^{-1}dr^{2}+r^{2}\left(\sin^{2}\theta d\phi^{2} + d\theta^{2}\right)\quad .
\eeq

De fato, se a métrica $\hat{g}_{\mu\nu}$ for solução das equações de Einstein $4-$dimensionais, isto é, se tivermos $\hat{R}_{\mu\nu}=0$, a equação (\ref{1.6.3}) será solução da equação ${}^{(5)}G_{AB}=-\Lambda_{5}{}^{(5)}g_{AB}$. Um observador no bulk verá uma singularidade em forma de linha perpendicular a brana para todos valores de $y$. Cada superfície $y=constante$ consiste no espaço-tempo de Schwarzschild $4-$dimensional. Esta solução é conhecida como {\it{corda negra de Schwarzschild}}, e como foi mostrado em \cite{hawking3} não está localizada na brana $y=0$. Como dissemos, o termo de Weyl nesta solução é nulo, o que leva a violação das correções em altas energias. O quadrado do tensor de curvatura para esse caso é 
\beq\label{1.6.4}
{}^{(5)}R_{ABCD}{}^{(5)}R^{ABCD}=40\kappa^{2}+\frac{48G^{2}M^{2}}{r^{6}}e^{4|y|\kappa}\quad ,
\eeq
que diverge no horizonte AdS $z=1/\kappa e^{y\kappa}=\infty$.

A corda negra ainda sofre da chamada {\it{instabilidade de Gregory-Laflame}} \cite{gregory1}, que torna o sistema instável para distâncias próximas do horizonte AdS, sendo estável apenas em regiões próximas a brana. No trabalho \cite{hawking3}, os autores conjecturaram que o horizonte de eventos da corda negra deve se fechar, formando algo chamado de {\it{cigarro negro}}, que não sofre instabilidades, sendo, segundo a conjectura, a única solução estável em $5$ dimensões que descreve o ponto final de um colapso gravitacional na brana. O fato é que essa solução não é localizada na brana e, portanto, falha na descrição de buracos negros realistas na brana.

Já que a solução mais óbvia falha, o próximo passo dado na busca da descrição desses objetos é considerar soluções em que $E_{\mu\nu}$ seja não nulo, isto foi feito em \cite{maartens2}, mas nesse caso só há o conhecimento da métrica induzida, não foi encontrada uma métrica que descreva esta objeto no bulk. 
Não há soluções tipo buraco negro localizados na brana. Sem essa característica, não podemos modelar buracos negros astrofísicos.  
 
Há um caso especial \cite{horowitz1} em que foi possível encontrar uma solução de buraco negro localizada. Tal solução é encontrada considerando uma $2-$brana imersa num bulk $4-$dimensional.

Pelo fato do termo de Weyl carregar a assinatura da dimensão extra, é de se es\-perar que no estado final do processo realista de colapso gravitacional na brana, apareça esta assinatura. Se isto acontecer, o buraco negro formado deve ter um {\it{cabelo extra}}, contrastando com o teorema do não-cabelo da Relatividade Geral em 4 dimensões.  

O que podemos concluir sobre a modelagem de buracos negros no mudo brana é que o assunto ainda está em aberto. Há muitos problemas referentes à estabilidade das soluções e localização, bem como um bom entendimento do processo de colapso gravitacional.


\chapter{Espaço-tempo de Kerr-Randall-Sundrum}

Neste capítulo apresentaremos alguns resultados encontrados no estudo da evolução de uma perturbação escalar no espaço-tempo de Kerr-Randall-Sundrum, que consiste em considerar, de forma análoga a \cite{hawking3}, a métrica induzida na $3-$brana como a solução de Kerr, imersa num bulk $5-$dimensional.

\section{Buracos Negros em rotação no mundo brana}

A extensão das soluções de Schwarzschild e Kerr da Relatividade Geral $4-$dimensional para $n$ dimensões foi realizada no trabalho de Meyers {\it{et al}} \cite{meyers}. Para o caso de Kerr, a métrica, considerando apenas um parâmetro de rotação, é dada por
\beq\label{3.1}\nonumber
ds^{2}_{n-Kerr}&=&-\left(\frac{\Delta - a^{2}\sin^{2}\theta}{\Sigma}\right)dt^{2}-\frac{2a\left(r^{2}+a^{2}-\Delta\right)}{\Sigma}dtd\phi\\\nonumber
&+&\left[\frac{\left(r^{2}+a^{2}\right)^{2}-\Delta a^{2}\sin^{2}\theta}{\Sigma}\right]\sin^{2}\theta d\phi^{2}+\frac{\Sigma}{\Delta}dr^{2}\\
&+&\Sigma d\theta^{2}+r^{2}\cos^{2}\theta d\Omega^{2}_{n}\quad ,
\eeq
sendo $\Sigma=r^{2}+a^{2}\cos^{2}\theta$, $\Delta=r^{2}+a^{2}-\mu r^{1-n}$ e $d\Omega^{2}$ a métrica da esfera unitária $n-$dimensional. A métrica (\ref{3.1}) descreve um buraco negro assintóticamente plano, no vácuo com massa e momento angular proporcionais a $\mu$ e $\mu a$, respectivamente, e supondo-se $\mu,a > 0$.

O horizonte de eventos é localizado em $r=r_{H}$. Se $n=1$, o horizonte de eventos existe apenas para $a< (\mu)^{\frac{1}{2}}$, com área  que se reduz a zero no limite do buraco negro de Kerr extremo, isto é , para o limite $a\rightarrow (\mu)^{\frac{1}{2}}$. No caso $n \geq 2$, a condição para o horizonte de eventos $\Delta=0$ possui apenas uma raíz exata para um valor arbitrário $a>0$. Não há buraco negro de Kerr extremo para $n \geq 2$.

Sengupta \cite{sengupta} utilizou a generalização (\ref{3.1}) como um mundo brana de $(n+1)$ dimensões no contexto de RS, considerando esta métrica como uma corda negra com rotação não nula que intercepta a brana de $(n-1)$ dimensões numa métrica que representa uma buraco negro em rotação. 

Em \cite{modgil1} foi considerado, de maneira análoga ao método de \cite{hawking3}, a métrica (\ref{3.1}) com $n=1$ sendo a métrica que representa a intersecção de uma brana com uma corda negra em rotação $5-$dimensional num bulk de RS. A intersecção descreve o buraco negro de Kerr $4-$dimensional. A métrica para esta corda negra em rotação é dada por
\beq\label{3.2}\nonumber
ds^{2}_{5-Kerr}&=&\frac{l^{2}}{z^{2}}\bigg\{-\left(\frac{\Delta-a^{2}\sin^{2}\theta}{\Sigma}\right)dt^{2}+\frac{\Sigma}{\Delta}dr^{2} +\Sigma d\theta^{2}\\
&+& \frac{4aMr\sin^{2}\theta}{\Sigma}dt d\phi +\left[\frac{\left(r^{2}+a^{2}\right)^{2}-\Delta a^{2}\sin^{2}\theta}{\Sigma}\right]\sin^{2}\theta d\phi^{2} + dz^{2}\bigg\}\quad ,
\eeq
com $z=le^{y\kappa}$ e $\Delta=r^{2}+a^{2}-2Mr$. Introduzindo a coordenada $\omega=z-z_{0}$, a brana estará localizada em $\omega=0$, com
\beq\label{3.3}\nonumber
ds^{2}_{5-Kerr}&=&\frac{l^{2}}{\left(|\omega|+z_{0}\right)^{2}}\bigg\{-\left(\frac{\Delta-a^{2}\sin^{2}\theta}{\Sigma}\right)dt^{2}+\frac{\Sigma}{\Delta}dr^{2} +\Sigma d\theta^{2}\\
&+& \frac{4aMr\sin^{2}\theta}{\Sigma}dt d\phi +\left[\frac{\left(r^{2}+a^{2}\right)^{2}-\Delta a^{2}\sin^{2}\theta}{\Sigma}\right]\sin^{2}\theta d\phi^{2} + d\omega^{2}\bigg\}\quad .
\eeq

O quadrado do tensor de Riemann para esta métrica é dado por \cite{modgil1}
\beq\label{3.4}
R_{ABCD}R^{ABCD}=40\kappa^{4}+\frac{48M^{2}z^{4}\kappa^{4}}{\Sigma^{6}}\left(r^{2}-a^{2}\cos^{2}\theta\right)\left(r^{2}-14a^{2}r^{2}\cos\theta +a^{4}\cos^{4}\theta\right)\quad .
\eeq
Desta equação podemos observar que para $r=0$ e $\theta=\pi/2$ obtemos a singularidade em forma de anel característica da solução de Kerr. Da mesma forma que na tentativa de modelar buracos negros em branas através da métrica de Schwarzschild, produz uma corda negra que é instável nas proximidades do horizonte AdS e sofre da instabilidade de Gregory-Laflamme \cite{hawking3}, neste caso com rotação ocorre o mesmo, sendo este sistema instável a distância próximas do horizonte AdS. 


\section{Perturbação escalar em Kerr-RS}

Apesar da métrica (\ref{3.3}) ser uma solução instável próxima ao horizonte AdS,  procuraremos nesta seção estudar a evolução de um campo escalar neste espaço-tempo. O campo escalar será considerado como um campo de teste, no sentido de que através do estudo de sua evolução poderemos inferir, talvez, sobre algumas propriedades dessa geometria (\ref{3.3}), que passaremos a chamar de Kerr-Randall-Sundrum (Kerr-RS).

Escrevendo (\ref{3.2}) como
\beq\label{3.5}
ds^{2}_{Kerr-RS}=f(z)\left[ds^{2}_{4-Kerr}+dz^{2}\right]\quad ,
\eeq
com $f(z)=l^{2}/z^{2}$, e 
\beq\label{3.6}\nonumber
ds^{2}_{4-Kerr}&=&-\left(1-\frac{2Mr}{\Sigma^{2}}\right)dt^{2}+\frac{\Sigma^{2}}{\Delta}dr^{2} +\Sigma^{2}d\theta^{2}-\frac{4Mr}{\Sigma}a\sin^{2}\theta d\phi dt\\&+&\left[\left(r^{2}+a^{2}\right)\sin^{2}\theta +\frac{2Mr}{\Sigma^{2}}a^{2}\sin^{2}\theta\right]d\phi^{2}\quad .
\eeq

As componentes contravariantes dessa métrica serão
\beq\label{3.7}
g^{00}&=&-\frac{1}{f(z)\Sigma^{2}}\left[\Delta^{-1}(r^{2}+a^{2})^{2}-a^{2}\sin^{2}\theta\right]\quad ,\\\nonumber
g^{03}&=&-\frac{1}{f(z)\Sigma^{2}}\left(\frac{4Mra}{\Delta}\right),\\\nonumber
g^{11}&=&\frac{\Delta}{f(z)\Sigma^{2}}\quad ,\\\nonumber
g^{22}&=&\frac{1}{f(z)\Sigma^{2}}\quad ,\\\nonumber
g^{33}&=&\frac{1}{f(z)\Sigma^{2}}\left(\frac{1}{\sin^{2}\theta}-\frac{a^{2}}{\Delta}\right)\quad ,\\
g^{44}&=&\frac{1}{f(z)}\quad ,
\eeq
que no caso $z=l$ se reduzem às componentes do caso padrão \cite{brill1}.
O determinante de (\ref{3.6}) é dado por
\beq\label{3.8}
det\left(g_{AB}\right)\equiv G=-f^{5}(z)\Sigma^{4}\sin^{2}\theta\quad .
\eeq

A equação que dá a dinâmica do campo escalar $\Phi(x^{\mu},z)$ ($x^{\mu}=t,r,\theta,\phi$) com massa $\mu^{2}$ é a equação de Klein-Gordon
\beq\label{3.9}
\frac{1}{\sqrt{-G}}\frac{\partial}{\partial
x^{A}}\left(g^{AB}\sqrt{-G}\frac{\partial\Phi}{\partial
x^{B}}\right)+\mu^{2}\Phi=0\quad .
\eeq
Tal equação, usando os resultados (\ref{3.7}) e (\ref{3.8}), assume a seguinte forma expandida
\beq\label{3.10}\nonumber
&-&\frac{1}{f\Sigma^{2}}\left[\frac{(r^{2}+a^{2})^{2}}{\Delta}-a^{2}\sin^{2}\theta\right]
\frac{\partial^{2}\Phi}{\partial t^{2}}-\frac{4Mra}{\Delta
f\Sigma^{2}}\frac{\partial^{2}\phi}{\partial t
\partial\phi}+\\\nonumber
&+&\frac{1}{f^{5/2}\Sigma^{2}\sin\theta}\frac{\partial}{\partial
r}\left(f^{5/2}\Sigma^{2}\sin\theta\frac{\Delta}{f\rho^{2}}\frac{\partial\Phi}{\partial
r}\right)+\frac{1}{f^{5/2}\Sigma^{2}\sin\theta}\frac{\partial}{\partial
\theta}\left(f^{5/2}\Sigma^{2}\sin\theta\frac{\Delta}{f\Sigma^{2}}\frac{\partial\Phi}{\partial
\theta}\right)+\\\nonumber
&+&\frac{1}{f\Sigma^{2}}\left(\frac{1}{\sin^{2}\theta}-\frac{a^{2}}{\Delta}\right)
\frac{\partial^{2}\Phi}{\partial\phi^{2}}+\frac{1}{f^{5/2}}\frac{\partial}{\partial z}\left(f^{3/2}\frac{\partial\Phi}{\partial
z}\right)+\mu^{2}\Phi=0\quad .
\eeq

Analogamente à separação de variáveis usada em \cite{brill1}, podemos escrever o campo $\Phi$ em termos dos harmônicos apropriados às condições de simetria
\beq\label{3.11}
\Phi(x^{\mu},z)=R(r)S(\theta)Y(z)e^{im\phi}e^{-i\omega t}\quad .
\eeq
Substituindo este resultado em (\ref{3.10}) e em seguida multiplicando os dois lados da equação resultante por $\Phi^{-1}(x^{\mu},z)$
\beq\label{3.12}\nonumber
&&\frac{1}{\Sigma^{2}}\left[\frac{(r^{2}+a^{2})^{2}}{\Delta}-a^{2}\sin^{2}\theta\right]\omega^{2}-
\frac{4Mra}{\Delta\Sigma^{2}}(m\omega)+\frac{1}{R(r)\Sigma^{2}}\frac{d}{dr}\left(\Delta\frac{dR(r)}{dr}
\right)\\\nonumber
&&+\frac{1}{\Sigma^{2}\sin\theta
S(\theta)}\frac{d}{d\theta}\left(\sin\theta\frac{dS(\theta)}{d\theta}\right)+\frac{1}{Y(z)f^{3/2}
}\frac{d}{dz}\left(f^{3/2}\frac{dY(z)}{dz}\right)\\\nonumber
&-&m^{2}\left(\frac{1}{\sin^{2}\theta}-\frac{a^{2}}{\Delta}
\right)\frac{1}{\Sigma^{2}}+\mu^{2}f=0\quad .
\eeq
Agora, podemos separar esta equação em duas outra equações: uma dependente somente da coordenada da dimensão extra, a coordenada $z$, e outra equação dependente das coordenadas $x^{\mu}$, sendo $Q$ a constante de separação,
\beq\label{3.13}
&&\frac{1}{\Sigma^{2}}\left[\frac{(r^{2}+a^{2})^{2}}{\Delta}-a^{2}\sin^{2}\theta\right]\omega^{2}-
\frac{4Mra}{\Delta\Sigma^{2}}(m\omega)+\frac{1}{R(r)\Sigma^{2}}\frac{d}{dr}\left(\Delta\frac{dR(r)}{dr}
\right)\\\nonumber
&&+\frac{1}{\Sigma^{2}\sin\theta
S(\theta)}\frac{d}{d\theta}\left(\sin\theta\frac{dS(\theta)}{d\theta}\right)
-m^{2}\left(\frac{1}{\sin^{2}}-\frac{a^{2}}{\Delta}
\right)\frac{1}{\Sigma^{2}}-Q=0\quad ,\\
\label{3.14}
&&\frac{d}{dz}\left(f^{3/2}\frac{dY(z)}{dz}\right)+QY(z)f^{3/2}+\mu^{2}f=0\quad .
\eeq

Separando a parte radial da parte angular na equação (\ref{3.13}), obtemos duas equações diferenciais ordinárias
\beq\label{3.15}
&&\frac{1}{R}\frac{d}{dr}\left(\Delta\frac{dR}{dr}\right)+\frac{(r^{2}+a^{2})^{2}}{\Delta}\omega^{2}
-a^{2}\omega^{2}-\frac{4Mra}{\Delta}m\omega
-\frac{(ma)^{2}}{\Delta}-Qr^{2}-P=0\quad ,\\
\label{3.16}
&&\frac{1}{\sin\theta
S(\theta)}\frac{d}{d\theta}\left(\sin\theta\frac{dS(\theta)}{d\theta}\right)+a^{2}\cos^{2}\theta-
\frac{m^{2}}{\sin^{2}\theta}+P=0\quad ,
\eeq
sendo $P$ a constante de separação. 

No fim do processo de separação de variáveis, obtemos três equações diferenciais ordinárias, com a dependência do campo na coordenada $z$ separada da dependência nas coordenadas $x^{\mu}$ na $3-$brana,
\beq\label{3.17}\nonumber
&&\frac{d}{dr}\left(\Delta\frac{dR}{dr}\right)+\frac{R}{\Delta}\left[(r^{2}+a^{2})^{2}-4Mram\omega 
+(ma)^{2}\right]\\
&&-{R}\left((a\omega)^{2}+Qr^{2}+P\right)=0\quad ,\\\label{3.18}
&&\frac{1}{\sin\theta}\frac{d}{d\theta}\left(\sin\theta\frac{dS(\theta)}{d\theta}\right)+S(\theta)\left[
(\omega^{2}-Q)a^{2}\cos^{2}\theta-\frac{m^{2}}{\sin^{2}\theta}+P\right]=0\quad ,\\\label{3.19}
&&\frac{d}{dz}\left(f^{3/2}\frac{dY(z)}{dz}\right)+QY(z)f^{3/2}+\mu^{2}f=0\quad .
\eeq

A equação radial (\ref{3.17}) pode ser reescrita utiliado a coordenada tartaruga $r_{*}$
\beq\label{3.20}
\frac{dr_{*}}{dr}=\frac{r^{2}+a^{2}}{\Delta}\quad ,
\eeq
e a definição
\beq\label{3.21}
\Psi(r)=\sqrt{r^{2}+a^{2}}R(r)\quad ,
\eeq
na forma
\beq\label{3.22}
\frac{d^{2}\Psi(r)}{dr_{*}^{2}}+\Psi(r)\left[\omega^{2}-V(r,\omega)\right]=0\quad ,
\eeq
sendo, para o campo sem massa,
\beq\label{3.22}\nonumber
V(r,\omega)&=&\frac{4Mra(m\omega)-(ma)^{2}-\Delta((a\omega)^{2}
+Qr^{2}+P)}{(r^{2}+a^{2})^{2}}\\
&+&\frac{\Delta(3r^{2}-4Mr+a^{2})}{(r^{2}+a^{2})^{3}}-\frac{3\Delta^{2}r^{2}}
{(r^{2}+a^{2})^{4}}\quad .
\eeq

A solução da equação angular (\ref{3.18}) são os harmônicos esferoidais oblatos  $S^{m}_{Ln}\left(a^{2}c^{2}_{n},\cos\theta\right)$ com $c^{2}_{n}=\omega^{2}-Q^{2}_{n}$. Estes harmônicos , no caso $a=0$, se reduzem aos harmônicos esféricos e o autovalor $P\rightarrow P_{Lmn}$ se torna $L(L+1)$. 

No caso geral, os autovalores $ P_{Lmn}$ s\~ao determinados formalmente por uma expans\~ao em pot\^encias de $(ac_{n})$ \cite{brill1}, tal como
\beq\label{3.23}
P_{lmn}=\sum_{i}f_{(2i)}^{lm}(ac_{n})^{2i}\quad ,
\eeq
 onde os coeficientes da expans\~ao s\~ao fun\c c\~oes de $L$ e $m$ apenas, os dois primeiros termos s\~ao dados por 
 \[
 f_{0}^{lm}=l(l+1)\quad ,
 \]
 
 \[
 f^{lm}_{2}=h(l+1,m)-h(l,m)-1\quad ,
 \]
 com
 \[
 h(l,m)=\frac{l(l-m)(l+m)}{2(l-1/2)(l+1/2)}\quad .
 \]

A dinâmica do campo escalar no bulk é dado pela equação (\ref{3.19}), que pode ser reescrita na forma
\beq\label{3.24}
\frac{d^{2}}{dz^{2}}Y(z)-\frac{3}{z}\frac{d}{dz}Y(z)+Y(z)\left[Q+\frac{k^{2}}{z^{2}}\right]=0\quad ,
\eeq
com $k^{2}=\mu^{2}l^{2}$.

Esta equa\c c\~ao \'e id\^entica \`a equa\c c\~ao de Bessel,

\beq\label{3.25}
\frac{d^{2}}{dx^{2}}W(x)+\frac{1}{x}\frac{dW(x)}{dx}+W(x)\left[1-\frac{m^{2}}{x^{2}}\right]=0\quad .
\eeq

A solu\c c\~ao para a equa\c c\~ao (\ref{3.24}) \'e
\beq\label{3.26}
Y(z)=z^{2}\left[A\mathcal{J}_{\sqrt{4-k^{2}}}(Qz)+B\mathcal{Y}_{\sqrt{4-k^{2}}}(Qz)\right]\quad ,
\eeq
sendo $A$ e $B$ constantes, e

$\mathcal{J}_{\sqrt{4-k^{2}}}=$ Função de Bessel do primeiro tipo de ordem $\sqrt{4-k^{2}}$,

$\mathcal{Y}_{\sqrt{4-k^{2}}}=$ Função de Bessel do segundo tipo de ordem $\sqrt{4-k^{2}}$.

Considerando o modelo RS com duas $3-$branas, as condições de contorno para o campo $Y(z)$ vem da continuidade da derivada primeira deste campo nos pontos onde se localizam as branas, que, neste caso, se localizam em $z=l (y=0)$ e $z=le^{d\kappa} (y=d)$, sendo $d$ a distância de separação entre as branas. Então, denotando a derivada em relação a $z$ por $Y'$, devemos ter \cite{walter}
\beq\label{3.27}
Y'(z)|_{z=l}=Y'(z)|_{z=le^{d\kappa}}=0\quad .
\eeq
Calculando a derivada primeira de $Y(z)$ temos
\beq\label{3.28}
Y'(z)=2z\left[A\mathcal{J}_{\gamma}(Qz)+B\mathcal{Y}_{\gamma}(Qz)\right]+z^{2}\left[A\mathcal{J}'_{\gamma}(Qz)+B\mathcal{Y}'_{\gamma}(Ql)\right]\quad ,
\eeq
com $\gamma=\sqrt{4-k^{2}}$. Em $z=l$ teremos
\beq\label{3.29}
Y'(z=l)=A\left[2\mathcal{J}_{\gamma}(Ql)+l\mathcal{J}'_{\gamma}(Ql)\right]+B\left[2\mathcal{Y}_{\gamma}(Ql)+l\mathcal{Y}'_{\gamma}(Ql)\right]\quad .
\eeq
Lan\c cando m\~ao das propriedades
\[
xJ'(x)=xJ_{n-1}-nJ_{n}(x)\quad ,
\]
\[
xY'(x)=xY_{n-1}-nY_{n}(x)\quad ,
\]
teremos
\beq\label{3.30}\nonumber
A\left[2\mathcal{J}_{\gamma}(Ql)+lQ\mathcal{J}_{\gamma-1}(Ql)-\gamma \mathcal{J}_{\gamma}(Ql) \right]+B\left[2\mathcal{Y}_{\gamma}(Ql)+lQ\mathcal{Y}_{\gamma-1}(Ql)-\gamma \mathcal{Y}_{\gamma}(Ql)\right]=0\quad ,
\eeq
\beq\label{3.31}
\frac{A}{B}=-\frac{\left[(2-\gamma)\mathcal{Y}_{\gamma}(Ql)+lQ\mathcal{Y}_{\gamma-1}(Ql)\right]}{\left[(2-\gamma)\mathcal{J}_{\gamma}(Ql)+lQ\mathcal{J}_{\gamma-1}(Ql)\right]}
\eeq
Em particular, se $\gamma=2$, isto \'e, para o campo escalar n\~ao massivo $\mu^{2}=0$, teremos
\beq\label{3.32}
A=-\frac{\mathcal{Y}_{1}(Ql)}{\mathcal{J}_{1}(Ql)}\quad ,
\eeq
ent\~ao
\beq\label{3.33}
Y(z)_{Q}=z^{2}B\left[-\frac{\mathcal{Y}_{1}(Ql)}{\mathcal{J}_{1}(Ql)}\mathcal{J}_{2}(Qy)+\mathcal{Y}_{2}(Qy)\right]\quad .
\eeq
Para o ponto onde se localiza a segunda brana, temos
\beq\label{3.34}\nonumber
Y'(z=le^{d\kappa})&=&2le^{d\kappa}\left[A\mathcal{J}_{\gamma}(Qle^{d\kappa})+\mathcal{Y}_{\gamma}(Qle^{d\kappa})\right]\\
&+&l^{2}e^{2d\kappa}\left[A\mathcal{J}'_{\gamma}(Qle^{d\kappa})
+\mathcal{Y}'_{\gamma}(Qle^{d\kappa})\right]=0\quad ,
\eeq

\beq\label{3.35}\nonumber
A\left[(2-\gamma)\mathcal{J}_{\gamma}(Qle^{d\kappa})+lQe^{d\kappa}\mathcal{J}_{\gamma-1}(Qle^{d\kappa})\right]+B\left[(2-\gamma)\mathcal{Y}_{\gamma}(Qle^{d\kappa})+lQe^{d\kappa}\mathcal{Y}_{\gamma-1}(Qle^{d\kappa})\right]=0\quad.
\eeq

Isolando $A$, teremos
\beq\label{3.36}
A=-B\frac{\left[(2-\gamma)\mathcal{Y}_{\gamma}(Qle^{d\kappa})+lQe^{d\kappa}\mathcal{Y}_{\gamma-1}(Qle^{d\kappa})\right]}{\left[(2-\gamma)\mathcal{J}_{\gamma}(Qle^{d\kappa})+lQe^{d\kappa}\mathcal{J}_{\gamma-1}(Qle^{d\kappa})\right]}\quad ,
\eeq

Igualando (\ref{3.36}) e (\ref{3.31}) temos,
\beq\label{3.37}
\frac{\left[(2-\gamma)\mathcal{Y}_{\gamma}(Qle^{d\kappa})+lQe^{d\kappa}\mathcal{Y}_{\gamma-1}(Qle^{d\kappa})\right]}{\left[(2-\gamma)\mathcal{J}_{\gamma}(Qle^{d\kappa})+lQe^{d\kappa}\mathcal{J}_{\gamma-1}(Qle^{d\kappa})\right]}=
\frac{\left[(2-\gamma)\mathcal{Y}_{\gamma}(Ql)+lQ\mathcal{Y}_{\gamma-1}(Ql)\right]}{\left[(2-\gamma)\mathcal{J}_{\gamma}(Ql)+lQ\mathcal{J}_{\gamma-1}(Ql)\right]}\quad .
\eeq
O autovalor $Q$ depende das raízes das funções de Bessel, portanto ele será, no caso do modelo de duas branas, discretizado, ou seja, $Q\rightarrow Q_{n}$. Fazendo as substituições $\kappa=1/l$, $Q_{n}e^{d/l}=x_{n}/l$ e particularizando para o caso $\gamma=2$, que representa o campo escalar não massivo, podemos reescrever (\ref{3.37}) como
\beq\label{3.38}
\mathcal{Y}_{1}(Q_{n})\mathcal{J}_{1}(x_{n})=\mathcal{Y}_{1}(x_{n})\mathcal{J}_{1}(Q_{n}l)\quad .
\eeq
Esta equação pode ser resolvida numericamente. Como o número de raízes das funções de Bessel é infinito, o número de autovalores $Q_{n}$, que chamaremos de modos massivos, pois {\it{emprestam}} um caráter de campo massivo ao campo escalar com $\gamma=2$; também será infinito. Nos limitamos a mostrar os dez primeiros modos $Q_{n}$ na tabela (\ref{modos massivos do bulk}).
\begin{table}[h]\label{modos massivos do bulk}
\begin{center}
\begin{tabular}{|l|l||l|l|}
\hline
$n$ & $Q_{n}$ & $n$ & $Q_{n}$\\
\hline
1 & 3.83171 & 6 & 19.6159  \\
\hline
2 & 7.01559  &  7& 22.7601\\
\hline
3 &10.1735 & 8& 25.9037  \\
\hline
4 & 13.3237  & 9 & 29.0468                    \\
\hline
5 &  16.4706  & 10 &    32.1897               \\
\hline
\end{tabular}\caption{Primeiros dez modos massivos $Q_{n}$.}
\end{center}
\end{table}
Substituindo os valores $Q_{n}$ e $P_{Lmn}$ no potencial efetivo (\ref{3.22}), teremos o potencial visto por um observador confinado na $3-$brana, que a este parecerá como o potencial efetivo de um campo escalar puramente $4-$dimensional {\it{com massa}}. De fato, a dimensão extra simula um termo de massa para o campo escalar, o que é esperado, devido a massividade dos modos de Kaluza-Klein \cite{maartens1}. 

Buracos negros em rotação são instáveis a perturbação escalar massiva \cite{detweiler1}. O que ocorre é a formação de um sistema altamente explosivo, que foi chamado de buraco negro bomba \cite{strafuss}. Esta instabilidade vem do fato que a massa do campo escalar age como se fosse um espelho, e se somarmos a isso o efeito do espalhamento super-radiante, que ocorre em buracos negros em rotação, teremos a amplitude da perturbação escalar crescendo indefinidamente \cite{cardoso1}. Este cenário é o que obtemos no caso de uma perturbação escalar não massiva no espaço-tempo de Kerr-RS. O cálculo das frequências quasi-normais para este caso fica impossibilitada, pois este sistema não nos parece fisicamente aceitável. Apesar disso é interesante calcular os coeficientes de transmissão e reflexão do espalhamento do campo escalar em Kerr-RS. Seguiremos de perto os traba\-lhos de Starobinski {\it{et al}} \cite{starobinski1}\cite{starobinski2}.

O perfil assintótico do termo potencial da equação (\ref{3.22}) é dado por
\beq\label{3.38}
\omega^{2}-V&\rightarrow& \omega^{2}-Q_{n}^{2}, \hspace{0.3cm} r_{*}\rightarrow +\infty\quad ,\\\label{3.39}
\omega^{2}-V&\rightarrow& \left(\omega-m\Omega_{H}\right)^{2}, \hspace{0.3cm} r_{*}\rightarrow -\infty\quad ,
\eeq
sendo
\beq\label{3.40}
\Omega_{H}=\frac{a}{2Mr_{+}}, \hspace{0.3cm} r_{+}=M+\left(M^{2}-a^{2}\right)^{\frac{1}{2}}\quad .
\eeq

Da equação (\ref{3.38}), vemos que quando $Q_{n}$ é maior que $\omega^{2}$ o potencial efetivo fica com sinal negativo no infinito espacial tornando a perturbação escalar, neste regime, instável. Se nos restringirmos ao caso $Q_{n}<\omega^{2}_{qn}$, sendo $\omega^{2}_{qn}$ a frequência quasi-normal fundamental, as soluções assintóticas da equação de onda (\ref{3.22}) tem a seguinte forma,
\beq\label{3.41}
\Psi(r_{*}\rightarrow -\infty)&\rightarrow& B_{Lm}e^{-i(\omega-m\Omega_{H})r_{*}}\quad ,\\\label{3.42}
\Psi(r_{*}\rightarrow +\infty)&\rightarrow& e^{-i(\omega^{2}-Q_{n}^{2})^{\frac{1}{2}}r_{*}}+A_{Lm}e^{i(\omega^{2}-Q_{n}^{2})^{\frac{1}{2}}r_{*}}\quad .
\eeq
Esta configuração exprime o fato de que consideraremos apenas ondas entrando no horizonte. A onda vem do infinito espacial, parcialmente atravessa a barreira de potencial entrando no horizonte de eventos, o resto é refletido para o infinito. 

Da constância do Wronskiano da equação (\ref{3.22}) com as soluções assintóticas (\ref{3.41}) e (\ref{3.42}), temos que
\beq\label{3.43}
1-|A_{Lm}|^{2}-|B_{Lm}|^{2}\frac{\omega-m\Omega_{H}}{\sqrt{\omega^{2}-Q_{n}^{2}}}=0\quad .
\eeq
Isto mostra, que da mesma forma que no caso da solução de Kerr padrão, $|A_{Lm}|>1$, ou seja, a amplitude da onda refletida é maior que da onda incidente se tivermos
\beq\label{3.44}
m\Omega_{H}>\omega\quad ,
\eeq
que é a condição de espalhamento super-radiante.

Podemos então, partir para o cálculo do coeficiente de reflexão da onda. Vamos supor que $1/\omega\gg M$, isto é, que o comprimento Compton da partícula escalar seja maior que o tamanho típico do buraco negro. Seguindo \cite{starobinski1}, dividimos o espaço-tempo exterior ao horizonte de eventos em duas regiões, a saber, região próxima, $r-r_{+}\ll 1/\omega$, e região distante, $r-r_{+}\gg M$. Resolvendo a equação de onda (\ref{3.22}) nestas duas regiões, e igualando os resultados onde a região distante e a região próxima se interceptam, isto é, na região $M\ll r-r_{+}\ll 1/\omega$, encontraremos o coeficiente de reflexão.

Na região próxima, podemos aproximar (\ref{3.22}) por
 \beq\label{3.45}
\Delta\frac{d}{dr}\left(\Delta\frac{dR(r)}{dr}\right)
+[r_{+}^{4} (\omega - m \Omega_{H})^{2} - L (L+ 1) \Delta] R(r)=0\quad ,
\eeq
cuja solução geral é dada por \cite{starobinski1}\cite{cardoso1} 
\beq\label{3.46}
R = A z ^{-i \chi} (1-z )^{L+1} F(a-c+1, b-c+1, 2-c, z)+ B z ^{i \chi}
(1-z )^{L+1} F(a, b, c, z)\quad , 
\eeq
sendo $F(a, b, c, z)$ e $ F(a-c+1, b-c+1, 2-c, z)$ funções hipergeométricas,
\beq\label{3.47}
\chi = (\omega - m \Omega-{H}) \frac{r_{+}^{2}} {r_{+}-r_{-}}\quad ,
\eeq
e
\beq\label{3.48}
z= \frac{r - r_{+}}{r - r_{-}}\quad .
\eeq
Segundo Starobinski \cite{starobinski1} para grandes valores de $r$  a solução acima tem a seguinte forma assintótica 
\beq\label{3.49}
R \sim A \Gamma (1- 2 i \chi) \left[\frac{(r_{+}-r_{-})^{-L} \Gamma (2
L + 1) r^{L}}{\Gamma (L+1) \Gamma (L+1- 2 i \chi) }+  \frac{(r_{+}-r_{-})^{L+1} \Gamma (-2
L - 1) r^{-L-1}}{\Gamma (-L) \Gamma (-L- 2 i \chi) }\right]\quad .
\eeq
Na região distante $r - r_{+} \gg M$ a equação de onda possui a forma
\beq\label{3.50}
\frac{d^2}{dr^2} (r R) + [\omega^2 - Q^{2}_{n} - \frac{L (L+1)} {r^2}]=0\quad ,
\eeq
com solução geral dada pela seguinte combinação de funções de Bessel
\beq\label{3.51}
R \sim  r^{-1/2} [a \mathcal{J}_{L + 1/2} (\sqrt{\omega^2 - Q^{2}_{n}}) + b \mathcal{J}_{-L - 1/2}(\sqrt{\omega^2 - Q^{2}_{n}})]\quad .
\eeq
Para pequeno valores de $r$ esta solução toma forma
\beq\label{3.52}
R \sim a \frac{(\sqrt{\omega^2 - Q^{2}_{n}}/2)^{L + 1/2} r^{L}}{\Gamma(L +
3/2)} + 
b \frac{(\sqrt{\omega^2 - Q^{2}_{n}}/2)^{-L - 1/2} r^{-L-1}}{\Gamma(-L + 1/2)}\quad .
\eeq

Combinando as soluções da região próxima (\ref{3.49}) e da região distante (\ref{3.52}) na região de intersecção $M \ll r - r_{+}
\gg 1/ \omega$, obtemos o coeficiente de reflexão $b / a$   
\beq\label{3.53}
\frac{b} {a} = 2 i (\omega^{2}-Q^{2}_{n})^{L +1/2}  \chi \frac{(1-)^L}{2L +1} (\frac{L!}{(2 L
-1)!})^{2} \frac{(r_{+}-r_{-})^{2 L+ 1}}{(2L)! (2 L +1)!} (k^2 + 4 \chi^2)\quad .
\eeq
Este coeficiente de reflexão difere da fórmula encontrada por Starobinski \cite{starobinski1} para o caso da solução de Kerr padrão pelo fator $Q_{n}$ que aparece no termo $\left(\omega^{2}-Q^{2}_{n}\right)^{L+1/2}$. De fato, isto não é apenas uma frequência recalibrada pelo fato $Q_{n}^{2}$, pois o termo $\chi$ definido por (\ref{3.47}) contém $\omega$. Esta influência da dimensão extra, representada pelos autovalores $Q_{n}$, no espalhamento super-radiante será importante apenas para o regime em que $Q_{n}$ é da mesma ordem que $\omega$. Estes resultados foram publicados em \cite{jeferson1}.


\chapter{Conclusão}

Vimos que nos modelos de mundo brana propostos por Randall e Sundrum, em particular o modelo com duas $3-$branas em um bulk $AdS_{5}$, há a possibilidade da resolução do problema da hierarquia entre a escala gravitacional e eletrofraca. O modelo com uma única brana contém o modo zero de excitação dos grávitons, que é a gravitação de Newton. Podemos considerar estes modelos como {\it{laboratórios}}  que nos permitem testar várias idéias da conjectura da teoria M, e também questões sobre a forma da Relatividade Geral neste contexto. 

A modelagem de buracos negros, como discutido no terceiro capítulo, é um problema em aberto, e não nos parece trivial. O fato da gravidade se {\it{espalhar}} pelo bulk, enquanto os outros campos físicos ficam confinados na $3-$brana, é um complicador para o estudo do colapso gravitacional de objetos astrofísicos. Além disso, as condições de contorno são complicadas de se implementar neste caso. Também não é claro como deve ser o limite assintótico do bulk, já que próximo ao horizonte AdS ocorrem problemas com a continuidade do horizonte de eventos de um buraco negro \cite{hawking3}, além da instabilidade de Gregory-Laflamme \cite{gregory1}.

Apesar da solução de corda negra em rotação estudada por Sengupta \cite{sengupta} apresentar as mesmas limitações do cigarro negro de Hawking \cite{hawking3}, estudamos a equação de Klein-Gordon neste cenário. Mostramos que perturbações escalares de spin 0 sem massa no modelo de Kerr-Randall-Sundrum com duas branas simula a perturbação escalar massiva no espaço-tempo de Kerr $4-$dimensional. O termo massivo para o campo escalar aparece devido unicamente a dimensão extra, como foi mostrado de maneira geral por Cardoso {\it{et al}} \cite{cardoso1}. A equação que governa a dinâmica no bulk do campo escalar em Kerr-Randall-Sundrum é a mesma encontrada na decomposição de Kaluza-Klein feita por Randall-Sundrum \cite{rs1} \cite{rs2} e a encontrada para campos escalares $5-$dimensionais no modelo de Randall-Sundrum com branas de Minkowski \cite{walter}. Este termo não massivo torna o espaço-tempo de Kerr-RS instável por perturbações escalares, já que devido ao espalhamento superradiante, a radiação espalhada para o infinito é refletida de volta para o buraco negro pelo termo de massa do campo, gerando assim um buraco negro bomba.

\bibliographystyle{utphys}
\bibliography{all_new}

\end{document}